\documentclass[prl,twocolumn,superscriptaddress]{revtex4-2}
\usepackage{epsfig}
\usepackage{graphicx}
\usepackage{palatino}
\usepackage[english]{babel}
\usepackage{hyphenat}
\usepackage{amsmath}
\usepackage{amssymb}
\usepackage{mathtools}
\usepackage{mathrsfs}
\usepackage{bbm} 
\usepackage{slashed}
\usepackage{epstopdf}
\usepackage{xcolor}
\usepackage{booktabs}
\usepackage{float}
\definecolor{lcolor}{rgb}{0.,0.0,0.}
\definecolor{citcolor}{rgb}{0,0.,0.5}
\usepackage[breaklinks,colorlinks,urlcolor=blue,citecolor=blue,linkcolor=blue]{hyperref}
\usepackage{multirow}
\usepackage{ltablex}
\usepackage{soul}

\newcommand{\rmL}{\textrm{L}}
\newcommand{\rmT}{\textrm{T}}
\newcommand{\Hcal}{\mathcal{H}}

\newcommand{\Ncal}{\mathcal{N}}

\newcommand{\Scal}{\mathcal{S}}

\newcommand{\Rcal}{\mathcal{R}}
\newcommand{\Ical}{\mathcal{I}}
\newcommand{\Jcal}{\mathcal{J}}
\newcommand{\vect}[1]{\boldsymbol{#1}_{\perp}}
\newcommand{\kt}{\vect{k}}

\newcommand{\Pt}{\vect{P}}

\newcommand{\qt}{\vect{q}}
\newcommand{\lt}{\vect{l}}

\newcommand{\bt}{\vect{b}}
\newcommand{\Bt}{\vect{B}}
\newcommand{\Kt}{\vect{K}}

\newcommand{\ruupt}{\boldsymbol{r}_{uu'}}
\newcommand{\ktone}{\boldsymbol{k_{1\perp}}}
\newcommand{\kttwo}{\boldsymbol{k_{2\perp}}}
\newcommand{\ltone}{\boldsymbol{l_{1\perp}}}
\newcommand{\lttwo}{\boldsymbol{l_{2\perp}}}
\newcommand{\ltthre}{\boldsymbol{l_{3\perp}}}

\newcommand{\xt}{\vect{x}}
\newcommand{\yt}{\vect{y}}
\newcommand{\zt}{\vect{z}}
\newcommand{\ut}{\vect{u}}

\newcommand{\rt}{\vect{r}}

\newcommand{\rxyt}{\boldsymbol{r}_{xy}}
\newcommand{\rbbpt}{\boldsymbol{r}_{bb'}}

\newcommand{\rzxt}{\boldsymbol{r}_{zx}}
\newcommand{\rzyt}{\boldsymbol{r}_{zy}}
\newcommand{\rzbt}{\boldsymbol{r}_{zb}}
\newcommand{\rzbpt}{\boldsymbol{r}_{zb'}}
\newcommand{\rbzt}{\boldsymbol{r}_{bz}}
\newcommand{\rbzpt}{\boldsymbol{r}_{b'z}}

\newcommand{\rbbptc}{\boldsymbol{r}_{b'b}}

\newcommand{\rxytp}{\boldsymbol{r}_{x'y'}}
\newcommand{\rxxtp}{\boldsymbol{r}_{xx'}}
\newcommand{\ryytp}{\boldsymbol{r}_{yy'}}

\newcommand{\QV}{\bar{Q}_{\rm V3}}
\newcommand{\DV}{\Delta_{\rm V3}}

\newcommand{\RtS}{\boldsymbol{R}_{\rm SE}}
\newcommand{\RtV}{\boldsymbol{R}_{\rm V}}

\newcommand{\der}{\mathrm{d}}

\newcommand{\Tr}{\mathrm{Tr}}
\newcommand{\deltatwo}{\delta(1-z_1-z_2)}

\begin{document}

\title{Back-to-back inclusive dijets in DIS at small $x$: 
Complete NLO results and predictions }
\author{Paul Caucal}
\email{caucal@subatech.in2p3.fr}
\affiliation{SUBATECH UMR 6457 (IMT Atlantique, Université de Nantes, IN2P3/CNRS), 4 rue Alfred Kastler,44307 Nantes, France}
\author{Farid Salazar}
\email{salazar@lbl.gov}
\affiliation{Nuclear Science Division, Lawrence Berkeley National Laboratory, Berkeley, California 94720, USA}
\affiliation{Physics Department, University of California, Berkeley, California 94720, USA}
\affiliation{Department of Physics and Astronomy, University of California, Los Angeles, California 90095, USA}
\affiliation{Mani L. Bhaumik Institute for Theoretical Physics, University of California, Los Angeles, California 90095, USA}
\author{Bj\"{o}rn Schenke}
\email{bschenke@bnl.gov}
\affiliation{Physics Department, Brookhaven National Laboratory, Upton, NY 11973, USA}
\author{Tomasz Stebel}
\email{tomasz.stebel@uj.edu.pl}
\affiliation{Institute of Theoretical Physics, Jagiellonian University, ul. Lojasiewicza 11, 30-348 Krak\'{o}w Poland}
\author{Raju Venugopalan}
\email{raju@bnl.gov}
\affiliation{Physics Department, Brookhaven National Laboratory, Upton, NY 11973, USA}

\begin{abstract}
 We compute the back-to-back dijet cross-section in deep inelastic scattering (DIS) at small $x$ to next-to-leading order (NLO) in the Color Glass Condensate effective field theory. Our result can be factorized into a convolution of the Weizs\"{a}cker-Williams gluon transverse momentum dependent distribution function (WW gluon TMD) with a universal soft factor and an NLO coefficient function. The soft factor includes both double and single logarithms in the ratio of the relative transverse momentum $P_\perp$ of the dijet pair to the dijet momentum imbalance $q_\perp$; its renormalization group (RG) evolution is resummed into the Sudakov factor. Likewise, the WW TMD obeys a nonlinear RG equation in $x$ that is kinematically constrained to satisfy both the lifetime and rapidity ordering of the projectile. Exact analytical expressions are obtained for the NLO coefficient function of transversely and longitudinally polarized photons. Our results allow for the first quantitative separation of the dynamics of Sudakov suppression from that of gluon saturation. They can be extended to other final states and provide a framework for precision tests of novel QCD many-body dynamics at the Electron-Ion Collider. \\
\end{abstract}

\maketitle

The transverse momentum dependent gluon Weizs\"{a}cker-Williams distribution (WW TMD) is an object of fundamental interest in QCD \cite{Boussarie:2023izj}. A particularly interesting feature is the prediction~\cite{McLerran:1993ni,McLerran:1993ka,Kovchegov:1998bi} that in contrast to the WW photon distribution in QED, strong nonlinear gluon self-interactions in QCD at small $x$ saturate its growth for transverse momenta $k_\perp \lesssim Q_s$, where $Q_s(x)$ is an emergent saturation scale~\cite{Gribov:1984tu,Mueller:1985wy}.

A golden channel to extract the  WW TMD is the inclusive 
measurement of back-to-back jets (or hadrons) in deeply inelastic electron-nucleus scattering (e+A DIS), \cite{Dominguez:2010xd,Dominguez:2011wm} characterized by the relative transverse momentum $P_\perp$ of the dijet pair and the dijet momentum imbalance $q_\perp$. For clean extraction of this quantity, perturbative QCD (pQCD) requires that  $P_\perp\gg q_\perp \gg \Lambda_{\rm QCD}$, where $\Lambda_{\rm QCD}$ is the intrinsic QCD scale. At leading order (LO), this process factorizes into the product of a hard factor computed in pQCD and the nonperturbative gluon WW TMD~\cite{Dominguez:2011wm}. At next-to-leading-order (NLO) in the QCD coupling $\alpha_s$, the mismatch between real and virtual soft gluon radiation leads to $\alpha_s\ln^2(P_\perp/q_\perp)$ contributions that significantly suppress the  back-to-back cross-section for  
$P_\perp \gg q_\perp$ \cite{Catani:1989ne,Catani:1990rp,Catani:1994sq,Balitsky:2015qba,Balitsky:2022vnb,Balitsky:2023hmh,Mueller:2012uf,Mueller:2013wwa,Xiao:2017yya}. NLO effects in gluon radiation also generate large small $x$ logarithms that drive gluon saturation \cite{Mueller:2012uf,Mueller:2013wwa,Xiao:2017yya}. It is therefore critical to understand the interplay of these effects in the extraction of the WW gluon TMD.

In this letter, we will demonstrate within the framework of the Color Glass Condensate effective field theory (CGC EFT) \cite{Gelis:2002nn,Albacete:2014fwa,Morreale:2021pnn} that TMD factorization of the inclusive back-to-back dijet cross-section in DIS at small $x$ persists at NLO~\footnote{A detailed derivation of the cross-section for longitudinally polarized virtual photons is provided in a previous paper~\cite{Caucal:2023nci}.}. Our derivation is valid to leading power (LP) in $q_\perp/P_\perp$, $Q_s/P_\perp$, and to all orders in  $Q_s/q_\perp$. We identify three key components that emerge from our NLO calculation: (i) The WW gluon TMD, satisfying a nonlinear renormalization group evolution 
equation (RGE) in $x$ that incorporates the dynamics of gluon saturation. (ii) Double and single Sudakov logarithms of  $P_\perp/q_\perp$. (iii) Perturbative hard factors (functions of $P_\perp$ and the photon virtuality $Q^2$).
 
 We will apply our results 
 to the kinematics of the future Electron-Ion Collider (EIC)~\cite{Accardi:2012qut,Aschenauer:2017jsk,AbdulKhalek:2021gbh}. While dijet studies are challenging at small $x$ \cite{Dumitru:2014vka,Dumitru:2018kuw,Zhao:2021kae,Mantysaari:2019hkq,Boussarie:2021ybe,vanHameren:2021sqc}, a window in the required phase space may exist for this theoretically robust final state. The extension of our study to the phenomenologically more accessible~\cite{Zheng:2014qfa}, if theoretically less robust, di-hadron final state is straightforward~\cite{Bergabo:2023wed}. Our computation is the first to include all above listed NLO components to determine the relative impact of Sudakov suppression and gluon saturation. 
 
In the dipole frame with virtual photon four-momentum $q^\mu=(-Q^2/(2q^-),q^-,\vect{0})$ and nucleon four-momentum  $P^\mu=(P^+,0,\vect{0})$,  the inclusive LO dijet cross-section $\der \sigma^\lambda_{\rm LO} = z_1 z_2 \frac{\der \sigma^\lambda_{\rm LO}}{\der^2 \Pt \der^2 \qt \der z_1 \der z_2 }$ can be expressed as \cite{Dominguez:2011wm,Caucal:2021ent} 
\begin{align}
    &\langle\der \sigma^{\lambda}_{\mathrm{LO}}\rangle= \int \der^2 \rt \der^2 \rt' \der^2 \bt \der^2 \bt' e^{-i \Pt \cdot(\rt-\rt') }  \nonumber \\
    & \times e^{-i \qt \cdot (\bt-\bt')} \Rcal^{\lambda}(\rt,\rt') \langle \Xi(\rt,\rt',\bt,\bt') \rangle \,,
    \label{eq:LO} 
\end{align}
where $\lambda=\rm L,T$ is the polarization of the photon, $\Pt = z_2 \ktone - z_1 \kttwo$, $\qt= \ktone + \kttwo$, and $z_i=k_i^-/q^-$. Here $\Rcal^{\lambda}$ is an analytic function representing 
the splitting of the virtual photon into the quark-antiquark dipole, and  $\Xi$ is an expression containing  
 two-point dipole and four-point quadrupole lightlike Wilson line correlators. In the CGC EFT, these describe the coherent multiple scattering of the dipole with the nonperturbative shockwave classical gauge field configurations of the nuclear target; the $\langle \dots \rangle$ is the average over static large $x$ color charge densities generating these configurations.

\begin{figure}[t]
    \centering
    \includegraphics[width=0.45\textwidth]{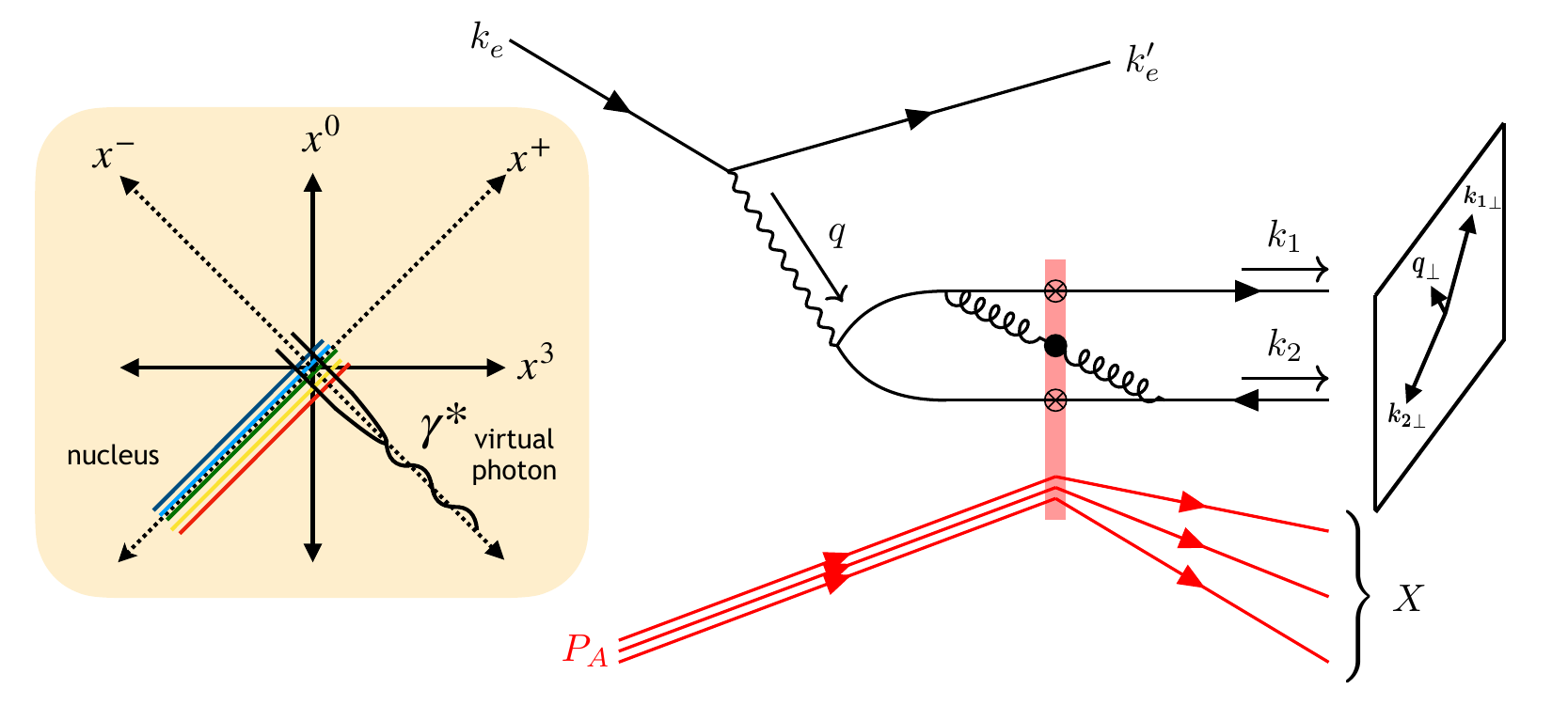}
    \caption{NLO DIS diagram for dijet production in dipole scattering off shockwave (red rectangle) in the CGC EFT. The nuclear target and the virtual photon move close to the lightcone with large $q^-$ and $P_A^+$ components respectively. }
    \label{fig:schematic}
\end{figure}

The one-loop (1L) correction to Eq.\,\eqref{eq:LO} consisting of real and virtual gluon emissions (with an example of the latter illustrated in Fig.~\ref{fig:schematic}) can be expressed as~\cite{Caucal:2021ent}
\begin{align}
    \alpha_s \der \sigma^{\lambda}_{1\mathrm{L}} = \alpha_s \int_{z_0}^1 \frac{\der z_g}{z_g} \int  \der^2 \zt \der \widetilde{\sigma}^{\lambda}_{1\mathrm{L}} (z_g,\zt)\,, 
    \label{eq:dijet_1-loop}
\end{align} 
where $z_g=k_g^-/q^-$, $k_g^-$ is the longitudinal momentum of the emitted (real or virtual) gluon, $\zt$ is the transverse coordinate of the scattered gluon off the shockwave, and $z_0 \propto 1/s$ is a physical cutoff regulating the $z_g$ divergence \cite{Caucal:2021ent}. The integrand  satisfies $\alpha_s\int_{\zt}\der \widetilde{\sigma}_{1\mathrm{L}} (0,\zt) = H_{\mathrm{LL}} \der \sigma_{\mathrm{LO}}$,
with the leading log (LL$x$) JIMWLK Hamiltonian
of the form~\cite{Balitsky:1995ub,JalilianMarian:1996xn,JalilianMarian:1997dw,Kovner:2000pt,Iancu:2000hn,Iancu:2001ad,Ferreiro:2001qy}
\begin{align}
	H_{\mathrm{LL}} = \frac{\alpha_s N_c}{2\pi^2} \int_{\zt} \mathcal{K}_{\mathrm{LL}} \,,
\end{align}
where $\mathcal{K}_{\mathrm{LL}}$ is the LL$x$ kernel. 
 Eq.\,\eqref{eq:dijet_1-loop} can be rewritten as

\begin{align}
    \alpha_s \der \sigma^{\lambda}_{1\mathrm{L}} &= \ln\left(\frac{z_f}{z_0} \right) H_{\mathrm{LL}} \der \sigma^{\lambda}_{\mathrm{LO}} + \alpha_s\int_{0}^1 \frac{\der z_g}{z_g} \int \der^2 \zt     \nonumber \\
    &  \Big[ \der \widetilde{\sigma}^{\lambda}_{1\mathrm{L}} (z_g,\zt) -  \der \widetilde{\sigma}^{\lambda}_{1\mathrm{L}} (0,\zt)  \Theta(z_f-z_g)  \Big]\,.
    \label{eq:dijet_1-loop-2}
\end{align} 
The first term isolates the large projectile-rapidity \footnote{We use the words projectile rapidity for $Y=\ln(k^-/q^-)$ and target rapidity for $\eta=\ln(P^+/k^+)$. These two scales are related to the physical rapidity $y=1/2\ln(k^+/k^-)$ via the kinematic identity $y=y_{\gamma^*}-1/2(\ln(x_{\rm Bj})+Y+\eta)$ with $y_{\gamma^*}$ the rapidity of the virtual photon.} logarithm by introducing an arbitrary  $z_f\equiv k_f^-/q^-$ factorization scale separating "fast" gluons with $z_g\ge z_f$ from "slow" ones with $z_g\le z_f$. 
Since the NLO cross-section should be independent of $z_f$, 
the LO cross-section satisfies the small-$x$ JIMWLK RGE resumming $\alpha_s\ln(z_f/z_0)$ to all orders, which for $Y_f=\ln(z_f)$ is 
\begin{align}
    \frac{\partial \langle \der\sigma_{\rm LO}\rangle_{Y_f}}{\partial Y_f}=\left\langle H_{\mathrm{LL}} \der\sigma_{\rm LO}\right\rangle_{Y_f}\,.
    \label{eq:JIMWLK}
\end{align}
The second term in Eq.\,\eqref{eq:dijet_1-loop-2} constitutes the NLO impact factor $\langle \der\sigma _{\rm NLO}\rangle_{Y_f}$. Here we have taken $z_0 \propto 1/s \to 0$; any  dependence 
is power suppressed in the energy $\sqrt{s}$. At NLO, the $Y_f$ dependence of the cross-section is determined by the action of the next-to-leading log $x$ (NLL$x$) JIMWLK Hamiltonian $H_{\mathrm{LL}}\to H_{\mathrm{LL}}+\alpha_sH_{\textrm{NLL}}$ \cite{Balitsky:2013fea,Kovner:2013ona,Kovner:2014lca,Lublinsky:2016meo} on the LO cross-section since the $\alpha_s^2\ln(z_f/z_0)$ terms it resums are of the same order as $\langle \der\sigma _{\rm NLO}\rangle_{Y_f}$. 

We turn now to the back-to-back limit of the NLO inclusive dijet result in \cite{Caucal:2021ent}. 
At LO, it was shown to have the TMD factorized form~\cite{Dominguez:2011wm}
\begin{align}
    \langle \der\sigma^{\lambda,\mathrm{b2b}}_{ \rm LO}\rangle _{Y_0} = \mathcal{H}_{\rm LO}^{\lambda,ij}\int_{\bt,\bt'}\frac{e^{-i\qt\cdot\rbbpt}}{(2\pi)^4} \hat G^{ij}_{Y_0}(\rbbpt,\mu)\,,
    \label{eq:diff_xsec_TMD_LO}
\end{align}
with $\rbbpt=\bt-\bt'$. The perturbative hard factor $\Hcal_{\rm LO}^{\lambda,ij}$ depends on $\Pt$, $Q$ and $z_i$ alone; it is specified in the supplemental material. The LO WW gluon TMD at the initial projectile rapidity scale $Y_0=\ln(z_0)$ is  
\begin{align}
    &\hat G^{ij}_{Y_0}(\rbbpt,\mu)\equiv\frac{-2}{\alpha_s(\mu)}\left\langle\Tr\left[ V_{\bt} \partial^iV^\dagger_{\bt}V_{\bt'} \partial^jV^\dagger_{\bt'}\right]\right\rangle_{Y_0} \!\! \,,
    \label{eq:WWTMD}
\end{align}
with $V_{\bt}$ the lightlike Wilson line in the fundamental representation of SU(3) with transverse coordinate $\bt$.
The separation of hard versus soft modes (as opposed to fast versus slow) is specified by the renormalization scale $\mu$. In the saturation regime, $\mu\sim Q_s$ since $q_\perp \sim Q_s$ is the typical momentum of the gluon polarization tensor in the shockwave background. 


As in the LO case~\cite{Altinoluk:2019wyu,Boussarie:2020vzf,Mantysaari:2019hkq,Boussarie:2021ybe,Rodriguez-Aguilar:2023ihz}, we can extract the NLO impact factor for inclusive back-to-back dijets from  the LP contributions ($q_\perp, Q_s\ll P_\perp$) to the fully inclusive NLO cross-section $\langle \der\sigma^{\lambda}_{\rm NLO}\rangle _{Y_f}$ \cite{Caucal:2021ent}. 
Remarkably, these  are 
also proportional to the WW gluon TMD \cite{Caucal:2023nci}, with dominant contributions to the impact factor from the Sudakov logarithms $\alpha_s\ln^2(P_\perp r_{bb'})$ and $\alpha_s\ln(P_\perp r_{bb'})$
when $P_\perp/q_\perp \gg 1$. The purely $\mathcal{O}(\alpha_s)$ corrections
are gathered in the NLO coefficient function. However in contrast to Eqs.~\eqref{eq:dijet_1-loop}  and \eqref{eq:dijet_1-loop-2}, the LP impact factor at NLO has additional $\ln^2(z_0)$ divergences. They are canceled by imposing a kinematic constraint (kc) in subtracting the $z_g\to 0$ divergence in the back-to-back cross-section~\cite{Caucal:2022ulg,Taels:2022tza}:
\begin{align}
    &\alpha_s \der \sigma^{\lambda,\mathrm{b2b}}_{\mathrm{NLO}} = \alpha_s \int_{0}^1 \frac{\der z_g}{z_g} \int \der^2 \zt \left[ \der \widetilde{\sigma}^{\lambda,\mathrm{b2b}}_{1\mathrm{L}} (z_g,\zt)   \right.\nonumber \\
	& \left. -  \der \widetilde{\sigma}^{\lambda,\mathrm{b2b}}_{1\mathrm{L}} (0,\zt)  \Theta(z_f-z_g)   \Theta\left(-\ln\left(z_g\right)-\Delta_{c}\right)  \right] \,,\label{eq:NLO-if-improved}
\end{align}
where $\Delta_c=\ln(\textrm{min}(\rzbt^2,\rzbpt^2)2k_c^+q^-)$, $\boldsymbol{r}_{zb}=\zt-\bt$, $k_c^+=x_cP^+$ and $x_c=\frac{1}{ec_0^2}\frac{M_{q\bar q}^2+Q^2}{W^2+Q^2}$ is a number fixed in terms of $Q^2$; $W$ is the nucleon-virtual photon center-of-mass energy, $M_{q\bar q} = P_\perp/\sqrt{z_1 z_2}$ the dijet invariant mass,  $c_0=2e^{-\gamma_E}$ and  $\gamma_E$ is the Euler constant.

Comparing Eq.\,\eqref{eq:NLO-if-improved} to Eq.\,\eqref{eq:dijet_1-loop-2}, 
the effect of the kinematic constraint on 
the LL$x$ RGE in Eq.\,\eqref{eq:JIMWLK} {\it specifically for the WW gluon TMD} is to modify the kernel 
in the JIMWLK Hamiltonian \cite{Caucal:2023nci,Taels:2022tza}:
\begin{align}
H^{kc}_{\mathrm{LL}}\der\sigma^{\lambda, \rm b2b}_{\rm LO}\equiv \frac{\alpha_s N_c}{2\pi^2} \int_{\zt}  \Theta(-Y_f-\Delta_{c}) \mathcal{K}_{\mathrm{LL}}\der\sigma^{\lambda, \rm b2b}_{\rm LO}\,. \label{eq:dmmx}
\end{align}
 
The $\Theta$-function 
enforces lifetime ordering $1/k_g^+\sim 2k_g^-/\kt^2\le 1/k_c^+$ 
in the evolution of the projectile since $1/k_c^+$ is of order of the dipole coherence time $1/|q^+|$~\cite{Gribov:1965hf,Ioffe:1969kf} and $\kt^2\sim 1/\textrm{min}(\rzbpt^2,\rzbpt^2)$ is the squared transverse momentum of the first emitted gluon. 
This constraint is long-understood~\cite{Andersson:1995ju,Kwiecinski:1996td,Kwiecinski:1997ee,Salam:1998tj,Ciafaloni:1998iv,Ciafaloni:1999yw,Ciafaloni:2003rd,SabioVera:2005tiv,Motyka:2009gi,Beuf:2014uia,Iancu:2015vea} to generate  all-order resummation of transverse double logarithms 
necessary to match small-$x$ with DGLAP collinear resummation. Eq.~\eqref{eq:dmmx} implements a piece of the NLL$x$ RGE; there are additional contributions that require a two-loop computation. 


On the surface, Eq.\,\eqref{eq:dmmx} appears process dependent since the kernel depends on the scale $k_c^+$ specific to inclusive back-to-back dijets. This is not the case: 
replacing $Y_f$ by $\eta_f=\ln(P^+/k_f^+)$ in Eq.\,\eqref{eq:dmmx} using the identity 
 \begin{equation}
     \eta_f  =Y_f+\ln\left(\frac{M_{q\bar q}^2+Q^2}{eP_\perp^2}\right)+\ln\left(\frac{\Pt^2\rbbpt^2}{c_0^2}\right)-\ln(x_c)\,,
    \label{eq:Y-eta-relation}
 \end{equation}
the $\Theta$-constraint in Eq.\,\eqref{eq:dmmx} 
is replaced by $\eta_f\le \eta_c=\ln(1/x_c)$, the maximal value for the target-rapidity factorization scale $\eta_f$. 
Eq.\,\eqref{eq:dmmx} is identical to the universal kinematically constrained dipole RGE in  $\eta_f$ of \cite{Hatta:2016ujq,Ducloue:2019ezk,Ducloue:2019jmy} described in the supplemental material.

Importantly, for scale choice  $\eta_f\sim\ln(1/x_c)$, the corresponding scale $Y_f\sim-\ln(\Pt^2\rbbpt^2)$  
in the NLO impact factor is now 
understood to be a Sudakov log due to gluons with $k_f^-\le k_g^-\lesssim q^-$. The target-rapidity ordered resummation scheme therefore clearly separates small-$x$ rapidity evolution from Sudakov logs (and other NLO corrections) in the NLO impact factor. 

The dijet cross-section is expanded in Fourier moments  as $\der \sigma^{\lambda} = \der \sigma^{(0),\lambda} + 2\sum_{n=1}^{\infty} \der \sigma^{(2n),\lambda} \cos(2n \phi)$, where $\phi$ is the angle between $\qt$ and $\Pt$. At LO, the coefficients of the zeroth and second moment of this expansion are respectively proportional to the unpolarized and linearly polarized WW TMD~\cite{Dumitru:2014vka, Dumitru:2018kuw,Mantysaari:2019hkq,Boussarie:2021ybe,Zhao:2021kae}. To NLO accuracy, the azimuthally averaged back-to-back dijet cross-section has the TMD factorized expression 
\begin{widetext}
    \begin{align}
    \left\langle\der\sigma^{(0),\lambda,\rm b2b}_{\rm LO}+ \alpha_s\der\sigma^{(0),\lambda,\rm b2b}_{\rm NLO}\right\rangle _{\eta_f} &=\Hcal_{\rm LO}^{0,\lambda}\int\frac{\der^2\Bt}{(2\pi)^2}\int\frac{\der^2\rbbpt}{(2\pi)^2}e^{-i\qt\cdot\rbbpt}\hat G^0_{\eta_f}(\rbbpt,\mu_0)\Bigg\{1+\frac{\alpha_s(\mu_R)}{\pi}\Big [\underbrace{-\frac{N_c}{4}\ln^2\left(\frac{\Pt^2\rbbpt^2}{c_0^2}\right)}_{\mathrm{Sudakov\ double\ log}} \nonumber\\
    &\hspace{-2.5cm}\underbrace{-s_L\ln\left(\frac{\Pt^2\rbbpt^2}{c_0^2}\right) +\beta_0\ln\left(\frac{\mu_R^2\rbbpt^2}{c_0^2}\right)}_{\mathrm{Sudakov\ single\ logs}}+\frac{N_c}{2}f^{\lambda}_1(\chi,z_1,R,\eta_f)+\frac{1}{2 N_c}f^{\lambda}_2(\chi,z_1,R)\Big]\Bigg\}\nonumber\\
    &\hspace{-2.5cm}+\frac{\alpha_s(\mu_R)}{\pi}\Hcal_{\rm LO}^{0,\lambda}\int\frac{\der^2\Bt}{(2\pi)^2}\int\frac{\der^2\rbbpt}{(2\pi)^2}e^{-i\qt\cdot\rbbpt}\hat h^0_{\eta_f}(\rbbpt,\mu_0)\left\{\frac{N_c}{2}\left[1+\ln(R^2)\right]-\frac{1}{2 N_c}\ln(z_1z_2R^2)\right\}\,.
    \label{eq:result-xsect}
\end{align}
\end{widetext}
Here $\hat G^0 = \delta^{ij} \hat G^{ij}$ and $\hat h^0 =  \left( \frac{2\rbbpt^i \rbbpt^j}{\rbbpt^2} -\delta^{ij} \right)\hat G^{ij}$ denote respectively the unpolarized and linearly polarized coordinate space  WW gluon TMD \cite{Dominguez:2011br}, 
with both distributions depending implicitly on the impact parameter $\Bt = \frac{1}{2}(\bt+\bt')$.  We emphasize that their CGC average is performed 
at the \textit{target} rapidity factorization scale $\eta_f=\ln(1/x_f)$. In our numerical study, we shall use $\eta_f=\ln(1/x_g)$ with $x_g=ec_0^2x_c$ as our central value. A novel feature at NLO is that the 
dependence of Eq.\,\eqref{eq:result-xsect} on $\hat h^0$. This dependence is absent at LO and arises from soft gluons emitted close to the jet cone boundary~\cite{Hatta:2020bgy,Hatta:2021jcd}.  

Further in Eq.~(\ref{eq:result-xsect}), 
 $\chi\equiv Q/M_{q\bar q}$ and $R$ is the anti-$k_t$ jet radius \cite{Cacciari:2008gp}. The coefficient of the single Sudakov logarithm is 
$s_L=-C_F\ln(z_1z_2R^2)+N_c\ln(1+\chi^2)$. The term proportional to the coefficient $\beta_0=(11N_c-2n_f)/12$ of the one-loop $\beta$-function arises from the RG evolution of the WW TMD as a function of $\mu$ from the initial scale $\mu_0^2=c_0^2/\rbbpt^2$ up to the renormalization scale $\mu_R^2$ \cite{Ayala:1995hx,Zhou:2018lfq,Caucal:2023nci}. Since $\mu_R\sim P_\perp$, this term can be combined with the Sudakov single logarithm. The NLO coefficient functions $f_1$ and $f_2$ have analytic expressions specified in the supplemental material (including Refs.~\cite{Dokshitzer:1997in,Wobisch:1998wt,Salam:2010nqg,Cacciari:2011ma,Salam:2007xv,Ivanov:2012ms,Kang:2017mda,abramowitz1964handbook,Cali:2021tsh}). 
  
 Eq.\,\eqref{eq:result-xsect} is the principal result of this letter. It demonstrates for the first time  TMD factorization of the back-to-back dijet cross-section at small $x$ at NLO. It is valid up to corrections of order $q_\perp^2/P_\perp^2$, $Q_s^2/P_\perp^2$, as well as $\alpha_sR^2$ and $\alpha_s^2$. All saturation effects are contained in the WW gluon TMD $\hat G^{ij}_{\eta_f}$ and its nonlinear kinematically constrained RGE, without introducing new operators in the NLO coefficient functions. Although TMD factorization could have been anticipated for $q_\perp\gg Q_s$ since the high-energy evolution equation of the WW gluon TMD has a closed form in this dilute limit \cite{Dominguez:2011gc}, it is remarkable that it persists when 
 $q_\perp\sim Q_s$, enabling precise extraction of saturation dynamics with back-to-back dijets. Explicit expressions for higher order even harmonics $\der\sigma^{(2n),\lambda}$ are provided in the supplemental material.

In Eq.\,\eqref{eq:result-xsect}, we obtained
the  $\mathcal{O}(\alpha_s)$ term for 
Sudakov logs for this process at small $x$. If we assume powers of these logarithms at all orders can be resummed and exponentiate \cite{Kang:2020xgk,delCastillo:2020omr,Gao:2023ulg}, as in the Collins-Soper-Sterman collinear factorization framework \cite{Collins:1981uk,Collins:1981uw,Collins:1984kg}, they can be absorbed in a Sudakov soft factor 
\begin{align}
\Scal=\exp\left(\!-\! \int_{\mu_0^2}^{P_\perp^2} \! 
 \frac{\der\mu^2}{\mu^2}\frac{\alpha_sN_c}{\pi}\left[\frac{1}{2}\ln\left(\frac{P_\perp^2}{\mu^2}\right)+\frac{\tilde{s}_{L}}{N_c}\right]\right)\,.\label{eq:soft-factor}
\end{align}
Here $\tilde{s}_{L} = s_L-\beta_0$ and $\alpha_s=\alpha_s(c\mu)$, with $c$ an arbitrary $\mathcal{O}(1)$ constant which can be varied to gauge the sensitivity to two-loop $\mathcal{O}(\alpha_s^2\ln^2(P_\perp/\mu_0))$ corrections not included in the resummation \cite{Collins:1984kg}. The  Sudakov factor agrees with previous collinear factorization calculations for this process \cite{Hatta:2021jcd}. This is noteworthy given the nontrivial interplay between Sudakov and small-$x$ factorization. The result of the resummation (shown in the supplemental material) is to remove the underlined Sudakov log terms in Eq.~(\ref{eq:result-xsect}) and multiply instead the WW TMD by $\mathcal{S}$.

We shall now discuss
the numerical evaluation of 
Eq.\,\eqref{eq:result-xsect} and higher Fourier moments to assess their predictive power at the EIC. We compute the differential yield $\der N =\Sigma_{\lambda} \phi^\lambda \der \sigma^{\lambda} / \der^2\Bt$ at the mean impact parameter, which represents minimum bias collisions. ($\phi^\lambda$ is the photon flux factor defined in the supplemental material.) For $\sqrt{s}=90$ GeV, $Q^2=4$ GeV$^2$ and 
$x_{\rm Bj}=0.55\cdot 10^{-3}$,  the smallest $x$ available for RG evolution is $x_c = (1.5-3.8) \cdot 10^{-3}$ for $P_\perp=(4-6)$ GeV. While not a large window, we will demonstrate that it may be sufficient to uncover clear evidence for gluon saturation. 

Using the Gaussian approximation~\cite{Blaizot:2004wv,Dumitru:2011vk,Iancu:2011nj,Metz:2011wb,Dominguez:2011br}, we first relate the WW gluon TMD to the dipole gluon distribution $\Ncal_{\eta_f}(\rt)$ satisfying the Balitsky-Kovchegov (BK) RGE~\cite{Balitsky:1995ub,Kovchegov:1999yj}
\footnote{The validity of this approximation relative to the full LL JIMWLK RGE was explored in \cite{Dumitru:2011vk} and shown to be quite good for a range of spatial configurations of dipoles.}. For the initial condition for the RGE, we use the McLerran-Venugopalan model~\cite{McLerran:1993ka,McLerran:1993ni},
\begin{equation}
\Ncal_{\eta_0}(\rt)=1-\exp\left[-\frac{\rt^2Q_{s0,A}^2}{4}\ln\left(\frac{1}{r_\perp\Lambda}+e\right)\right]\,,
\end{equation}
with the rapidity scale $\eta_0=\ln(1/x_0)$, $x_0=2.5\cdot 10^{-2}$ and the minimum bias $Q_{s0,A}^2=A^{1/3}\times 0.1$ GeV$^2$ \cite{Kowalski:2007rw}. The infrared regulator of the Coulomb logarithm is  $\Lambda=0.24$ GeV.
 Fits to e+A fixed target data~\cite{Lappi:2013zma} and electron-proton (e+p) HERA data~\cite{H1:2009pze} dictate these choices, which can be constrained from future global analyses. 

For full NLO accuracy at small $x$, one should evolve $\Ncal_{\eta_0}(\rt)$ with the NLL$x$ BK equation \cite{Balitsky:2007feb}. While this equation is challenging to solve \cite{Lappi:2015fma}, the dominant contributions to its kernel are accounted for by contributions from the running coupling and the lifetime ordering constraint \cite{Lappi:2016fmu}. We employ the minimal dipole size prescription for the 
former~\cite{Kovchegov:2006vj,Balitsky:2007feb,Iancu:2015joa}. For the Sudakov soft factor in Eq.\,\eqref{eq:soft-factor}, we employ one-loop running of the coupling. Its sensitivity to nonperturbative physics at large $r_{bb'}$ is modeled by freezing it at a maximal value (which is varied) of $\alpha_{s,{\rm max}}/\pi=0.24$. 

\begin{figure}[htp]
    \centering
        \includegraphics[width=0.49\textwidth]{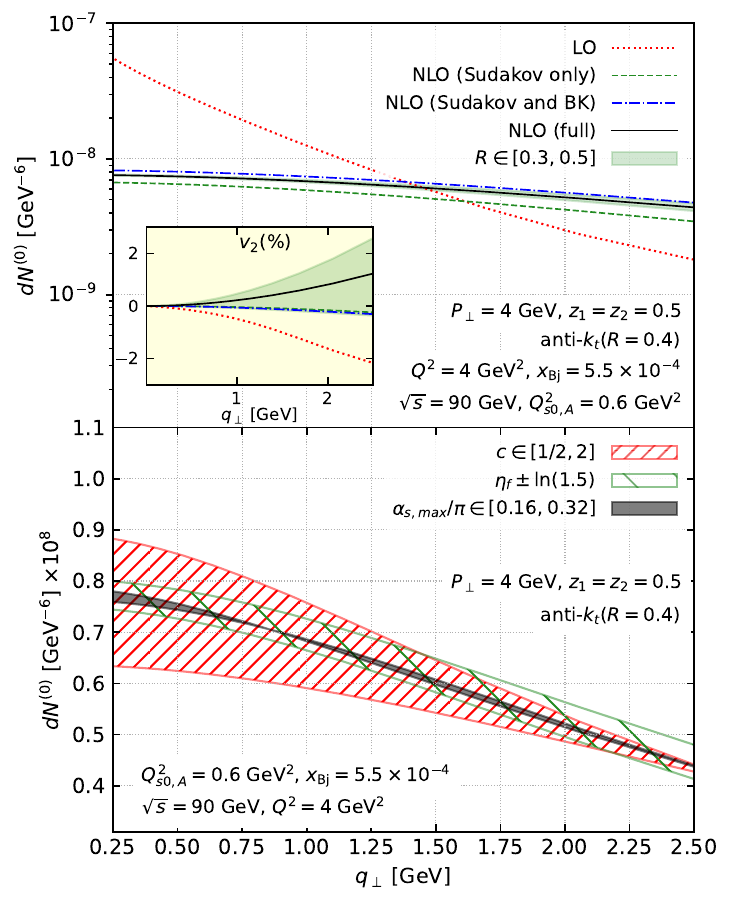} 
    \caption{(Top) Azimuthally averaged LO and NLO back-to-back dijet yield as a function of dijet momentum imbalance $q_\perp$. Inset shows $q_\perp$ dependence of the $v_2$ coefficient. The light green band corresponds to varying anti-$k_t$ parameter $R=0.3-0.5$. (Bottom) Theory uncertainties from renormalization and target rapidity scale variations and freezing of $\alpha_s$. Details in text.
    \label{fig:qt-xsection}}
\end{figure}

Results in EIC kinematics for the azimuthally averaged back-to-back dijet yield versus $q_\perp$  are shown in Fig.\,\ref{fig:qt-xsection} (Top). It illustrates the effects of three types of NLO corrections: i) Sudakov suppression (green dashed curve), ii) Sudakov suppression and small-$x$ evolution (dot-dashed blue curve), and iii) all NLO corrections labeled NLO (full) -- this includes as well the NLO coefficient function computed for the first time here. In comparing the dotted red and dashed green curves, one observes that Sudakov suppression is very significant for $q_\perp\lesssim 1.5$ GeV.
Small-$x$ resummation leads to an overall increase of the yield due to the proliferation of slow gluons. This increase is slowed down by nonlinear saturation corrections. The NLO coefficient function yields a small additional contribution to the yield, which varies depending on kinematic choices. The light green band shows the dependence of the full NLO result on the jet radius $R$ which decreases slightly with increasing  $R$. 

The inset shows the $q_\perp$ dependence of the $v_2=\der\sigma^{(2)}/\der\sigma^{(0)}$ coefficient, whose magnitude is very small, $ < 2\%$.  NLO corrections on the sign of $v_2$ 
are significant because they flip the LO (negative) value due to the preferential emission of soft gluons close to the jet boundary. The sign of $v_2$ is therefore sensitive to $R$; however as shown in the supplemental material, $v_2^{\lambda=L} \propto \hat h^0$ is unambiguously positive for the $R$ range studied.

Theory uncertainties in the NLO result can be divided into four classes; three of these are displayed in Fig.\,\ref{fig:qt-xsection} (Bottom). We first show uncertainties from the unknown order N$^2$LO contributions beyond the NLO impact factor; they are estimated by varying the running coupling scale $c=0.5-2$ both in the NLO coefficient function where $\mu_R=cP_\perp$ and in the Sudakov factor. 
Since they are parametrically of order $\alpha_s^2\ln^2(P_\perp/\mu_0)$, 
the band width grows with decreasing $q_\perp$. This illustrates the importance of controlling powers of $\alpha_s\ln(P_\perp/\mu_0)$ for future precision studies. 

The second source of uncertainty is from 
the target-rapidity factorization scale $\eta_f$, 
obtained by varying $x_f$ by a factor of 1.5 around the central value $x_g$. This dependence 
decreases from LO to NLO (see supplemental material);  it remains however the dominant source of uncertainty for $q_\perp\gtrsim Q_s$. 
Uncertainties from missing contributions in the full NLL BK kernel are subleading in comparison.
Variations with respect to $\alpha_{s,{\rm max}}$ are shown by the gray band in Fig.\,\ref{fig:qt-xsection} (Bottom). 
This sensitivity is mitigated
in large nuclei 
because the minimal transverse size controlling the coupling is set by the large $Q_s$. Not shown here are 
power correction  ($q_\perp^2/P_\perp^2, Q_s^2/P_\perp^2$) uncertainties  discussed at LO in \cite{Mantysaari:2019hkq,Boussarie:2021ybe}, 
of $\mathcal{O}(10\%)$ for $q_\perp\lesssim 1.5$ GeV and $P_\perp=4$ GeV.

    \begin{figure}[htp]
    \centering
        \includegraphics[width=0.48\textwidth,page=1]{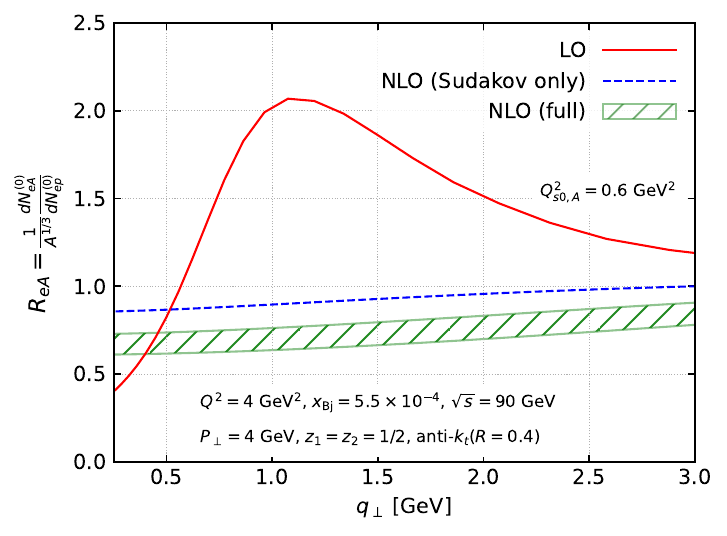} 
\hfill
        \includegraphics[width=0.48\textwidth,page=2]{ReA-qt-A-dep-x0=0.025.pdf}
    \caption{$q_\perp$ and $A$ dependence (Top and Bottom respectively) of the nuclear modification factor $R_{eA}$ for the azimuthally averaged back-to-back dijet yield.}\label{fig:ReA-qt-A-dep}
\end{figure}

Fig.\,\ref{fig:ReA-qt-A-dep} displays $R_{eA}$, the ratio of the azimuthally averaged back-to-back dijet yield in e+A to e+p collisions. Such ratios minimize 
theory uncertainties as well as experimental ones. The top plot shows the $q_\perp$ dependence of $R_{eA}$ for a large nucleus; for simplicity, we take $A^{1/3}=6$. 
At LO, it has a ``Cronin" peak well-known from the corresponding ratio in proton-nucleus (p+A) collisions~\cite{Antreasyan:1978cw}; in the CGC, it is generated by coherent multiple scattering that shifts the typical momentum imbalance to larger $q_\perp$ in heavier nuclei~\cite{Jalilian-Marian:2003rmm}. At NLO, we see that the Cronin enhancement is washed out by Sudakov corrections alone. A further strong effect is seen from the NLO contributions dominantly caused by the WW gluon TMD RGE which suppresses $R_{eA}$ 
as in the $R_{pA}$ case \cite{Kharzeev:2003wz,Albacete:2003iq}. 

A qualitative interpretation is that Sudakov logs suppress configurations corresponding to small $q_\perp$ (or large $\rbbpt$) in the projectile. However since a fundamental consequence of gluon saturation is that even configurations with small $\rbbpt$ are sensitive to nonlinear RG $x$ evolution, its precocious onset in large nuclei~\cite{Mueller:2003bz} leads to increasing suppression of $R_{eA}$ with $A^{1/3}$, as demonstrated by the bottom plot for fixed $q_\perp=1.5$ GeV.
The systematics of this suppression with $A^{1/3}$ and  $q_\perp$ are sensitive to the WW TMD RGE. Plots with different kinematic choices are provided in the supplemental material. 

 While more detailed studies are necessary, our results are suggestive that inclusive back-to-back dijets in e+A collisions show strong potential to be a golden channel for gluon saturation at the EIC when combined 
 with other processes 
 that constrain the initial condition for the WW small-$x$ RGE. Our conclusions can be strengthened by minimizing the stated theory uncertainties and by extending the comprehensive NLO study here to the di-hadron channel. Global analyses incorporating other e+A small $x$ final states \cite{Boussarie:2016bkq,Boussarie:2016ogo,Roy:2019cux,Roy:2019hwr,Beuf:2020dxl,Beuf:2021srj,Mantysaari:2022kdm,Bergabo:2022zhe,Beuf:2022kyp,Tong:2022zwp,Fucilla:2022wcg,Lipatov:2023ypn} and analogous studies~\cite{Chirilli:2011km,Ivanov:2012iv,Altinoluk:2014eka,Celiberto:2017ptm,Stasto:2018rci,Albacete:2018ruq,Liu:2020mpy,Shi:2021hwx,Hentschinski:2021lsh,Liu:2022ijp,Wang:2022zdu,vanHameren:2022mtk,Hentschinski:2020tbi,Celiberto:2022fgx,vanHameren:2023oiq,Ganguli:2023joy,Altinoluk:2023hfz} in p+A collisions at RHIC and the LHC will further enable unambiguous determination of the dynamics of gluon saturation.
 
\smallskip
\noindent{\bf Acknowledgements.} We are grateful to Bertrand Duclou\'{e} and Feng Yuan for valuable discussions. P.C., F.S. and T.S. thank the EIC theory institute at BNL for its support during the final stages of this work. F.S. is supported by the National Science Foundation under grant No. PHY-1945471, and the UC Southern California Hub, with funding from the UC National Laboratories division of the University of California Office of the President.
T.S. kindly acknowledges support of the Polish National Science Center (NCN) Grants No.\,2019/32/C/ST2/00202 and 2021/43/D/ST2/03375.
B.P.S. and R.V. are supported by the U.S. Department of Energy, Office of Science, Office of Nuclear Physics, under DOE Contract No.~DE-SC0012704 and within the framework of the Saturated Glue (SURGE) Topical Theory Collaboration. R.V.'s work is also supported in part by an LDRD grant from Brookhaven Science Associates.

\let\oldaddcontentsline\addcontentsline
\renewcommand{\addcontentsline}[3]{}
\bibliographystyle{apsrev4-1}
\bibliography{dijet-factorization}
\let\addcontentsline\oldaddcontentsline

\clearpage
\appendix

\begin{widetext}

\let\oldaddcontentsline\addcontentsline
\renewcommand{\addcontentsline}[3]{}
\section{Supplemental material}
\let\addcontentsline\oldaddcontentsline

\tableofcontents
\vspace{10 mm}

The supplemental material is divided into five sections. The first section gathers the analytic expressions for the NLO coefficient functions for both longitudinally and transversely polarized virtual photons. The second section presents additional figures, complementing those shown in the letter. Supplemental Material 3 explains how the NLO coefficient function is matched to the small-$x$ evolution ordered in target rapidity $\eta=\ln(1/x)$. The fourth supplemental material is the proof of NLO TMD factorization for transversely polarized photons. Though it largely follows the derivation provided in \cite{Caucal:2023nci} for longitudinally polarized photons, there are novel features of interest. Finally, the last section gathers useful integrals relevant for the calculation of the NLO coefficient function.

\section{Supplemental Material 1: Summary of analytic results for the NLO coefficient functions}

The inclusive back-to-back dijet differential cross-section in DIS is decomposed according to the virtual photon polarization (longitudinal or transverse):
\begin{equation}
    \frac{\der\sigma^{e+A\to e'+q\bar q+X}}{\der x_{\rm Bj} \der Q^2\der^2\Pt\der^2\qt\der\eta_1\der\eta_{2}}=\sum_{\lambda=\mathrm{L,T}}\phi_{\lambda}(x_{\rm Bj}, Q^2) \ \left.\frac{\der \sigma^{\gamma_{\lambda}^*+A\to q\bar{q}+X}}{ \der^2 \Pt \der^2 \qt \der \eta_1 \der \eta_{2}}\right|_{\eta_f} \,,
\end{equation}
with the photon flux factors $\phi_\lambda$ given by
\begin{align}
    \phi_{\lambda=\mathrm{L}}(x_{\rm Bj}, Q^2)&=\frac{\alpha_{\mathrm{em}}}{\pi Q^2 x_{\rm Bj}}(1-y)\,,\\
    \phi_{\lambda=\mathrm{T}}(x_{\rm Bj}, Q^2)&=\frac{\alpha_{\mathrm{em}}}{2\pi Q^2 x_{\rm Bj}}[1+(1-y)^2]\,,
\end{align}
where $y$ is the inelasticity and satisfies the relation $y \,x_{\mathrm{Bj}} = Q^2/s$. The hadronic $\gamma_{\lambda}^*+A\to q\bar{q}+X$ component of the cross-section depends on the target rapidity factorization scale $\eta_f=\ln(1/x_f)$ which should be chosen such that $\eta_f\le \eta_c=\ln(1/x_c)$ with
\begin{equation}
    x_c=\frac{1}{ec_0^2}\frac{M_{q\bar q}^2+Q^2}{W^2+Q^2}\,.\label{eq:xf_identity}
\end{equation}
A typical choice for $x_f$ is $x_g=ec_0^2x_c$ which is the value one gets from plus light cone momentum conservation in the LO process $\gamma^*+g\to q+\bar q$.

As is customary, we decompose the hadronic component of the differential cross-section into Fourier modes in the angle $\phi$ between $\Pt$ and $\qt$:
\begin{align}
    \left.\frac{\der \sigma^{\gamma_{\lambda}^*+A\to q\bar{q}+X}}{ \der^2 \Pt \der^2 \qt \der \eta_1 \der \eta_{2}} \right|_{\eta_f}= \der \sigma^{(0),\lambda} + 2 \sum_{n=1}^{\infty} \cos(2n \phi) \der \sigma^{(2n),\lambda} \,.\label{eq:Fourier-phi}
\end{align}
We present results both for longitudinally and transversely polarized photons. The leading order hard factors are
\begin{align}
    \Hcal_{\rm LO}^{\lambda=\rmL,ij}=\alpha_{\rm em}e_f^2\alpha_s\delta(1-z_1-z_2) 16 \,z_1^3 z_2^3 \,Q^2\, \frac{\Pt^i \Pt^j}{(\Pt^2+z_1z_2 Q^2)^4} \,,\label{eq:HLOij_L}
\end{align}
\begin{align}
    \Hcal_{\rm LO}^{\lambda=\rmT,ij}&= \alpha_{\rm em}e_f^2\alpha_s\delta(1-z_1-z_2) z_1z_2(z_1^2+z_2^2) \left[\frac{\delta^{ij}}{(\Pt^2+z_1z_2Q^2)^2}-\frac{4z_1z_2Q^2\Pt^i\Pt^j}{(\Pt^2+z_1 z_2Q^2)^4}\right] \,,\label{eq:HLOij_T}
\end{align}
for transversely and longitudinally polarized photons respectively. Here $\alpha_{\rm em}$ is the QED fine structure constant, $e_f^2$ the sum of squares of the light quark fractional charges, $i,j=1,2$ are transverse coordinates. We also define a shorthand for the trace of the hard factor:
\begin{align}
    \Hcal_{\rm LO}^{0,\lambda} = \frac{1}{2} \delta^{ij} \Hcal_{\rm LO}^{\lambda,ij} \,.
\end{align}

At NLO, for longitudinal photons, the calculation has been detailed in \cite{Caucal:2023nci}. With respect to Eqs.\,(5.2)-(5.3) in this paper, the only modification to the NLO coefficient function comes from the matching with $\eta$-ordered rapidity evolution explained in Supplemental Material 3. This is equivalent to setting $Y_f=\ln(z_f)$ in Eqs.\,(5.2)-(5.3) to the value given by Eq.\,\eqref{eq:Yf_values} (and absorb the large logarithm $-\ln(\Pt^2\rbbpt^2/c_0^2)$ in the Sudakov) and to add the contribution given by Eq.\,\eqref{eq:virtual-dmmx-finite-piece}.

For transversely polarized photons, we combine the results obtained in Supplemental Material 4. More precisely, starting from Eq.\,\eqref{eq:transverse-b2b-start}, we express the two hard factors $\Hcal_{\rm NLO,1}^{\lambda=\rmT}$ and $\Hcal_{\rm NLO,2}^{\lambda=\rmT}$ using Eqs.\,\eqref{eq:kappa-def}-\eqref{eq:kappa1-final}-\eqref{eq:kappa2-final} and Eqs.\,\eqref{eq:tau-def}-\eqref{eq:tau1-final}-\eqref{eq:tau2-final} respectively. For the contribution $\der \sigma^{(0),\lambda=\rm T}_{\rm other}$ in which the gluon scatters off the shockwave, one combines Eqs.\,\eqref{eq:inst-def}-\eqref{eq:Hinst-final}-\eqref{eq:HSE1-def}-\eqref{eq:HNLO3-final}-\eqref{eq:HNLO4_a-def}-\eqref{eq:HNLO4_b_def}-\eqref{eq:HNLO4_b_final}. Then one performs the remaining $z_g$ integral properly regulated in the limit $z_g\to0$ thanks to the kinematically constrained slow gluon counterterm given by Eq.\,\eqref{eq:slow-def}-\eqref{eq:slow-from-analytic}. The resulting expression depends on the scale $Y_f=\ln(z_f)$ which is expressed in terms of $\eta_f=\ln(1/x_f)$ thanks to the matching to $\eta$-ordered rapidity evolution by Eq.\,\eqref{eq:Yf_values}, as in the longitudinal case.

The NLO coefficient functions are provided for jets defined with the anti-$k_t$ algorithm \cite{Cacciari:2008gp} (or any jet clustering algorithm among the generalized $k_t$ family \cite{Dokshitzer:1997in,Wobisch:1998wt,Salam:2010nqg,Cacciari:2011ma}) with jet parameter $R$. For cone-jet definitions \cite{Salam:2007xv,Ivanov:2012ms,Kang:2017mda}, one should modify the NLO coefficient functions $f_1$ and $f_2$ as:
\begin{equation}
    f_1^{\lambda}\to  f_1^{\lambda}-\left(3-\frac{\pi^2}{3}-3\ln(2)\right)\,,\quad 
     f_2^{\lambda}\to  f_2^{\lambda}+\left(3-\frac{\pi^2}{3}-3\ln(2)\right)\,.
\end{equation}

In all expressions below, we assume exponentiation of the Sudakov double and single logarithms into a soft factor $\Scal(P_\perp^2,\mu_0^2)$ defined as
\begin{equation}
    \Scal(\Pt^2,\mu_0^2)\equiv \exp\left(-\int_{\mu_0^2}^{P_\perp^2}\frac{\der\mu^2}{\mu^2}\frac{\alpha_s(\mu)N_c}{\pi}\left[\frac{1}{2}\ln\left(\frac{P_\perp^2}{\mu^2}\right)+\frac{s_L-\beta_0}{N_c}\right]\right)\,,\label{eq:soft-factor-appendix}
\end{equation}
with $\mu_0=c_0/r_{bb'}$ and $s_L=-C_F\ln(z_1z_2R^2)+N_c\ln(1+Q^2/M_{q\bar q}^2)$.

\subsection{Longitudinally polarized cross-section}

For longitudinally polarized virtual photons, the zeroth moment of the Fourier decomposition Eq.\,\eqref{eq:Fourier-phi} reads
\begin{align}
 \der \sigma^{(0),\lambda=\rm L}&=\Hcal_{\rm LO}^{0,\lambda=\rm L}\int\frac{\der^2\Bt}{(2\pi)^2}\int\frac{\der^2\rbbpt}{(2\pi)^2}e^{-i\qt\cdot\rbbpt}\hat G^0_{\eta_f}(\rbbpt,\mu_0)\mathcal{S}(\Pt^2,\mu_0^2) \nonumber\\
    &\times\left\{1+\frac{\alpha_s(\mu_R)N_c}{2\pi}f_1^{\lambda=\rmL}(\chi,z_1,R,\eta_f)+\frac{\alpha_s(\mu_R)}{2\pi N_c}f_2^{\lambda=\rmL}(\chi,z_1,R)+\frac{\alpha_s(\mu_R)}{\pi}\beta_0\ln\left(\frac{\mu_R^2}{P_\perp^2}\right)\right\}\nonumber\\
    &+\Hcal_{\rm LO}^{0,\lambda=\rm L}\int\frac{\der^2\Bt}{(2\pi)^2}\int\frac{\der^2\rbbpt}{(2\pi)^2}e^{-i\qt\cdot\rbbpt}\hat h^0_{\eta_f}(\rbbpt,\mu_0)\mathcal{S}(\Pt^2,\mu_0^2) \nonumber\\
    &\times\left\{\frac{\alpha_s(\mu_R)N_c}{2\pi}\left[1+\ln(R^2)\right]+\frac{\alpha_s(\mu_R)}{2\pi N_c}\left[-\ln(z_1z_2R^2)\right]\right\}+\mathcal{O}\left(\frac{q_\perp}{P_\perp},\frac{Q_s}{P_\perp},\alpha_sR^2,\alpha_s^2\right) \,,
\label{result_xsect_L}
\end{align}
with $\chi=Q/M_{q\bar q}=\bar Q/P_\perp $, and $\bar Q^2=z_1z_2Q^2$.

The $f_1$ and $f_2$ functions contribute to the NLO coefficient function at leading and subleading $N_c$, with
\begin{align}
 f^{\lambda=\rmL}_1(\chi,z_1,R,\eta_f)&=9-\frac{3\pi^2}{2}-\frac{3}{2}\ln\left(\frac{z_1z_2R^2}{\chi^2}\right)-\ln(z_1)\ln(z_2)-\ln(1+\chi^2)\ln\left(\frac{1+\chi^2}{z_1z_2}\right)\nonumber\\
    &+\left\{\textrm{Li}_2\left(\frac{z_2-z_1\chi^2}{z_2(1+\chi^2)}\right)-\frac{1}{4(z_2-z_1\chi^2)}\right.\nonumber\\
    &\left.+\frac{(1+\chi^2)(z_2(2z_2-z_1)+z_1(2z_1-z_2)\chi^2)}{4(z_2-z_1\chi^2)^2}\ln\left(\frac{z_2(1+\chi^2)}{\chi^2}\right)+(1\leftrightarrow2)\right\}   \nonumber\\
    &+\ln^2\left(\frac{x_c}{x_f}\right)+2\ln\left(\frac{x_c}{x_f}\right)-\mathcal{I}_{\rm kc}\left(\sqrt{\frac{x_c}{x_f}}\right)\,,    \label{f1_def} \\
    f^{\lambda=\rmL}_2(\chi,z_1,R)&=-8+\frac{19\pi^2}{12}+\frac{3}{2}\ln(z_1z_2R^2)-\frac{3}{4}\ln^2\left(\frac{z_1}{z_2}\right)-\ln(\chi) \,, \nonumber\\
    &+\left\{\frac{1}{4(z_2-z_1\chi^2)}+\frac{(1+\chi^2)z_1(z_2-(1+z_1)\chi^2)}{4(z_2-z_1\chi^2)^2}\ln\left(\frac{z_2(1+\chi^2)}{\chi^2}\right)\right.\nonumber\\
    &\left.+\frac{1}{2}\textrm{Li}_2(z_2-z_1\chi^2)-\frac{1}{2}\textrm{Li}_2\left(\frac{z_2-z_1\chi^2}{z_2}\right) +(1\leftrightarrow2)\right\}  \,.  \label{f2_def} 
\end{align}
We define the dilogarithm or Spence's function as
\begin{align}
    \textrm{Li}_2(x)=-\int_0^x\frac{\ln(1-z)}{z}\der z\,,
\end{align}
for $x\le 1$. The function $\mathcal{I}_{\rm kc}(X)$ coming from the kinematic constraint on gluons which do not cross the shock-wave is defined as (for $X\ge 0$) \cite{Caucal:2022ulg}
\begin{align}
    \mathcal{I}_{\rm kc}(X)= \left\{
\begin{array}{ll}
 4\ln^2(X) +2\textrm{Li}_2\left(\frac{1}{X^2}\right)& \mbox{if } X \ge 2 \\
 \int_{1/X}^{\infty}\der u \, \frac{16}{u(u^2-1)}\arctan\left(\frac{u-1}{u+1}\sqrt{\frac{2u+1}{2u-1}}\right)\ln(uX) & \mbox{else} 
\end{array}
\right.
\end{align}
For $X\le 2$, we were not able to find an analytic expression, except for $X=1$ where one has $\mathcal{I}_{\rm kc}(1)= 2\pi^2/27$. Since one must choose $x_f\ge x_c$ in order to have factorization as explained in supplemental material 3, the relevant values of $X$ in the NLO coefficient function $f_1$ are $X\le 1$.

Similarly, the second harmonic is given by
\begin{align}
 \der \sigma^{(2),\lambda=\rm L}&=\Hcal_{\rm LO}^{0,\lambda=\rm L}\int\frac{\der^2\Bt}{(2\pi)^2}\int\frac{\der^2\rbbpt}{(2\pi)^2}e^{-i\qt\cdot\rbbpt}\frac{\cos(2\theta)}{2}\hat h^0_{\eta_f}(\rbbpt,\mu_0)\mathcal{S}(\Pt^2,\mu_0^2) \nonumber\\
    &\times\left\{1+\frac{\alpha_s(\mu_R)N_c}{2\pi}\left[f_1^{\lambda=\rmL}(\chi,z_1,R,\eta_f)-\frac{5}{4}-\ln(R)\right]+\frac{\alpha_s(\mu_R)}{2\pi N_c}\left[f_2^{\lambda=\rmL}(\chi,z_1,R)+\frac{1}{2}\ln(z_1z_2R^2)\right]\right.\nonumber\\
    &\left.+\frac{\alpha_s(\mu_R)}{\pi}\beta_0\ln\left(\frac{\mu_R^2}{P_\perp^2}\right)\right\}\nonumber\\
    &+\Hcal_{\rm LO}^{0,\lambda=\rm L}\int\frac{\der^2\Bt}{(2\pi)^2}\int\frac{\der^2\rbbpt}{(2\pi)^2}e^{-i\qt\cdot\rbbpt}\frac{\cos(2\theta)}{2}\hat G^0_{\eta_f}(\rbbpt,\mu_0)\mathcal{S}(\Pt^2,\mu_0^2) \nonumber\\
    &\times\left\{\frac{\alpha_s(\mu_R)N_c}{\pi}\left[1+\ln(R^2)\right]+\frac{\alpha_s(\mu_R)}{\pi N_c}\left[-\ln(z_1z_2R^2)\right]\right\}+\mathcal{O}\left(\frac{q_\perp}{P_\perp},\frac{Q_s}{P_\perp},\alpha_sR^2,\alpha_s^2\right) \,.
\label{result_xsect_2_L}
\end{align}
Lastly, the expression for the higher modes $n\ge 4$ in the Fourier series Eq.\,\eqref{eq:Fourier-phi} is
\begin{align}
     \der \sigma^{(n=2p),\lambda=\rm L}&=\Hcal_{\rm LO}^{0,\lambda=\rm L}\int\frac{\der^2\Bt}{(2\pi)^2}\int\frac{\der^2\rbbpt}{(2\pi)^2}e^{-i\qt\cdot\rbbpt}\cos(n\theta) \hat G^0_{\eta_f}(\rbbpt,\mu_0)\mathcal{S}(\Pt^2,\mu_0^2)\nonumber\\
     &\times\frac{\alpha_s(\mu_R)(-1)^{p+1}}{n\pi}\left\{2N_c\left(\mathfrak{H}(p)-\frac{1}{n}\right)+2C_F\ln(R^2)-\frac{1}{N_c}\ln(z_1z_2)\right\}\nonumber\\
     &+\Hcal_{\rm LO}^{0,\lambda=\rm L}\int\frac{\der^2\Bt}{(2\pi)^2}\int\frac{\der^2\rbbpt}{(2\pi)^2}e^{-i\qt\cdot\rbbpt}\cos(n\theta) \hat h^0_{\eta_f}(\rbbpt,\mu_0)\mathcal{S}(\Pt^2,\mu_0^2)\nonumber\\
     &\times\frac{\alpha_s(\mu_R)(-1)^p}{n^2-4}\left\{N_c\left((n+2)\mathfrak{H}(p-1)+(n-2)\mathfrak{H}(p+1)-\frac{2(n^2+4)}{n^2-4}\right)\right.\nonumber\\
     &\left.+n\left(2C_F\ln(R^2)-\frac{1}{N_c}\ln(z_1z_2)\right)\right\}+\mathcal{O}\left(\frac{q_\perp^2}{P_\perp^2},\frac{Q_s^2}{P_\perp^2},\alpha_sR^2,\alpha_s^2\right)\,,
     \label{eq:R2R2b2b-cn-final-L}
\end{align}
where $\mathfrak{h}(p)= \sum_{k=1}^p \frac{1}{k}$ is the p$^{\rm th}$ harmonic number.

\subsection{Transversely polarized cross-section} 

We turn now to the hadronic part of the back-to-back dijet cross-section in the case of a transversely polarized virtual photon. The azimuthally averaged cross-section reads
\begin{align}
 \der \sigma^{(0),\lambda=\rmT}&=\Hcal_{\rm LO}^{0,\lambda=\rm T}\int\frac{\der^2\Bt}{(2\pi)^2}\int\frac{\der^2\rbbpt}{(2\pi)^2}e^{-i\qt\cdot\rbbpt}\hat G^0_{\eta_f}(\rbbpt,\mu_0)\mathcal{S}(\Pt^2,\mu_0^2) \nonumber\\
    &\times\left\{1+\frac{\alpha_s(\mu_R)N_c}{2\pi}f_1^{\lambda=\rmT}(\chi,z_1,R,\eta_f)+\frac{\alpha_s(\mu_R)}{2\pi N_c}f_2^{\lambda=\rmT}(\chi,z_1,R)+\frac{\alpha_s(\mu_R)}{\pi}\beta_0\ln\left(\frac{\mu_R^2}{P_\perp^2}\right)\right\}\nonumber\\
    &+\Hcal_{\rm LO}^{0,\lambda=\rmT}\int\frac{\der^2\Bt}{(2\pi)^2}\int\frac{\der^2\rbbpt}{(2\pi)^2}e^{-i\qt\cdot\rbbpt}\hat h^0_{\eta_f}(\rbbpt,\mu_0)\mathcal{S}(\Pt^2,\mu_0^2) \nonumber\\
    &\times\frac{-2\chi^2}{1+\chi^4}\left\{\frac{\alpha_s(\mu_R)N_c}{2\pi}\left[1+\ln(R^2)\right]+\frac{\alpha_s(\mu_R)}{2\pi N_c}\left[-\ln(z_1z_2R^2)\right]\right\}+\mathcal{O}\left(\frac{q_\perp}{P_\perp},\frac{Q_s}{P_\perp},\alpha_sR^2,\alpha_s^2\right) \,.
\label{result_xsect_T}
\end{align}

For transverse photons, the tensor structure of the hard factor is more complicated, as shown in Eq.\,(3) in the letter, and the contraction with the WW gluon TMD results in the following expressions for the $f_1$ and $f_2$ functions in the NLO coefficient function:
\begin{align}
    f_1^{\lambda=\rmT}(\chi,z_1,R,\eta_f)&=\frac{(1+\chi^2)^2}{1+\chi^4}f_{A,1}^{\lambda=\rmT}(\chi,z_1,R,\eta_f)-\frac{2\chi^2}{1+\chi^4}f_{B,1}^{\lambda=\rmT}(\chi,z_1,R,\eta_f) \,, \\
    f_2^{\lambda=\rmT}(\chi,z_1,R)&=\frac{(1+\chi^2)^2}{1+\chi^4}f_{A,2}^{\lambda=\rmT}(\chi,z_1,R)-\frac{2\chi^2}{1+\chi^4}f_{B,2}^{\lambda=\rmT}(\chi,z_1,R) \,,
\end{align}
with $f_{A,1}$, $f_{A,2}$  given by
\begin{align}
    f_{A,1}^{\lambda=\rmT}(\chi,z_1,R,\eta_f)&=\frac{19}{2}-\frac{3\pi^2}{2}-\frac{3}{2}\ln\left(\frac{z_1z_2R^2}{\chi^2}\right)+\frac{z_1z_2}{z_1^2+z_2^2}-\ln\left(1+\chi^2\right)\ln\left(\frac{1+\chi^2}{z_1z_2}\right)-\ln(z_1)\ln(z_2)\nonumber\\
    &+\left\{\frac{z_1^2\chi^2}{2(z_2-z_1\chi^2)(z_1^2+z_2^2)}+\textrm{Li}_2\left(\frac{z_2-z_1\chi^2}{z_2(1+\chi^2)}\right)\right.\nonumber\\
    &-\left.\frac{z_1z_2[4\chi^2-3z_1z_2(1+\chi^2)^2]}{2(z_1^2+z_2^2)(z_2-z_1\chi^2)^2}\ln\left(\frac{z_2(1+\chi^2)}{\chi^2}\right)+(1\leftrightarrow 2)\right\}\nonumber\\
    &+\ln^2\left(\frac{x_c}{x_f}\right)+2\ln\left(\frac{x_c}{x_f}\right)-\mathcal{I}_{\rm kc}\left(\sqrt{\frac{x_c}{x_f}}\right)\,, \label{fA1_def}\\
    f_{A,2}^{\lambda=\rmT}(\chi,z_1,R)&=-7+\frac{3\pi^2}{2}+\frac{3}{2}\ln(z_1z_2R^2)-\frac{1}{2}\left[1+\frac{2z_2+z_1^2(1+\chi^2)}{(1+\chi^2)(z_1^2+z_2^2)}\right]\ln^2\left(\frac{z_1}{z_2}\right)\nonumber\\
    &+\frac{1+z_1^2+z_2^2\chi^2}{(1+\chi^2)(z_1^2+z_2^2)}\frac{\pi^2}{6}+\frac{(1-\chi^2)(z_1-z_2)}{2(1+\chi^2)(z_1^2+z_2^2)}\left[\ln\left(\frac{z_2}{z_1}\right)\ln(z_1z_2)+2\textrm{Li}_2\left(-\frac{z_1}{z_2}\right)\right]\nonumber\\
    &-\frac{z_1z_2(1-\chi^2)}{2(z_1-z_2\chi^2)(z_2-z_1\chi^2)(z_1^2+z_2^2)}-\frac{1}{2(1+\chi^2)}\left[1+3\chi^2+\frac{2}{z_1^2+z_2^2}\right]\ln(\chi^2)\nonumber\\
    &+\left\{\frac{2z_2+z_1^2(1+\chi^2)}{(1+\chi^2)(z_1^2+z_2^2)}\left[\textrm{Li}_2(z_2-z_1\chi^2)-\textrm{Li}_2\left(\frac{z_2-z_1\chi^2}{z_2}\right)\right]\right.\nonumber\\
    &\left.+\frac{z_1z_2[-z_2(2+z_1)+2(1+z_1^2)\chi^2-z_1z_2\chi^4]}{2(z_2-z_1\chi^2)^2(z_1^2+z_2^2)}\ln\left(\frac{z_2(1+\chi^2)}{\chi^2}\right)+(1\leftrightarrow 2)\right\} \,, 
\end{align}
and $f_{B,1}$, $f_{B,2}$  given by
\begin{align}
    f_{B,1}^{\lambda=\rmT}(\chi,z_1,R,\eta_f)&=9-\frac{3\pi^2}{2}-\frac{3}{2}\ln\left(\frac{z_1z_2R^2}{\chi^2}\right)+\frac{(1-\chi^2)(z_1z_2-(z_1-z_2)^2\chi^2+z_1z_2\chi^4)}{4\chi^2(z_1-z_2\chi^2)(z_2-z_1\chi^2)(z_1^2+z_2^2)}\nonumber\\
    &-\ln(z_1)\ln(z_2)-\ln(1+\chi^2)\ln\left(\frac{1+\chi^2}{z_1z_2}\right)+\left\{\textrm{Li}_2\left(\frac{z_2-z_1\chi^2}{z_2(1+\chi^2)}\right)\right.\nonumber\\
    &\left.-\frac{z_1z_2(1+\chi^2)(z_2(2z_2-z_1)+z_1(2z_1-z_2)\chi^2)}{2(z_2-z_1\chi^2)^2(z_1^2+z_2^2)}\ln\left(\frac{z_2(1+\chi^2)}{\chi^2}\right)+(1\leftrightarrow2)\right\}\nonumber\\
    &+\ln^2\left(\frac{x_c}{x_f}\right)+2\ln\left(\frac{x_c}{x_f}\right)-\mathcal{I}_{\rm kc}\left(\sqrt{\frac{x_c}{x_f}}\right)\,, \label{fB1_def}\\
    f_{B,2}^{\lambda=\rmT}(\chi,z_1,R)&=-\frac{15}{2}+\frac{3\pi^2}{2}+\frac{3}{2}\ln(z_1z_2R^2)+\frac{(1+\chi^2)[z_1z_2(z_2-z_1)^2-(1-2z_1z_2)^2\chi^2+z_1z_2\chi^4]}{4\chi^2(z_1-z_2\chi^2)(z_2-z_1\chi^2)(z_1^2+z_2^2)}\nonumber\\
    &-\left[1+\frac{1}{2(z_1^2+z_2^2)}\right]\ln(\chi^2)+\frac{1-z_1z_2}{z_1^2+z_2^2}\frac{\pi^2}{6}+\frac{-2+3z_1z_2}{2(z_1^2+z_2^2)}\ln^2\left(\frac{z_1}{z_2}\right)\nonumber\\
    &+\left\{\frac{1-z_1z_2}{z_1^2+z_2^2}\left[\textrm{Li}_2\left(z_2-z_1\chi^2\right)-\textrm{Li}_2\left(\frac{z_2-z_1\chi^2}{z_2}\right)\right]\right.\nonumber\\
    &\left.-\frac{z_1^2z_2(1+\chi^2)(z_2-(1+z_1)\chi^2)}{2(z_2-z_1\chi^2)^2(z_1^2+z_2^2)}\ln\left(\frac{z_2(1+\chi^2)}{\chi^2}\right)+(1\leftrightarrow2)\right\} \,.
\end{align}
Similarly, the second harmonic coefficient in the Fourier decomposition of the transversely polarized cross-section is given by
\begin{align}
 \der \sigma^{(2),\lambda=\rm T}&=\Hcal_{\rm LO}^{0,\lambda=\rm T}\left(\frac{-2\chi^2}{1+\chi^4}\right)\int\frac{\der^2\Bt}{(2\pi)^2}\int\frac{\der^2\rbbpt}{(2\pi)^2}e^{-i\qt\cdot\rbbpt}\frac{\cos(2\theta)}{2}\hat h^0_{\eta_f}(\rbbpt,\mu_0)\mathcal{S}(\Pt^2,\mu_0^2) \nonumber\\
    &\times\left\{1+\frac{\alpha_s(\mu_R)N_c}{2\pi}\left[f_{B,1}^{\lambda=\rmT}(\chi,z_1,R,\eta_f)-\frac{5}{4}-\ln(R)\right]+\frac{\alpha_s(\mu_R)}{2\pi N_c}\left[f_{B,2}^{\lambda=\rmT}(\chi,z_1,R)+\frac{1}{2}\ln(z_1z_2R^2)\right]\right.\nonumber\\
    &\left.+\frac{\alpha_s(\mu_R)}{\pi}\beta_0\ln\left(\frac{\mu_R^2}{P_\perp^2}\right)\right\}\nonumber\\
    &+\Hcal_{\rm LO}^{0,\lambda=\rm T}\int\frac{\der^2\Bt}{(2\pi)^2}\int\frac{\der^2\rbbpt}{(2\pi)^2}e^{-i\qt\cdot\rbbpt}\frac{\cos(2\theta)}{2}\hat G^0_{\eta_f}(\rbbpt,\mu_0)\mathcal{S}(\Pt^2,\mu_0^2) \nonumber\\
    &\times\left\{\frac{\alpha_s(\mu_R)N_c}{\pi}\left[1+\ln(R^2)\right]+\frac{\alpha_s(\mu_R)}{\pi N_c}\left[-\ln(z_1z_2R^2)\right]\right\}+\mathcal{O}\left(\frac{q_\perp}{P_\perp},\frac{Q_s}{P_\perp},\alpha_sR^2,\alpha_s^2\right) \,.
\label{result_xsect_2_T}
\end{align}
Finally, the other Fourier modes read
\begin{align}
     \der \sigma^{(n=2p),\lambda=\rm T}&=\Hcal_{\rm LO}^{0,\lambda=\rm T}\int\frac{\der^2\Bt}{(2\pi)^2}\int\frac{\der^2\rbbpt}{(2\pi)^2}e^{-i\qt\cdot\rbbpt}\cos(n\theta) \hat G^0_{\eta_c}(\rbbpt,\mu_0)\mathcal{S}(\Pt^2,\mu_0^2)\nonumber\\
     &\times\frac{\alpha_s(\mu_R)(-1)^{p+1}}{n\pi}\left\{2N_c\left(\mathfrak{H}(p)-\frac{1}{n}\right)+2C_F\ln(R^2)-\frac{1}{N_c}\ln(z_1z_2)\right\}\nonumber\\
     &+\Hcal_{\rm LO}^{0,\lambda=\rm T}\left(\frac{-2\chi^2}{1+\chi^4}\right)\int\frac{\der^2\Bt}{(2\pi)^2}\int\frac{\der^2\rbbpt}{(2\pi)^2}e^{-i\qt\cdot\rbbpt}\cos(n\theta) \hat h^0_{\eta_c}(\rbbpt,\mu_0)\mathcal{S}(\Pt^2,\mu_0^2)\nonumber\\
     &\times\frac{\alpha_s(\mu_R)(-1)^p}{n^2-4}\left\{N_c\left((n+2)\mathfrak{H}(p-1)+(n-2)\mathfrak{H}(p+1)-\frac{2(n^2+4)}{n^2-4}\right)\right.\nonumber\\
     &\left.+n\left(2C_F\ln(R^2)-\frac{1}{N_c}\ln(z_1z_2)\right)\right\}+\mathcal{O}\left(\frac{q_\perp^2}{P_\perp^2},\frac{Q_s^2}{P_\perp^2},\alpha_sR^2,\alpha_s^2\right)\,.
     \label{eq:R2R2b2b-cn-final-T}
\end{align}

\subsection{The photo-production limit}

We expect a smooth $Q^2\to0$ limit of our result. We have indeed
\begin{align}
    f_{A,1}^{\lambda=\rmT}(0,z_1,R,\eta_f)&=\frac{19}{2}-\frac{7\pi^2}{6}-\frac{3}{2}\ln(z_1z_2R^2)+\frac{z_1z_2}{z_1^2+z_2^2}-\ln(z_1)\ln(z_2) \,, \nonumber\\
    &+\frac{3}{2(z_1^2+z_2^2)}\left[z_1^2\ln(z_2)+z_2^2\ln(z_1)\right]+\ln^2\left(\frac{x_c}{x_f}\right)+2\ln\left(\frac{x_c}{x_f}\right)-\mathcal{I}_{\rm kc}\left(\sqrt{\frac{x_c}{x_f}}\right) \,, \\
    \chi^2f_{B,1}^{\lambda=\rmT}(\chi,z_1,R)&=\frac{1}{4(z_1^2+z_2^2)}+\mathcal{O}(\chi^2) \,, \\
   f_{A,2}^{\lambda=\rmT}(0,z_1,R)&= -7+\frac{3\pi^2}{2} +\frac{3}{2}\ln(z_1z_2R^2)-\frac{1}{2}\ln^2\left(\frac{z_1}{z_2}\right)-\frac{1+z_1^2}{2(z_1^2+z_2^2)}\ln^2(z_1)\nonumber  \\
   &-\frac{1+z_2^2}{2(z_1^2+z_2^2)}\ln^2(z_2)-\frac{z_1(2+z_1)\ln(z_2)}{2(z_1^2+z_2^2)}-\frac{z_2(2+z_2)\ln(z_1)}{2(z_1^2+z_2^2)}-\frac{1}{2(z_1^2+z_2^2)} \,, \\
   \chi^2f_{B,2}^{\lambda=\rmT}(\chi,z_1,R)&=\frac{(z_2-z_1)^2}{4(z_1^2+z_2^2)}+\mathcal{O}(\chi^2) \,.
\end{align}
The fact that $\chi^2f_{B,1}^{\lambda=\rmT}$ and $\chi^2f_{B,2}^{\lambda=\rmT}$ converges towards a nonzero constant has an important consequence: the $\langle \cos(2\phi)\rangle$ anisotropy or $v_2$ coefficient proportional to the linearly polarized WW gluon TMD does not vanish in the photo-production limit at NLO (although it is zero at leading order).

\section{Supplemental Material 2: Further illustrative results}

The first set of plots shows the differential back-to-back dijet yield as a function of $q_\perp$ separately for longitudinally and transversely polarized virtual photons. As in Fig.\,(2) of the letter, the dotted red curve is the LO result, the dashed green curve is the NLO calculation 
including only the Sudakov factor, the dot-dashed blue curve is the NLO calculation including both Sudakov and small-$x$ evolution, and the solid black line is the full NLO calculation which includes in addition the NLO coefficient function. 

    \begin{figure}[htp]
    \centering
        \includegraphics[width=0.48\textwidth]{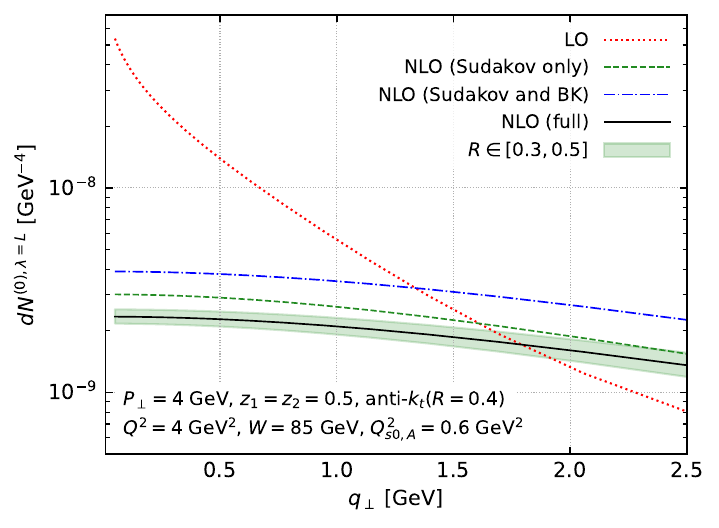} 
\hfill
        \includegraphics[width=0.48\textwidth]{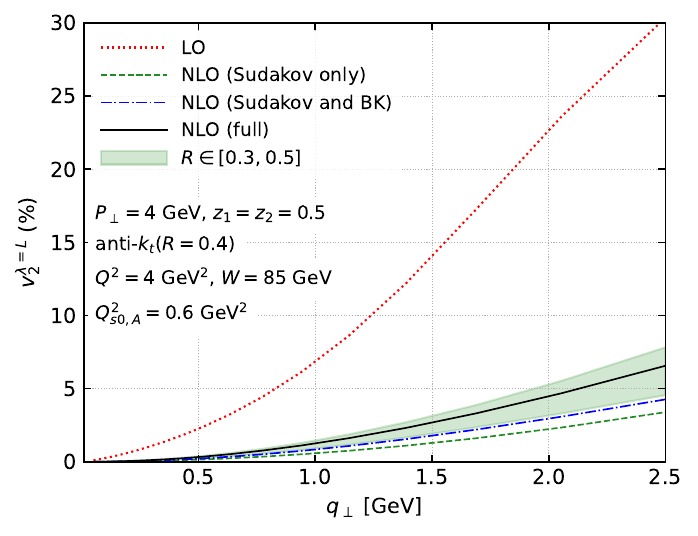}
    \caption{Azimuthally average back-to-back yield (left) and $v_2$ coefficient (right) as a function of the dijet momentum imbalance $q_\perp$ for a longitudinally polarized virtual photon.}\label{fig:longidudinal-qt-dep}
\end{figure}

Comparing Figs.\,\ref{fig:longidudinal-qt-dep} and \ref{fig:transverse-qt-dep} for these kinematics, the yield is observed to be dominated by the transversely polarized photon component. We also note that the $v_2$ is significantly larger for longitudinal photons compared to transverse photons, and its sensitivity to the jet radius is overall smaller, which could potentially help in the extraction of the linearly polarized WW gluon distribution. 

For longitudinal photons, the NLO coefficient function has a sizeable effect on the azimuthally averaged yield as seen in Fig.\,\ref{fig:longidudinal-qt-dep}. For transverse photons, the NLO coefficient functions have a negligible impact, but this is specific to our choice of kinematics in Fig.\,\ref{fig:transverse-qt-dep}. On the other hand, we see in Fig.\,\ref{fig:transverse-qt-dep-2} that the NLO coefficient function can give a significant contribution to the transversely polarized cross-section depending on the ratio between $Q^2$ and $M_{q\bar q}^2$. 

    \begin{figure}[htp]
    \centering
        \includegraphics[width=0.48\textwidth]{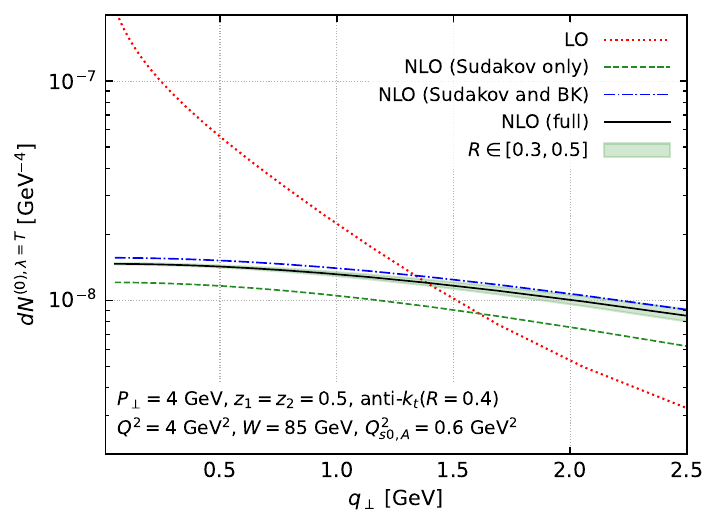} 
\hfill
        \includegraphics[width=0.48\textwidth]{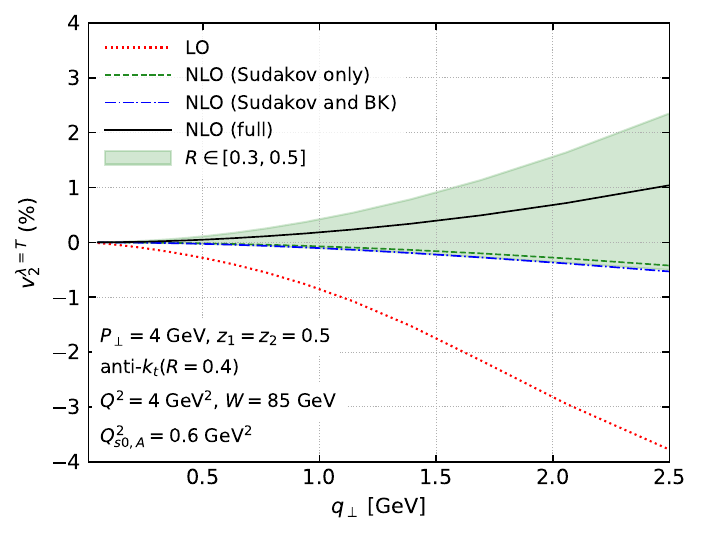}
    \caption{Azimuthally averaged back-to-back yield (left) and $v_2$ coefficient (right) as a function of the dijet momentum imbalance $q_\perp$ for a transversely polarized virtual photon.}\label{fig:transverse-qt-dep}
\end{figure}

    \begin{figure}[htp]
    \centering
        \includegraphics[width=0.48\textwidth]{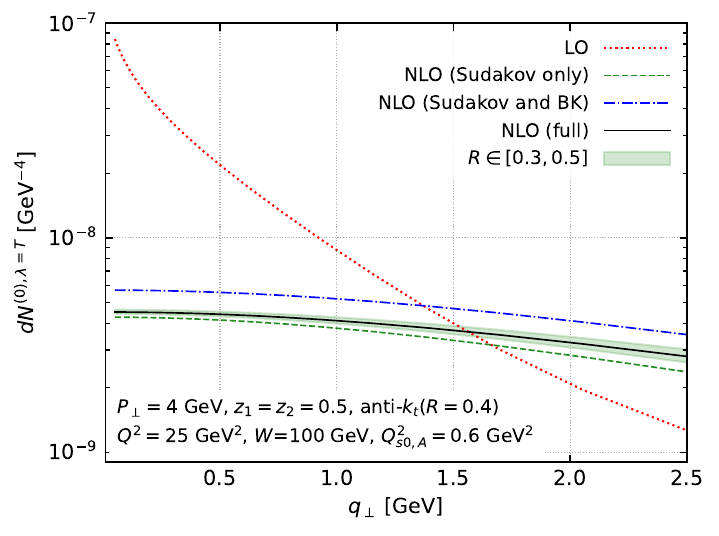} 
    \caption{Azimuthally averaged back-to-back yield as a function of the dijet momentum imbalance $q_\perp$ for a transversely polarized virtual photon for a different choice of DIS kinematics.}\label{fig:transverse-qt-dep-2}
\end{figure}

We discuss now the nuclear modification factor for the transversely polarized cross-section, for a larger jet $P_\perp (5$ GeV) than that considered in the main text. Although the $x_g$ value is larger, which results in a smaller range for small-$x$ evolution, jets with larger transverse momenta should be easier to measure at the EIC. As one can see from Fig.\,\ref{fig:ReA-pt=6}, the suppression for the curve labeled NLO (full) is less significant for $P_\perp=5$ GeV than for $P_\perp=4$ GeV because of the smaller amount of evolution $\Delta\eta=\eta_f-\eta_0\simeq 0.56$ (for the central value $\eta_f=\ln(1/x_g)$) relative to $\Delta\eta\simeq 1.0$ for $P_\perp=4$ GeV.
Nevertheless, it is important to note that the signal from nonlinear small-$x$ evolution can still be disentangled from Sudakov suppression, although the $\eta_f$ scale uncertainties are larger for smaller values of $\Delta\eta$.

\begin{figure}[htp]
    \centering
        \includegraphics[width=0.48\textwidth,page=1]{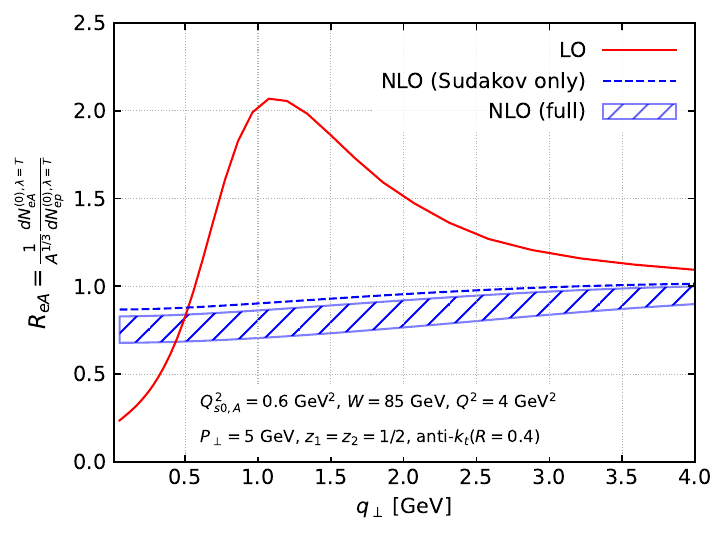} 
\hfill
        \includegraphics[width=0.48\textwidth,page=2]{ReA-qt-A-dep-2.pdf}
    \caption{Nuclear modification factor for the back-to-back dijet cross-section for transversely polarized photons, as a function of $q_\perp$ (left) and $A$ (right). The relative dijet momentum $P_\perp$ is set to $P_\perp=6$ GeV here.}\label{fig:ReA-pt=6}
\end{figure}

Another important aspect of our NLO calculation is that it enables one to address the convergence of QCD perturbation theory, improved by the resummation of small-$x$ logarithms. In Fig.\,\ref{fig:etaf-variations}, we show how the theory uncertainties coming from $\eta_f$ variations shrink from LO to NLO. For the LO curve, the LO BK equation (with fixed coupling) is used to evolve the WW gluon TMD and for the NLO curve, we use running coupling BK with kinematic constraint and NLO DGLAP single transverse logarithms corrections as in \cite{Ducloue:2019ezk,Ducloue:2019jmy}. Running coupling, kinematic constraint and NLO DGLAP single transverse logarithms are the dominant pieces of the NLO BK kernel. The reduction of the uncertainty bands comes from two effects: (i) the high energy evolution of the WW gluon TMD is slower at NLO, and (ii) the $x_f$ dependence of the NLO coefficient function partially compensates the $x_f$ dependence of the WW gluon TMD.

\begin{figure}[htp]
    \centering
    \includegraphics[width=0.48\textwidth]{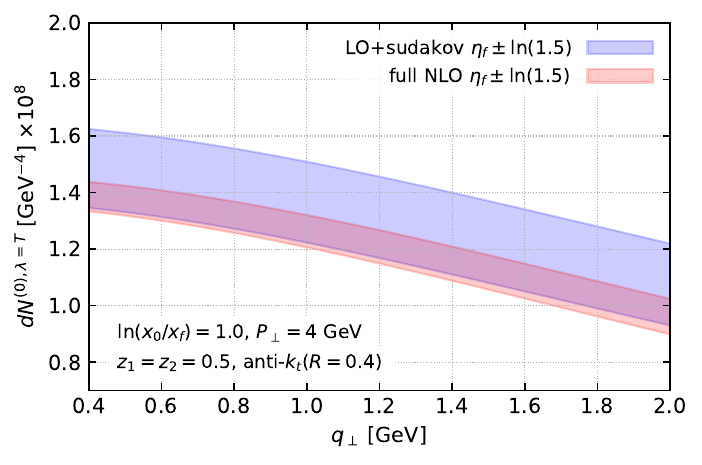}
    \caption{$\eta_f$ sensitivity of the back-to-back inclusive dijet cross-section at LO and NLO for a transversely polarized virtual photon. For the LO curve, we have also included the Sudakov factor (albeit formally a NLO effect) in order to stabilize the numerical evaluation of the WW gluon TMD.}
    \label{fig:etaf-variations}
\end{figure}

\section{Supplemental Material 3: Matching the WW gluon TMD evolution in $\eta_f=\ln(P^+/k_f^+)$ with the NLO impact factor}

In this Supplemental Material, we show how to perform the matching between our NLO impact factor derived from projectile rapidity factorization with target rapidity evolution ordered in $\eta=\ln(P^+/k_g^+)$. The final result is given by Eqs.\,\eqref{eq:xf_identity} and \eqref{eq:Yf_values} which respectively set the final value of the rapidity evolution in $\eta$ and the value of $Y_f$ to be used in our NLO impact factor.

For the sake of simplicity and also, in order to highlight the connection of our results with \cite{Ducloue:2019ezk}, we shall present the arguments for the kinematically constrained $Y$ evolution of the dipole $S$-matrix (related to the dipole gluon distribution $\Ncal$ by $S=1-\Ncal$) which satisfies the BK equation instead of the more complicated evolution equation in Eq.\,(6) in the letter. Then, we will provide the RGE for the WW gluon distribution in the $\eta$ representation.

\subsection{Kinematically constrained BK equation}

With the step function in Eq.\,(6) of the letter, the kinematically constrained $Y=\ln(k_g^-/q^-)$ evolution of the dipole S-matrix $S_Y(\rbbpt)$ reads
\begin{equation}
    \frac{\partial S_Y(\rbbpt)}{\partial Y}=\frac{\alpha_sN_c}{\pi} \int\frac{\der^2\zt}{2\pi}\Theta\left(-Y-\ln(\textrm{min}(\rzbt^2,\rzbpt^2)\mu_\perp^2)\right)\frac{\rbbpt^2}{\rzbt^2\rzbpt^2}\left[S_Y(\rzbt)S_Y(\rzbpt)-S_Y(\rbbpt)\right]\,, \label{eq:dipole-Yevol}
\end{equation}
with
\begin{equation}
    \mu_\perp^2=\frac{M_{q\bar q}^2+Q^2}{ec_0^2} \,.
\end{equation}
Eq.~\eqref{eq:dipole-Yevol} is solved between the initial rapidity scale $Y_0$ and the (arbitrary) rapidity factorization scale $Y_f$. The first question is how to relate $Y_0$ with the kinematics of the dijet process. Gluons with momenta $k_g^+\ge x_0 P^+$ belong to the target, and are accounted for by the classical background field. On the other hand, gluons in the projectile satisfy $k_g^+\le x_0 P^+$, which translates into the following constraint on $k_g^-$ for on-shell gluons:
\begin{equation}
    k_g^-\ge \frac{\boldsymbol{k_{g\perp}^2}}{2x_0P^+}\Leftrightarrow z_g \ge \frac{\boldsymbol{k_{g\perp}^2}}{W^2 + Q^2} 
 =\frac{x_c}{x_0}\frac{\boldsymbol{k_{g\perp}^2}}{\mu_\perp^2} \equiv z_0 \,,
\end{equation}
where $x_c$ is given by Eq.\,\eqref{eq:xf_identity}.

The minimal value of $z_g$ therefore depends on the transverse momentum of the emitted gluon from the projectile, that we relate to the transverse coordinate $\zt$ as  \cite{Caucal:2022ulg},
\begin{equation}
    \boldsymbol{k_{g\perp}^2}=\frac{1}{r_<^2}\,,\quad r_<^2\equiv \textrm{min}(\rzbt^2,\rzbpt^2)\label{eq:kt_rt-identity} \,.
\end{equation}
Note that Eq.\,\eqref{eq:kt_rt-identity} is only parametric, but prefactors of order one can always be absorbed in a redefinition of the arbitrary initial scale $x_0$.

Hence, the initial value $Y_0$ of the $Y$-evolution reads
\begin{equation}
    Y_0=-\ln\left(\frac{x_0}{x_c}\right)-\ln\left(\mu_\perp^2 r_<^2\right) \,,
\end{equation}
and therefore, \textit{depends on $\zt$}. It means that the Eq.\,\eqref{eq:dipole-Yevol} should be solved as a boundary value problem with initial boundary condition $S^{(0)}(\rbbpt)$, as first noticed in \cite{Ducloue:2019ezk}.  In integral form, it reads
\begin{equation}
    S_{Y_f}(\rbbpt)=S^{(0)}(\rbbpt)+\frac{\alpha_sN_c}{\pi}\int\frac{\der^2\zt}{2\pi}\int_{-\ln\left(\frac{x_0}{x_c}\right)-\ln(\mu_\perp^2 r_<^2)}^{Y_f} \!\!\!\!\!\!\!\!\!\!\!\!\!\!\!\!\!\!\!\!  \der Y \  \Theta\left(-Y-\ln(r_<^2\mu_\perp^2)\right)\frac{\rbbpt^2}{\rzbt^2\rzbpt^2}\left[S_{Y}(\rzbt)S_{Y}(\rzbpt)-S_{Y}(\rbbpt)\right] \,.
\end{equation}
We now formulate the evolution in terms of the rapidity with respect to the target $\eta = \ln( P^+ / k_g^+)$. This amount to performing the change of variables
\begin{equation}
\eta = Y+\ln(\mu_\perp^2 r_<^2)+\ln\left(\frac{1}{x_c} \right) \,,
\end{equation}
so that it now reads
\begin{align}
    S_{Y_f}(\rbbpt)&=S^{(0)}(\rbbpt)+\frac{\alpha_sN_c}{\pi}\int\frac{\der^2\zt}{2\pi}\int_{\ln\left(\frac{1}{x_0} \right)}^{Y_f+\ln(\mu_\perp^2 r_<^2)+\ln\left(\frac{1}{x_c} \right)}\der \eta \ \Theta\left(\ln\left(\frac{1}{x_c}\right) - \eta \right)\frac{\rbbpt^2}{\rzbt^2\rzbpt^2}\nonumber\\
    &\times\left[S_{\eta-\ln(\mu_\perp^2 r_<^2)-\ln\left(\frac{1}{x_c} \right)}(\rzbt)S_{\eta-\ln(\mu_\perp^2 r_<^2)-\ln\left(\frac{1}{x_c} \right)}(\rzbpt)-S_{\eta-\ln(\mu_\perp^2 r_<^2)-\ln\left(\frac{1}{x_c} \right)}(\rbbpt)\right] \,.
\end{align}
We now define
\begin{align}
    \eta_f&\equiv Y_f+\ln(\mu_\perp^2 \rbbpt^2)+\ln(1/x_c)\,, \label{eq:Yf_etaf_relation}\\
    \overline{S}_{\eta_f}(\rbbpt)&\equiv S_{\eta_f-\ln(\mu_\perp^2 \rbbpt^2)+\ln(x_c)}(\rbbpt) \,.
\end{align}
The resulting evolution equation for $\overline{S}_{\eta_f}(\rbbpt)$ is given by
\begin{align}
    \overline{S}_{\eta_f}(\rbbpt)&=S^{(0)}(\rbbpt)+\frac{\alpha_sN_c}{\pi}\int\frac{\der^2\zt}{2\pi}\int_{\ln\left(\frac{1}{x_0} \right)}^{\ln\left(\frac{1}{x_c} \right)}\der \eta \ \Theta\left(\eta_f-\ln\left(\frac{\rbbpt^2}{r_<^2}\right)-\eta\right)\frac{\rbbpt^2}{\rzbt^2\rzbpt^2}\nonumber\\
    &\times\left[\overline{S}_{\eta+\ln(\rzbt^2/r_<^2)}(\rzbt)\overline{S}_{\eta+\ln(\rzbpt^2/r_<^2)}(\rzbpt)-\overline{S}_{\eta+\ln(\rbbpt^2/r_<^2)}(\rbbpt)\right]\,, \label{eq:dipole-etaevol-integral}
\end{align}
where now the problem has become a well-defined initial value problem with initial condition $S^{(0)}(\rbbpt)$.
Finally, differentiating with respect to $\eta_f$ leads to
\begin{align}
    \frac{\partial \overline{S}_{\eta_f}(\rbbpt)}{\partial \eta_f}&=\frac{\alpha_sN_c}{\pi}\int\frac{\der^2\zt}{2\pi}\Theta\left(\ln\left(\frac{1}{x_c}\right) + \ln\left(\frac{\rbbpt^2}{r_<^2}\right) - \eta_f \right)\Theta\left(\eta_f -\ln\left(\frac{\rbbpt^2}{r_<^2}\right) -\ln\left(\frac{1}{x_0}\right)\right)\frac{\rbbpt^2}{\rzbt^2\rzbpt^2}\nonumber\\
    &\times\left[\overline{S}_{\eta_f-\ln(\rbbpt^2/\rzbt^2)}(\rzbt)\overline{S}_{\eta_f-\ln(\rbbpt^2/\rzbpt^2)}(\rzbpt)-\overline{S}_{\eta_f}(\rbbpt)\right]\,.\label{eq:dipole-etaevol}
\end{align}
This equation is the nonlocal version of the kinematically constrained RGE in $\eta$ of the dipole S-matrix derived in \cite{Ducloue:2019ezk}. The only difference is due to the presence of the first $\Theta$-function which depends on $x_c$. Such a dependence is worrisome, as it seems to imply a violation of the universality of the factorization of the high-energy logarithms since the $\Theta$-function depends on the kinematic of the specific process we are looking at (inclusive back-to-back dijet). In fact, it is easy to check that this $x_c$-dependence constraint has no effect on the evolution equation, as long as $\eta_f\le \ln(1/x_c)$. This is because the rapidity shift $\ln(\rbbpt^2/r_<^2)$ must be positive so when $\ln(1/x_c)\ge \eta_f$, the condition in the $\Theta$-function is automatically satisfied. Therefore, one should interpret the number $\ln(1/x_c)$ as the upper limit for the allowed values of $\eta_f$ in order to preserve consistent universal factorization of the small-$x$ logarithms.

In summary, we can either solve Eq.\,\eqref{eq:dipole-Yevol} or solve Eq.\,\eqref{eq:dipole-etaevol} and use the relation
\begin{equation}
    S_{Y_f}(\rbbpt)=\overline{S}_{Y_f+\ln(\mu_\perp^2 \rbbpt^2)+\ln(1/x_c)}(\rbbpt) \,.
    \label{eq:Y_vs_eta}
\end{equation}
If we choose $Y_f \sim 0$ (e.g. corresponding to $z_f = z_1 z_2 \sim \mathcal{O}(1)$ as in \cite{Caucal:2022ulg,Taels:2022tza}), it is evident from the evolution in the $\eta$ representation in Eq.\,\eqref{eq:Y_vs_eta} that we resum a Sudakov logarithm $\ln(\mu_\perp^2 \rbbpt^2)$, in addition to small-$x$ logarithm $\ln(1/x_c)$. This Sudakov single logarithm hidden in the evolution equation will precisely cancel the ``$-1$" coefficient of the Sudakov single logarithm obtained in \cite{Caucal:2023nci} which now restores the agreement with the Sudakov factor obtained in collinear factorization.

Alternatively, we can choose the factorization scale $Y_f\sim-\ln(\mu_\perp^2 \rbbpt^2)$, so that the evolution in the $\eta$ representation resums only the large small-$x$ logarithms $\eta_c = \ln(1/x_c)$. 
With this choice, a Sudakov single logarithm is generated in the NLO coefficient function. We recall that in \cite{Caucal:2023nci}, the NLO coefficient function has a term
\begin{align}
    \frac{\alpha_s N_c}{\pi} Y_f &=-\frac{\alpha_s N_c}{\pi} \ln\left(\frac{\rbbpt^2\Pt^2}{c_0^2}\right)+ \frac{\alpha_s N_c}{\pi} \left[1-\ln\left(\frac{1+\chi^2}{z_1z_2}\right) \right] +\frac{\alpha_sN_c}{\pi}\ln\left(\frac{x_c}{x_f}\right) \,.
    \label{eq:Yf_values}
\end{align}
which is expressed on the right-hand side in terms of $\eta_f-\eta_c=\ln(x_c/x_f)$ thanks to Eq.\,\eqref{eq:Yf_etaf_relation}.
The first term in the NLO coefficient function is then resummed into the Sudakov soft factor, canceling the $-1$ coefficient of the Sudakov single logarithm. The next two terms contain finite $\mathcal{O}(\alpha_s)$ pieces contributing to the NLO coefficient function. Finally, the last term is the residual factorization scale $x_f$ dependence coming from the transition from projectile to target rapidity factorization. In this work, we choose this latter approach where the factorization scale $\eta_f \sim \ln(1/x_c)$ or equivalently $Y_f \sim -\ln(\mu_\perp^2 \rbbpt^2)$.

The transition from $Y_f$ to $\eta_f$ and the role of the kinematic constraint can be better understood using a phase space diagram like the one shown in Fig.\,\ref{fig:phase-space}. A point in this phase space represents a gluon emission in the high-energy evolution with a given squared transverse momentum $\boldsymbol{k}^2_{g\perp}$ on the $x$-axis and a given minus light-cone relative momentum $z_g=k_g^-/q^-$ on the $y$-axis. The kinematic constraint in the $Y_f$ evolution equation and $z_f$ chosen of order 1 (meaning that the red line goes up to the top of the figure) enforces a gluon emission in the high-energy evolution to belong to the blue phase space, corresponding to $x_0\ge x \ge x_c$ with $x=k_g^+/P^+$. By choosing $z_f$ of order $\qt^2/\Pt^2$ as in Fig.\,\ref{fig:phase-space}, the effective area for the $\eta_f$-ordered high energy evolution shrinks to the hatched area, while the blue triangle above the red line gives a single Sudakov logarithm contribution.

\begin{figure}[htp]
    \centering
    \includegraphics[width=0.48\textwidth]{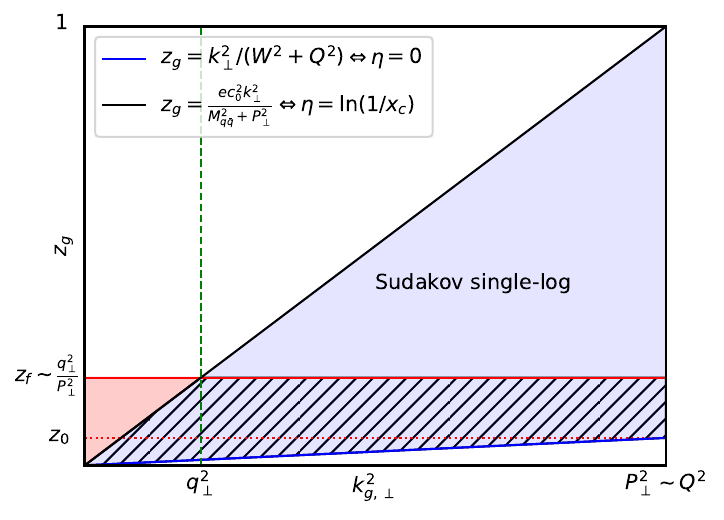}
    \caption{The hatched region represents the phase-space for high-energy evolution when a kinematic constraint is imposed on the BK kernel and $z_f$ is taken of order $\qt^2/\Pt^2$.}
    \label{fig:phase-space}
\end{figure}

Last but not least, the choice Eq.\,\eqref{eq:Yf_values} implies that $z_f\sim q_\perp/P_\perp$. This forces us to revise the calculation of the subtraction term from the unconstrained virtual contribution in the small-$x$ evolution of the WW gluon TMD, given by Eq.\,(4.50) in \cite{Caucal:2022ulg} which assumed $z_f=\mathcal{O}(1)$. We have now (with $\xi=z_g/z_f$)
\begin{align}
    -\frac{\alpha_sN_c}{2\pi^2}\hat G^{ij}_\eta(\rbbpt)\int_0^{1}\frac{\der \xi}{\xi}\int\der^2\zt\Theta\left(\textrm{min}(\rzbt^2,\rzbpt^2)Q_f^2-\frac{1}{\xi}\right)\frac{\rbbpt^2}{\rzbt^2\rzbpt^2}=-\frac{\alpha_sN_c}{2\pi}\times \mathcal{I}_{\rm kc}(\mu_\perp|\rbbpt|\sqrt{z_f})\hat G^{ij}_\eta(\rbbpt)\,.\label{eq:virtual-dmmx-finite-piece}
\end{align}
with
\begin{equation}
    \mathcal{I}_{\rm kc}(X)\equiv \int_0^{1}\frac{\der \xi}{\xi}\int\frac{\der^2\zt}{\pi}\Theta\left(\textrm{min}(\rzbt^2,\rzbpt^2)X^2\xi-\rbbpt^2\right)\frac{\rbbpt^2}{\rzbt^2\rzbpt^2}
\end{equation}
After the change of variable from $Y_f$ to $\eta_f$, the argument of the function $\mathcal{I}_{\rm kc}$ becomes simply $(x_c/x_f)^{1/2}$.
This additional finite piece is included in the NLO coefficient function $f_1$ (see Eqs.\,\eqref{f1_def} and \eqref{fA1_def}-\eqref{fB1_def}).

We close this section by pointing out that the choice of factorization scale $Y_f \sim -\ln(\mu_\perp^2 \rbbpt^2)$ was employed in \cite{Balitsky:2022vnb, Balitsky:2023hmh} to separate the Sudakov and small-$x$ regions within the rapidity-only factorization approach. 

\subsection{Kinematically constrained evolution for the WW gluon distribution}
The kinematically constrained $Y$ evolution of the unpolarized WW gluon distribution $ \hat G^{(0)}_{Y_f}(\rbbpt)$ reads
\begin{align}
   & \frac{\partial \hat G^{(0)}_{Y_f}(\rbbpt)}{\partial Y_f}  = -\frac{\alpha_s N_c}{\pi} \int \frac{\der^2 \zt}{2\pi} \frac{\rbbpt^2}{\rzbt^2 \rzbpt^2} \Theta\left(-Y_f-\ln(\textrm{min}(\rzbt^2,\rzbpt^2)\mu_\perp^2)\right) \nonumber \\
   & \times \left\{  \hat G^{(0)}_{Y_f}(\rbbpt) + \frac{2}{\rbbpt^2} \left[1 - \frac{2 (\rzbt\cdot\rzbpt)^2}{\rzbt^2 \rzbpt^2} \right] \hat G^{(2)}_{Y_f}(\rzbpt,\rzbt)   + \left[ \frac{\rbbpt^i}{\rbbpt^2} + \frac{\rzbt^i}{\rzbt^2} \right] \hat G^{(1),i}_{Y_f}(\rzbt,\rzbpt) +  \left[ \frac{\rbbptc^i}{\rbbptc^2} + \frac{\rzbpt^i}{\rzbpt^2} \right] \hat G^{(1),i}_{Y_f}(\rzbpt,\rzbt) \right\} \,,
   \label{eq:kc-Y--DMMX}
\end{align}
where the WW gluon distribution is given by
\begin{align}
    \alpha_s \hat G^{(0)}_{Y_f}(\rzbt,\rzbpt) & = -2\left \langle \Tr\left[V(\bt) \left(\partial^i V^\dagger(\bt) \right) 
    V(\bt') \left(\partial^i V^\dagger(\bt') \right) \right] \right \rangle_{Y_f} \,,
\end{align}
and the new operators appearing in the evolution $\hat G^{(1),i}_{Y_f}(\rzbt,\rzbpt)$ and $\hat G^{(2)}_{Y_f}(\rbzt,\rbzpt)$ are defined as
\begin{align}
    \alpha_s \hat G^{(1),i}_{Y_f}(\rzbt,\rzbpt) & = -\frac{2}{N_c} \left \langle \Tr\left[V(\zt) V^\dagger(\bt)V(\bt') \left(\partial^i V^\dagger(\bt') \right) \right] \Tr\left[V(\bt) V^\dagger(\zt)\right] \right \rangle_{Y_f} \nonumber \\
    & + \frac{2}{N_c} \left \langle \Tr\left[V(\bt) V^\dagger(\zt)V(\bt') \left(\partial^i V^\dagger(\bt') \right) \right] \Tr\left[V(\zt) V^\dagger(\bt)\right] \right \rangle_{Y_f}\,, \\
    \alpha_s \hat G^{(2)}_{Y_f}(\rzbt,\rzbpt) & = 
    -\frac{2}{N_c}\left \langle  \Tr\left[V(\zt)V^\dagger(\bt') \right] \Tr\left[V(\bt')V^\dagger(\zt) \right] \right \rangle_{Y_f}  - \frac{2}{N_c}\left \langle \Tr\left[V(\zt)V^\dagger(\bt) \right] \Tr\left[V(\bt)V^\dagger(\zt) \right] \right \rangle_{Y_f} \nonumber \\
    & + \frac{2}{N_c}\left \langle \Tr\left[V(\bt)V^\dagger(\bt') \right] \Tr\left[V(\bt')V^\dagger(\bt) \right] \right \rangle_{Y_f}  + 2N_c \,.
\end{align}
Following the same steps as in the previous subsection, we find that the nonlocal $\eta$-ordered kinematically constrained RGE for the unpolarized WW gluon distribution is given by
\begin{align}
   \frac{\partial \overline{\hat G^{(0)}}_{\eta_f}(\rbbpt)}{\partial {\eta_f}}  = -\frac{\alpha_s N_c}{\pi} & \int \frac{\der^2 \zt}{2\pi} \frac{\rbbpt^2}{\rzbt^2 \rzbpt^2} \Theta\left(\ln\left(\frac{1}{x_c}\right) + \ln\left(\frac{\rbbpt^2}{r_<^2}\right) - {\eta_f} \right)\Theta\left({\eta_f} -\ln\left(\frac{\rbbpt^2}{r_<^2}\right) -\ln\left(\frac{1}{x_0}\right)\right) \nonumber \\
   & \times \left\{  \overline{\hat G^{(0)}}_{\eta_f}(\rbbpt) + \frac{2}{\rbbpt^2} \left[1 - \frac{2 (\rzbt\cdot\rzbpt)^2}{\rzbt^2 \rzbpt^2} \right] \overline{\hat G^{(2)}}_{{\eta_f}-\ln(\rbbpt^2/r_{<}^2)}(\rzbt,\rzbpt)   \right. \nonumber \\
   & \left. + \left[ \frac{\rbbpt^i}{\rbbpt^2} + \frac{\rzbt^i}{\rzbt^2} \right] \overline{\hat G^{(1),i}}_{{\eta_f}-\ln(\rbbpt^2/r_{<}^2)}(\rzbt,\rzbpt) +  \left[ \frac{\rbbptc^i}{\rbbptc^2} + \frac{\rzbpt^i}{\rzbpt^2} \right]\overline{ \hat G^{(1),i}}_{{\eta_f}-\ln(\rbbpt^2/r_{<}^2)}(\rzbpt,\rzbt) \right\} \,,
   \label{eq:non-local-eta-DMMX}
\end{align}
with the change of variables
\begin{align}
    \overline{\hat G^{(0)}}_{\eta_f}(\rbbpt)&=\hat G^{(0)}_{Y_f-\ln(\mu_\perp^2\rbbpt^2)+\ln(x_c)}(\rbbpt)\,\\
     \overline{\hat G^{(1)}}_{\eta_f}(\rzbt,\rzbpt)&=\hat G^{(1)}_{Y_f-\ln(\mu_\perp^2r_<^2)+\ln(x_c)}(\rzbt,\rzbpt)\,,\\
      \overline{\hat G^{(2)}}_{\eta_f}(\rzbt,\rzbpt)&=\hat G^{(2)}_{Y_f-\ln(\mu_\perp^2r_<^2)+\ln(x_c)}(\rzbt,\rzbpt)\,.
\end{align}
(In the letter, and throughout the supplemental material, we use the notation $\overline{\hat G_{\eta_f}}(\rbbpt)\to \hat G_{\eta_f}(\rbbpt)$ for simplicity.)
A similar evolution equation can be derived for the linearly polarized WW gluon distribution. Unlike the BK equation, the evolution equations in Eqs.\,\eqref{eq:kc-Y--DMMX} and \eqref{eq:non-local-eta-DMMX} are not closed (not even at large $N_c$), and are a subset of the B-JIMWLK hierarchy. In practice, the evolution of the WW gluon distribution can be obtained from the Langevin formulation of the JIMWLK equation \cite{Hatta:2016ujq,Cali:2021tsh}. 

\section{Supplemental Material 4: Proof of WW gluon TMD factorization for the transversely polarized photon cross-section and calculation of the NLO coefficient function}

The NLO impact factor for back-to-back dijets, first computed in \cite{Caucal:2023nci} for longitudinally polarized photons, and in this Supplemental Material for transversely polarized photons, relies on rapidity factorization with respect to the projectile $Y=\ln(k_g^-/q^-)$. In a nutshell, the rapidity singularity as $k_g^-\to0$ is removed by subtracting the kinematically constrained slow gluon divergence up to some \textit{arbitrary} factorization scale $Y_f=\ln(k_f^-/q^-)$. The resulting finite NLO correction constitutes the NLO impact factor. It depends on $Y_f$, while the singular counterterm cancels with the $\mathcal{O}(\alpha_s)$ contribution to the kinematically constrained rapidity evolution of the WW gluon TMD from some initial rapidity scale $Y_0$ up to $Y_f$ (with $|Y_0|\gg |Y_f|$).

The kinematically constrained slow gluon divergence also appears to depend on a target rapidity scale $k_c^+$ which enforces lifetime ordering $k_g^+>k_c^+$. A remarkable outcome of our calculation is that the back-to-back limit of virtual diagrams with the gluon crossing the shock-wave is well defined {\it if and only if}
\begin{equation}
    k_c^+\equiv \frac{M_{q\bar q}^2+Q^2}{2ec_0^2q^-}\,,\quad c_0=2e^{-\gamma_E}\,.\label{eq:kfplus}
\end{equation}
thereby fixing the scale $k_c^+$ in the kinematic constraint imposing life-time ordering.

We follow the notation of \cite{Caucal:2022ulg,Caucal:2023nci} and write the NLO cross-section for transversely polarized photons as
\begin{align}
        \der \sigma^{(0),\lambda=\rmT}&=\Hcal_{\rm LO}^{0,\lambda=\rmT}\int\frac{\der^2\Bt}{(2\pi)^2}\int\frac{\der^2\rbbpt}{(2\pi)^2}e^{-i\qt\cdot\rbbpt}\hat G^0_{Y_f}(\rbbpt)
    \times\left\{1+\frac{\alpha_s N_c}{4\pi}\ln^2\left(\frac{\Pt^2\rbbpt^2}{c_0^2}\right)\right.\nonumber\\
    &-\frac{\alpha_s }{\pi}\left[C_F\ln\left(\frac{1}{z_1 z_2R^2}\right)-N_c \ln\left(\frac{z_f}{z_1z_2}\right)\right]\ln\left(\frac{\Pt^2\rbbpt^2}{c_0^2}\right)\nonumber\\
    &+\frac{\alpha_sC_F}{\pi}\left[\frac{3}{2}\ln(c_0^2)-3\ln(R)+\frac{1}{2}\ln^2\left(\frac{z_1}{z_2}\right)+\frac{17}{2}-\frac{5\pi^2}{6}\right]+\frac{\alpha_sN_c}{2\pi}\ln\left(\frac{z_f^ 2}{z_1z_2}\right)\ln(c_0^2)\,\nonumber\\
    &\left.+\frac{\alpha_sN_c}{2\pi}\left[\frac{1}{2}\ln\left(\frac{z_1z_2}{z_f^2}\right)-\frac{3C_F}{2N_c}\right]\frac{\Hcal_{\rm NLO,1}^{\lambda=\textrm {T},ii}(\Pt)}{2\Hcal_{\rm LO}^{0,\lambda=\rmT}(\Pt)}-\frac{\alpha_s}{2\pi N_c}\frac{\Hcal_{\rm NLO,2}^{\lambda=\textrm {T},ii}(\Pt)}{2\Hcal_{\rm LO}^{0,\lambda=\rmT}(\Pt)}\right\}\nonumber\\
    &+\frac{\alpha_sN_c}{2\pi}\Hcal_{\rm LO}^{0,\lambda=\rmT}\left(\frac{-2\chi^2}{1+\chi^4}\right)\int\frac{\der^2\Bt}{(2\pi)^2}\int\frac{\der^2\rbbpt}{(2\pi)^2}e^{-i\qt\cdot\rbbpt}\hat h^0_{Y_f}(\rbbpt)\times\left\{1+\frac{2C_F}{N_c}\ln(R^2)-\frac{1}{N_c^2}\ln(z_1z_2)\right\}\nonumber\\
    &-\frac{\alpha_sN_c}{2\pi}\Hcal_{\rm LO}^{0,\lambda=\rmT}\int\frac{\der^2\Bt}{(2\pi)^2}\int\frac{\der^2\rbbpt}{(2\pi)^2}e^{-i\qt\cdot\rbbpt}\hat G^{0}_{Y_f}(\rbbpt)\int_0^{z_f}\frac{\der z_g}{z_g}\int\frac{\der^2\zt}{\pi}\Theta\left(\textrm{min}(\rzbt^2,\rzbpt^2)Q_f^2-\frac{z_f}{z_g}\right)\frac{\rbbpt^2}{\rzbt^2\rzbpt^2}\nonumber\\
    &+\der \sigma^{(0),\lambda=\rm T}_{\rm other}\,,\label{eq:transverse-b2b-start}
\end{align}
with $Q_f^2\equiv 2k_c^+ k_f^-$.
This expression is valid for jets defined with anti-$k_t$ \cite{Cacciari:2008gp} or any definition from the generalized $k_t$ family \cite{Dokshitzer:1997in,Wobisch:1998wt,Salam:2010nqg,Cacciari:2011ma}. The terms in the first two lines gather contributions that give explicit double and single Sudakov logarithms for back-to-back kinematics. However, since in the end, $z_f$ must be chosen of order $1/(\rbbpt^2\Pt^2)$, any $\ln(z_f)$ is a Sudakov logarithm.

The hard factors $\Hcal_{\rm NLO,1}$ and $\Hcal_{\rm NLO,2}$ are computed analytically in the next two subsections. They come from virtual graphs in which the gluon does not interact with the target background field, and therefore simply factorize in terms of the WW gluon TMD.  The term proportional to $\hat h^0$ is the contribution from soft real gluon emissions emitted close to the jet cone boundary. 
The term proportional to $\hat G^{0}_{Y_f}$ above the last line is the difference between the kinematically unconstrained and constrained virtual term of the rapidity evolution of the WW gluon TMD, which is now added to the impact factor. 

Using the constraint Eq.\,\eqref{eq:kfplus} for $k_c^+$, i.e. $z_f/Q_f^2\sim 1/\Pt^2$ (derived in the following subsection by inspection of the back-to-back limit of the virtual graphs with gluons crossing the shockwave), and $z_f\sim 1/(\rbbpt^2\Pt^2)$ given by Eq.\,\eqref{eq:Yf_values} derived in the previous section, we find
\begin{equation}
  \int_0^{z_f}\frac{\der z_g}{z_g}\int\frac{\der^2\zt}{\pi}\Theta\left(\textrm{min}(\rzbt^2,\rzbpt^2)Q_f^2-\frac{z_f}{z_g}\right)\frac{\rbbpt^2}{\rzbt^2\rzbpt^2}= \int_0^{1}\frac{\der \xi}{\xi}\int\frac{\der^2\zt}{\pi}\Theta\left(\textrm{min}(\rzbt^2,\rzbpt^2) \xi-\rbbpt^2\right)\frac{\rbbpt^2}{\rzbt^2\rzbpt^2}=\frac{2\pi^2}{27} \,. \label{eq:sub_term_dmmx}
\end{equation}
Therefore this term does not give rise to Sudakov logarithms in the back-to-back limit for $z_f$ of order $q_\perp^2/P_\perp^2$. This should be contrasted to what was claimed in \cite{Taels:2022tza,Caucal:2022ulg}: in these papers, $z_f$ was considered to be of order one (more precisely, $z_f = z_1 z_2 $) and $Q_f$ of order $P_\perp$, meaning that Eq.\,\eqref{eq:sub_term_dmmx} would contain double Sudakov logarithms as shown in \cite{Caucal:2022ulg} (cf.\,Eq.\,(4.50)-(4.51) in this reference). 

Yet, there is no contradiction. The way Sudakov double logarithms appear in the calculation depends on the scheme one uses to resum small-$x$ logarithms and its associated natural factorization scale $z_f$. In the $Y$-ordered scheme, with $z_f=\mathcal{O}(1)$, the  Sudakov double logarithm comes from the sum of the double log in the first line in Eq.\,\eqref{eq:transverse-b2b-start} and the double log hidden in Eq.\,\eqref{eq:sub_term_dmmx}. In the $\eta$-ordered scheme with $z_f=\mathcal{O}(1/(\rbbpt^2\Pt^2))$, the Sudakov double logarithms comes from the first two lines of Eq.\,\eqref{eq:transverse-b2b-start} and the $\der \sigma^{(0),\lambda=\rm T}_{\rm other}$ contribution. Likewise, the origin of the Sudakov single logarithms is also scheme dependent.

The term $\der \sigma^{(0),\lambda=\rmT}_{\rm other}$ gathers all virtual diagrams with the gluon crossing the shockwave. Demonstrating that this term factorizes at leading power in $q_\perp/P_\perp$ in terms of the WW gluon TMD is the main goal of this supplemental material. This is not obvious as this term depends on a CGC operator which does not manifestly collapse to the WW gluon TMD if one performs the same correlation expansion as for the LO CGC operator. The formula for $\der \sigma^{(0),\lambda=\rmT}_{\rm other}$ was derived in \cite{Caucal:2022ulg}. It reads
\begin{align}
    &\der \sigma^{(0),\lambda=\rm T}_{\rm other}=\frac{\alpha_{\rm em}e_f^2N_c\deltatwo}{(2\pi)^6}\int\der^2\xt\der^2\xt'\der^2\yt\der^2\yt' e^{-i\ktone\cdot\rxxtp-i\kttwo\cdot\ryytp} \ 2z_1^2z_2^2\frac{QK_1(\bar Qr_{x'y'})}{r_{x'y'}}\nonumber\\
    &\times\frac{\alpha_s}{\pi}\Bigg\{ \int_0^{z_1}\frac{\der z_g}{z_g}\int\frac{\der^2\zt}{\pi}\left\{e^{-i\frac{z_g}{z_1}\ktone\cdot\rzxt}\frac{\bar Q K_1(QX_V)}{X_V}\Xi_{\rm NLO,1}\left[-\frac{z_g(z_g-z_1)^2z_2}{2 z_1^3}\frac{\rzxt\cdot\rxytp}{\rzxt^2}\right.\right.\nonumber\\
    &\left.+(z_1^2+z_2^2)\left(1-\frac{z_g}{z_1}+\frac{z_g^2}{2z_1^2}\right)\frac{\RtS\cdot\rxytp}{\rzxt^2}\right]-(z_1^2+z_2^2)e^{-\frac{\rzxt^2}{\rxyt^2e^{\gamma_E}}}\left(1-\frac{z_g}{z_1}+\frac{z_g^2}{2z_1^2}\right)\frac{\rxyt\cdot\rxytp}{\rzxt^2}\nonumber\\
    &\times QK_1(\bar Qr_{xy})C_F\Xi_{\rm LO}-e^{-i\frac{z_g}{z_1}\ktone\cdot\rzxt}\frac{\bar QK_1( Q X_V)}{X_V}\Xi_{\rm NLO,1}\left[\frac{z_g(z_1-z_g)}{2(z_g+z_2)}\frac{\rzxt\cdot\rxytp}{\rzxt^2}\right.\nonumber\\
    &+[z_1(z_1-z_g)+z_2(z_2+z_g)]\left(1-\frac{z_g}{z_1}\right)\left(1+\frac{z_g}{z_2}\right)\left(1-\frac{z_g}{2z_1}-\frac{z_g}{2(z_2+z_g)}\right)\nonumber\\
    &\left.\left.\times\frac{(\RtV\cdot\rxytp)(\rzxt\cdot\rzyt)}{\rzxt^2\rzyt^2}+\frac{z_g(z_1-z_g)(z_g+z_2-z_1)^2}{2z_1^2z_2}\frac{(\RtV\times\rxytp)(\rzxt\times\rzyt)}{\rzxt^2\rzyt^2}\right]\right\}+(1\leftrightarrow 2)\Bigg\}+c.c.\,,\label{eq:V-no-sud-other-full}
\end{align}
where we introduce the two transverse vectors:
\begin{align}
    \RtS&=\rxyt+\frac{z_g}{z_1}\rzxt \,,\\
    \RtV&=\rxyt-\frac{z_g}{z_{2}+z_g}\rzyt \,.
\end{align}
The argument $X_V$ of the NLO light-cone wave-function $K_1(QX_V)$ (note $K_n(x)$ is the modified Bessel functions of the second kind and order $n$)  is defined as 
\begin{align}
X_{\rm V}^2=z_{2}(z_1-z_g)\rxyt^2+z_g(z_1-z_g)\rzxt^2+z_{2}z_g\rzyt^2\,,
\end{align}
and the CGC operator $\Xi_{\rm NLO,1}$ reads
\begin{align}
   \Xi_{\rm NLO,1}(\xt,\yt\zt;\xt',\yt')&=\frac{1}{N_c}\left \langle \textrm{Tr}[t^aV(\xt)V^\dagger(\zt)t_aV(\zt)V^\dagger(\yt)-C_F][V(\yt')V^\dagger(\xt')-1] \right\rangle_{Y_f}\,.\label{eq:Xi-NLO1}
\end{align}
This CGC operator has an additional dependence on the transverse coordinate $\zt$ of the gluon crossing the shock-wave, which complicates the extraction of the leading power in $q_\perp/P_\perp$ contribution in Eq.\,\eqref{eq:V-no-sud-other-full}. The extraction of the back-to-back limit of Eq.\,\eqref{eq:Xi-NLO1} and Eq.\,\eqref{eq:V-no-sud-other-full} is discussed in the subsection~\textit{Hard factor from dressed self-energy and vertex corrections}.

\subsection{Hard factor from initial state self-energy, vertex corrections and unresolved real emissions}

We start with the calculation of the hard factor $\Hcal_{\rm NLO,1}$ defined in \cite{Caucal:2022ulg}, and it reads
\begin{equation}
    \Hcal_{\rm NLO,1}^{\textrm{T},ij}(\Pt)\equiv\frac{1}{2}\alpha_{\rm em}\alpha_se_f^2\deltatwo\int\frac{\der^2\ut}{(2\pi)}\int\frac{\der^2\ut'}{(2\pi)}e^{-i\Pt\cdot\ruupt}\ut^i\ut'^j\Rcal_{\rm LO}^{\lambda=\rmT}(\ut,\ut')\ln(\Pt^4\ut^2\ut'^2)\,,\label{eq:hard-NLO1}
\end{equation}
with the LO light-cone wave-function $\Rcal_{\rm LO}^{\lambda=\rmT}$ given by
\begin{align}
      \Rcal_{\mathrm{LO}}^{\mathrm{T}}(\ut,\ut') &=  2 z_1 z_{2} \left[z_1^2 + z_{2}^2 \right]  \frac{\ut \cdot \ut'}{u_\perp u_\perp'}  \bar{Q}^2K_1(\bar{Q}u_\perp) K_1(\bar{Q}u_\perp')\,.\label{eq:dijet-NLO-TLO} 
\end{align}
Introducing the function 
\begin{equation}
k_1(\chi)=\int_0^\infty \der u \ u J_1(u)K_1(\chi u)\ln(u)\,,
\end{equation}
it is relatively straightforward to write $\Hcal_{\rm NLO,1}$ in terms of $k_1$ as
\begin{align}
\Hcal_{\rm NLO,1}^{\textrm{T},ij}(\Pt)
&=\alpha_{\rm em}\alpha_se_f^2\deltatwo\times z_1z_2(z_1^2+z_2^2)\left\{\frac{\delta^{ij}}{(\Pt^2+\bar Q^2)^2}\kappa_{T,1}\left(\chi\right)-\frac{4\bar Q^2\Pt^i\Pt^j}{(\Pt^2+\bar Q^2)^4}\kappa_{T,2}\left(\chi\right)\right\}\,,
\label{eq:kappa-def}
\end{align}
where we define the two functions $\kappa_{T,1}$ and $\kappa_{T,2}$ as
\begin{align}
    \kappa_{T,1}(\chi)&=4\chi(1+\chi^2)k_1(\chi) \,,\\
    \kappa_{T,2}(\chi)&=\frac{(1+\chi^2)^2}{\chi}\left((3\chi^2-1)k_1(\chi)-\chi(1-\chi^2)k_1'(\chi)\right)-\frac{1}{\chi^2}+\chi^2 \,.
\end{align}
Let us now find an analytic expression for $k_1(\chi)$ using the identity
\begin{equation}
    \ln(x)=\lim\limits_{\alpha\to 0}\partial_\alpha x^\alpha\,,
\end{equation}
as in the longitudinally polarized case \cite{Caucal:2023nci}. Introducing the ordinary hypergeometric function ${}_2F_1(a,b,c;z)$, we have
\begin{align}
    k_1(\chi)&=\lim\limits_{\alpha\to 0}\partial_\alpha  \int_0^\infty \der u \ u^{1+\alpha}J_1(u\chi) K_1(u\chi) \nonumber  \\
    &=\lim\limits_{\alpha\to 0}\partial_\alpha\left\{\frac{2^\alpha}{\chi^{3+\alpha}}\Gamma\left(1+\frac{\alpha}{2}\right)\Gamma\left(2+\frac{\alpha}{2}\right){}_2F_1\left(1+\frac{\alpha}{2},2+\frac{\alpha}{2},2;-\chi^{-2}\right)\right\} \nonumber  \\
    &=-\frac{1}{2\chi(1+\chi^2)}\left(1-\ln(\chi^2/c_0^2)\right)+\frac{1}{2\chi^3}\left(\partial_a\,{}_2F_1(a,2,2;-\chi^{-2}) \Big|_{a=1} +\partial_b\,{}_2F_1(1,b,2;-\chi^{-2}) \Big|_{b=2}\right)\,.
\end{align}
Then, using ${}_2F_1(a,b,b;z)=(1-z)^{-a}$, one gets the derivative with respect to $a$ at $a=1$:
\begin{equation}
    \partial_a\,{}_2F_1(a,2,2;z) \Big|_{a=1}=-\frac{\ln(1-z)}{1-z}\,.
\end{equation}
Thanks to the identity \cite{abramowitz1964handbook}
\begin{equation}
    {}_2F_1(a,b,c;z)=(1-z)^{-b}{}_2F_1\left(b,c-a,c;\frac{z}{z-1}\right)\,, \label{eq:ab-2F1}
\end{equation}
we change the $b$ derivative into an $a$ derivative. Furthermore, using 
\begin{equation}
    {}_2F_1(b,1,2;z)=\frac{(1-z)^b(1-z-(1-z)^b)}{(b-1)z} \,,
\end{equation}
the derivative with respect to $b$ at $b=2$ gives
\begin{equation}
    \partial_b\,{}_2F_1(1,b,2;z) \Big |_{b=2}=-\frac{z+\ln(1-z)}{z(1-z)}\,.
\end{equation}
The integral $k_1(\chi)$ reads then
\begin{equation}
k_1(\chi)=-\frac{1}{2\chi(1+\chi^2)}\left[\ln\left(\frac{\chi^2}{c_0^2}\right)+(1-\chi^2)\ln\left(1+\frac{1}{\chi^2}\right)\right]\,.
\end{equation}
From this result, one can obtain analytic formulas for $\kappa_{T,1}$ and $\kappa_{T,2}$:
\begin{align}
\kappa_{T,1}(\chi)&=-2\ln\left(\frac{\chi^2}{c_0^2}\right)-2(1-\chi^2)\ln\left(1+\frac{1}{\chi^2}\right)\,,\label{eq:kappa1-final}\\
\kappa_{T,2}(\chi)&=2-\frac{1}{\chi^2}-\chi^2-(1-\chi^2)(3+\chi^2)\ln\left(1+\frac{1}{\chi^2}\right)-2\ln\left(\frac{\chi^2}{c_0^2}\right)\,,\label{eq:kappa2-final}
\end{align}
which defined the hard factor $\Hcal_{\rm NLO,1}^{\lambda=\rmT}$ in Eq.\,\eqref{eq:kappa-def}.

\subsection{Hard factor from final state vertex correction}

For transversely polarized virtual photons, the hard factor $\Hcal_{\rm NLO,2}^{\lambda=\textrm{T},ij}$ is defined by \cite{Caucal:2022ulg}:
\begin{align}
    &\Hcal_{\rm NLO,2}^{\lambda=\textrm{T},ij}(\Pt)=\frac{1}{2}\alpha_{\rm em}\alpha_se_f^2\deltatwo\int\frac{\der^2\ut}{(2\pi)}\int\frac{\der^2\ut'}{(2\pi)}e^{-i\Pt\ruupt} \ \ut^i\ut'^j\Rcal_{\rm LO}^{\lambda=\rm T}(\ut,\ut')\nonumber\\
    &\times\int_0^{z_1}\frac{\der z_g}{z_g}\left\{\frac{\bar Q_{\rm V3}K_1(\bar{Q}_{\mathrm{V3}} u_\perp)}{\bar QK_1(\bar Q u_\perp)}\right.\nonumber\\
    &\times \left[\frac{\left[z_1(z_1-z_g)+z_2(z_2+z_g)\right]}{[z_1^2+z_2^2]}(1+z_g)\left(1-\frac{z_g}{z_1}\right)e^{i\Pt\cdot\ut} K_0(-i\Delta_{\rm V3}u_\perp)\right.\nonumber\\
    &-\frac{\left[z_1(z_1-z_g)+z_2(z_2+z_g)\right]}{[z_1^2+z_2^2]}\left(1-\frac{z_g}{2z_1}+\frac{z_g}{2z_2}-\frac{z_g^2}{2z_1z_2}\right)e^{i\frac{z_g}{z_1}\Pt\cdot\ut}\Jcal_{\odot}\left(\ut,\left(1-\frac{z_g}{z_1}\right)\Pt,\Delta_{\rm V3}\right)\nonumber\\
    &\left.-i\frac{z_g(z_g+z_2-z_1)^2}{z_1z_2(z_1^2+z_2^2)}e^{i\frac{z_g}{z_1}\Pt\cdot\ut}\frac{(\ut\times\ut')}{\ut\cdot\ut'}\Jcal_{\otimes}\left(\ut,\left(1-\frac{z_g}{z_1}\right)\Pt,\Delta_{\rm V3}\right)\right]\nonumber\\
    &+\ln\left(\frac{z_gP_\perp u_\perp}{c_0z_1z_2}\right)\Bigg\}+(1\leftrightarrow 2)\,,\label{eq:HNLO2-transverse}
\end{align}
with the $\Jcal_{\odot}$ and $\Jcal_{\otimes}$ functions (computed in \cite{Caucal:2021ent}):
\begin{align}
        \Jcal_{\odot}(\rt,\Kt,\Delta)&=\int\frac{\der^2 \lt}{(2\pi)}\frac{2\lt \cdot \Kt \ e^{i\lt \cdot \rt}}{\lt^2\left[(\lt-\Kt)^2-\Delta^2- i \epsilon\right]}\,,    \label{eq:Jdot-def}\\
        \Jcal_{\otimes}(\rt,\Kt,\Delta)&=\int\frac{\der^2 \lt}{(2\pi)}\frac{(-i)\lt \times \Kt \ e^{i\lt \cdot \rt}}{\lt^2\left[(\lt-\Kt)^2-\Delta^2- i \epsilon\right]}\,.
    \label{eq:Jtimes-def}
\end{align}
The kinematic variables $\bar Q_{\rm V3}$ and $\Delta_{\rm V3}$ in Eq.\,\eqref{eq:HNLO2-transverse} are defined as  $\bar Q_{\rm V3}^2=z_1z_2(1-z_g/z_1)(1+z_g/z_2)Q^2$ and $\Delta_{\rm V3}^2=(1-z_g/z_1)(1+z_g/z_2)\Pt^2$.

Contrary to the longitudinally polarized case \cite{Caucal:2023nci}, the first term proportional to $ K_0(-i\Delta_{\rm V3}u_\perp)$ does contribute because of the more complicated tensorial structure. Let us then divide the calculation and first compute the new terms proportional to $K_0(-i\DV u)$ and $\Jcal_{\otimes}$.

\paragraph{Computation of the $K_0(-i\DV u)$ term.} We define the integral
\begin{align}
    \Ical_{K}^{ik}&=\int\frac{\der^2\ut}{(2\pi)}e^{-i\Pt\cdot\ut}\frac{\ut^i\ut^k}{u_\perp}K_1(\bar{Q}_{\mathrm{V3}} u_\perp)e^{i\Pt\cdot\ut}K_0(-i\DV u) \nonumber  \\
    &=\int\frac{\der^2\ut}{(2\pi)}\frac{\ut^i\ut^k}{u_\perp}K_1(\bar{Q}_{\mathrm{V3}} u_\perp)K_0(-i\DV u) \,.
\end{align}
Based on our experience with the longitudinally polarized case \cite{Caucal:2023nci}, the trick is to write the second Bessel function in momentum space as an integral over $\lt$ and then switch the order of integration:
\begin{align}
    \Ical_{K}^{ik}&=\int\frac{\der^2\ut}{(2\pi)}\frac{\ut^i\ut^k}{u_\perp}K_1(\bar{Q}_{\mathrm{V3}} u_\perp)\int\frac{\der^2\lt}{(2\pi)}\frac{e^{i\lt\cdot\ut}}{\lt^2-\DV^2-i\epsilon} \nonumber  \\
    & = \int\frac{\der^2\lt}{(2\pi)}\frac{1}{\lt^2-\DV^2-i\epsilon}\int\frac{\der^2\ut}{(2\pi)}\frac{\ut^i\ut^k}{u_\perp}K_1(\bar{Q}_{\mathrm{V3}} u_\perp)e^{i\lt\cdot\ut} \nonumber  \\
    &=\frac{1}{\QV}\int\frac{\der^2\lt}{(2\pi)}\frac{1}{\lt^2-\DV^2-i\epsilon}\left[ \frac{\delta^{ik}}{(\lt^2+\bar{Q}_{\mathrm{V3}}^2)}-\frac{2\lt^i\lt^k}{(\lt^2+\bar{Q}_{\mathrm{V3}}^2)^2}\right] \nonumber  \\
    &=-\frac{\delta^{ik}}{2\QV(\DV^2+\QV^2)}\left[1+\frac{\QV^2}{\DV^2+\QV^2}\ln\left(\frac{\DV^2}{\QV^2}\right)-\frac{i\pi\QV^2}{\DV^2+\QV^2}\right] \,.
\end{align}
We now write the contribution of this integral to $\Hcal^{\textrm{T},ij}_{\rm NLO,2}$ as
\begin{equation}
    \Hcal_{\textrm{NLO},2,K}^{\textrm{T},ij}(\Pt)=\alpha_{\rm em}\alpha_se_f^2\deltatwo\times z_1z_2(z_1^2+z_2^2)\left\{\frac{\delta^{ij}}{(\Pt^2+\bar Q^2)^2}\tau_{T,1,K}\left(\chi\right)-\frac{4\bar Q^2\Pt^i\Pt^j}{(\Pt^2+\bar Q^2)^4}\tau_{T,2,K}\left(\chi\right)\right\} +(1\leftrightarrow 2) \,,
\end{equation}
with $\chi=Q/M_{q\bar q}=\bar Q/P_\perp$. After replacing the values of $\bar Q_{\rm V3}$ and $\Delta_{\rm V3}$, the corresponding factor $\tau_{T,1,K}$ reads
\begin{align}
    \tau_{T,1,K}(\chi)&=\left[\frac{\chi^2}{1+\chi^2}\ln(\chi^2)-1\right]\int_0^{z_1}\frac{\der z_g}{z_g}\frac{z_1(z_1-z_g)+z_2(z_2+z_g)}{z_1^2+z_2^2}\frac{z_2(1+z_g)}{z_2+z_g}\,.
\end{align}
The $z_g$ integral is divergent but the divergence disappears once all terms are summed.
Similarly, the factor proportional to the $\Pt^i\Pt^j$ term gives
\begin{align}
\tau_{T,2,K}(\chi)
&=\frac{1}{2}\left[\ln(\chi^2)-1-\frac{1}{\chi^2}\right]\int_0^{z_1}\frac{\der z_g}{z_g}\frac{z_1(z_1-z_g)+z_2(z_2+z_g)}{z_1^2+z_2^2}\frac{z_2(1+z_g)}{z_2+z_g} \nonumber \\
&=\frac{\chi^2+1}{2\chi^2}\times\tau_{T,1,K}(\chi) \,.
\end{align}

\paragraph{Computation of the $\Jcal_{\otimes}$ term.} We now compute the term proportional to $\Jcal_{\otimes}$ in Eq.\,\eqref{eq:HNLO2-transverse}. The integral we want to compute is then
\begin{equation}
    \Ical_{\otimes}^{ik}=i\int\frac{\der^2\ut}{(2\pi)}e^{-i\Pt\cdot\ut}\frac{\ut^i\ut^k}{u_\perp}K_1(\bar{Q}_{\mathrm{V3}} u_\perp)e^{i\frac{z_g}{z_1}\Pt\cdot\ut}\Jcal_{\otimes}\left(\ut,\left(1-\frac{z_g}{z_1}\right)\Pt,\Delta_{\rm V3}\right) \,.
\end{equation}
After the change of variable $\lt\to\lt-\Kt$ this integral becomes
\begin{align}
    \Ical^{ik}_{\otimes}&=\int\frac{\der^2\ut}{(2\pi)}\frac{\ut^i\ut^k}{u_\perp}K_1(\bar{Q}_{\mathrm{V3}} u_\perp)\int\frac{\der^2 \lt}{(2\pi)}\frac{\lt \times \Kt \  e^{i\lt \cdot \ut}}{(\lt+\Kt)^2\left[\lt^2-\Delta_{\rm V3}^2- i \epsilon\right]} \nonumber \\
    &=\int\frac{\der^2 \lt}{(2\pi)}\frac{\lt \times \Kt}{(\lt+\Kt)^2\left[\lt^2-\Delta_{\rm V3}^2- i \epsilon\right]}\int\frac{\der^2\ut}{(2\pi)}\frac{\ut^i\ut^k}{u_\perp}K_1(\bar{Q}_{\mathrm{V3}} u_\perp)e^{i\lt\cdot\ut} \nonumber \\
    &=\frac{1}{\bar{Q}_{\mathrm{V3}}}\int\frac{\der^2 \lt}{(2\pi)}\frac{\lt \times \Kt}{(\lt+\Kt)^2\left[\lt^2-\Delta_{\rm V3}^2- i \epsilon\right]}\left[ \frac{\delta^{ik}}{(\lt^2+\bar{Q}_{\mathrm{V3}}^2)}-\frac{2\lt^i\lt^k}{(\lt^2+\bar{Q}_{\mathrm{V3}}^2)^2}\right] \,.
\end{align}
The integral of the first term $\propto\delta^{ik}$ vanishes because the integrand is odd. The second term has a more complicated tensor structure, since it vanishes when $i=k$, but does not vanish when $i=1$, $k=2$ or $i=2$, $k=1$. As it is a symmetric tensor, we have
\begin{align}
\Ical^{ik}_{\otimes}
    &=\frac{-2(\Kt^i\epsilon^{km}+\Kt^k\epsilon^{im})\Kt^m}{\bar{Q}_{\mathrm{V3}}\Kt^2}\int_0^\infty\der \ell\frac{\ell^4 K_\perp}{(\ell^2+K_\perp^2)\left[\ell^2-\Delta_{\rm V3}^2- i \epsilon\right](\ell^2+\bar{Q}_{\mathrm{V3}}^2)^2}\int_0^{2\pi}\frac{\der\theta}{2\pi}\frac{\sin^2(\theta)\cos(\theta)}{1+\frac{2\ell K_\perp}{\ell^2+K_\perp^2}\cos(\theta)} \,.
\end{align}
Computing the $\theta$ integral thanks to the formula
\begin{equation}
    \int_0^{2\pi}\frac{\der \theta}{2\pi}\frac{\cos(\theta)\sin^2(\theta)}{1+a\cos(\theta)}=\frac{-2+a^2+2\sqrt{1-a^2}}{2a^3}
\end{equation}
for $0<a\le 1$, we find
\begin{equation}
    \int_0^{2\pi}\frac{\der\theta}{2\pi}\frac{\sin^2(\theta)\cos(\theta)}{1+\frac{2\ell K_\perp}{\ell^2+K_\perp^2}\cos(\theta)}=-\frac{\ell^2+K_\perp^2}{4}\left[\Theta(K_\perp-\ell)\frac{\ell}{K_\perp^3}+\Theta(\ell-K_\perp)\frac{K_\perp}{\ell^3}\right]\,.
\end{equation}
We have then two integrals over $\ell$ that need to be computed
\begin{align}
    \Ical^{ik}_{\otimes,1}&=\int_0^{K_\perp}\der \ell\frac{\ell^5}{\left[\ell^2-\Delta_{\rm V3}^2\right](\ell^2+\bar{Q}_{\mathrm{V3}}^2)^2} \,, \\
    \Ical^{ik}_{\otimes,2}&=\int_{K_\perp}^{\infty}\der \ell\frac{\ell}{\left[\ell^2-\Delta_{\rm V3}^2- i \epsilon\right](\ell^2+\bar{Q}_{\mathrm{V3}}^2)^2} \,.
\end{align}
We find
\begin{align}    \Ical^{ik}_{\otimes,1}&=-\frac{\Kt^2\QV^2}{2(\DV^2+\QV^2)(\Kt^2+\QV^2)}\nonumber\\
    &+\frac{1}{2(\DV^2+\QV^2)^2}\left[\DV^4\ln\left(1-\frac{\Kt^2}{\DV^2}\right)+\QV^2(2\DV^2+\QV^2)\ln\left(1+\frac{\Kt^2}{\QV^2}\right)\right] \,, \label{eq:l5-int} \\
    \Ical^{ik}_{\otimes,2}&=-\frac{1}{2(\Delta_{\rm V3}^2+\bar Q_{\rm V3}^2)(\Kt^2+\bar Q_{\rm V3}^2)}-\frac{1}{2(\Delta_{\rm V3}^2+\bar Q_{\rm V3}^2)^2}\ln\left(\frac{\Delta_{\rm V3}^2-\Kt^2}{\Kt^2+\bar Q_{\rm V3}^2}\right)\nonumber\\
    &+\frac{i\pi}{2(\Delta_{\rm V3}^2+\bar Q_{\rm V3}^2)^2}\,.
\end{align}
We now write the contribution of this integral to $\Hcal^{\textrm{T},ij}_{\rm NLO,2}$ as
\begin{align}
    \Hcal^{\textrm{T},ij}_{\rm NLO,2,\otimes}&=\alpha_{\rm em}\alpha_se_f^2\deltatwo\times z_1z_2(z_1^2+z_2^2)\int_0^{z_1}\frac{\der z_g}{z_g}\frac{-z_g(z_g+z_2-z_1)^2}{z_1z_2(z_1^2+z_2^2)}\QV\Ical_{\otimes}^{ik}\epsilon^{kl}\left[\frac{\delta^{lj}}{\Pt^2+\bar Q^2}-\frac{2\Pt^l\Pt^j}{(\Pt^2+\bar Q^2)^2}\right]\nonumber\\
    &+(1\leftrightarrow 2)+c.c. \,.
\end{align}
The tensor structure can be simplified using the identities $\epsilon^{km}\epsilon^{kj}=\delta^{mj}$ and $\epsilon^{im}\epsilon^{jk}=\delta^{ij}\delta^{mk}-\delta^{ik}\delta^{mj}$:
\begin{align}
    (\Pt^i\epsilon^{km}+\Pt^k\epsilon^{im})\Pt^m\epsilon^{kl}\delta^{lj}&=2\Pt^i\Pt^j-\delta^{ij}\Pt^2 \,, \\
    (\Pt^i\epsilon^{km}+\Pt^k\epsilon^{im})\Pt^m\epsilon^{kl}\Pt^l\Pt^j&=\Pt^i\Pt^j\Pt^2 \,.
\end{align}
The term proportional to $\delta^{ij}$, that we express as
\begin{equation}
    \alpha_{\rm em}\alpha_se_f^2\deltatwo\times z_1z_2(z_1^2+z_2^2)\frac{\delta^{ij}}{(\Pt^2+\bar Q^2)^2}\times\tau_{T,1,\otimes}\left(\chi\right) \,,
\end{equation}
reads
\begin{align}
    &\tau_{T,1,\otimes}=-2\int_0^{z_1}\der z_g\frac{(z_g+z_2-z_1)^2}{z_1z_2(z_1^2+z_2^2)}\left(1-\frac{z_g}{z_1}\right)^2\frac{\Pt^2(\Pt^2+\bar Q^2)}{4(\Delta_{\rm V3}^2+\bar Q_{\rm V3}^2)(\Kt^2+\bar Q_{\rm V3}^2)}\left\{1+\frac{\QV^2}{\Kt^2}\right.\nonumber\\
    &\left.+\frac{\Kt^2+\bar Q_{\rm V3}^2}{\DV^2+\QV^2}\left[\frac{\DV^4}{\Kt^4}\ln\left(\frac{\DV^2}{\QV^2}\right)+\left(1-\frac{\DV^4}{\Kt^4}\right)\ln\left(\frac{\DV^2-\Kt^2}{\QV^2+\Kt^2}\right)-\frac{(\DV^2+\QV^2)^2}{\Kt^4}\ln\left(1+\frac{\Kt^2}{\QV^2}\right)\right]\right\}\nonumber\\
    &+(1\leftrightarrow 2) \,.
\end{align}
Replacing the values of $\QV$, $\Kt$ and $\DV$ in this expression we find
\begin{align}
    \tau_{T,1,\otimes}(\chi)=\int_0^{z_1}&\der z_g  \ \frac{z_1(z_g+z_2-z_1)^2}{2z_2(z_1^2+z_2^2)(z_1-z_g)^2}\left\{-\frac{z_2(z_1-z_g)}{z_1(z_2+z_g)}+\frac{1}{1+x^2}\ln(\chi^2)+(1+\chi^2)\ln\left(1+\frac{z_2(z_1-z_g)}{z_1(z_2+z_g)\chi^2}\right)\right.\nonumber\\
    &\left.+\frac{z_g(z_2(z_1-z_g)+z_1(z_2+z_g))}{z_1^2(z_2+z_g)^2(1+\chi^2)}\ln\left(\frac{z_g}{z_2(z_1-z_g)+z_1(z_2+z_g)\chi^2}\right)\right\}+(1\leftrightarrow 2) \,.
\end{align}
Notice that even though there is an apparent double pole in $z_g=z_1$ from the prefactor of the curly bracket, the integral is convergent in $z_g=z_1$ because the other factor behaves like $(z_1-z_g)^2$ in the vicinity of $z_1$. This integral is therefore well defined.

Let us conclude with the simplification of $\tau_{T,2,\otimes}$ defined such that the $\Pt^i\Pt^j$ tensor structure of the "$\otimes$” term reads
\begin{equation}
    \alpha_{\rm em}\alpha_se_f^2\deltatwo\times z_1z_2(z_1^2+z_2^2)\frac{-4\bar Q^2\Pt^i\Pt^j}{(\Pt^2+\bar Q^2)^4}\times\tau_{T,2,\otimes}\left(\chi\right) \,.
\end{equation}
Adding the two contributions coming from the $\delta^{lj}$ and $\Pt^l\Pt^j$ terms, we get
\begin{align}
    \tau_{T,2,\otimes}(\chi)&=-2\int_0^{z_1}\der z_g\frac{(z_g+z_2-z_1)^2}{z_1z_2(z_1^2+z_2^2)}\left(1-\frac{z_g}{z_1}\right)^2\frac{(\Pt^2+\bar Q^2)^2}{8(\Delta_{\rm V3}^2+\bar Q_{\rm V3}^2)(\Kt^2+\bar Q_{\rm V3}^2)}\left\{1+\frac{\QV^2}{\Kt^2}\right.\nonumber\\
    &\left.+\frac{\Kt^2+\bar Q_{\rm V3}^2}{\DV^2+\QV^2}\left[\frac{\DV^4}{\Kt^4}\ln\left(\frac{\DV^2}{\QV^2}\right)+\left(1-\frac{\DV^4}{\Kt^4}\right)\ln\left(\frac{\DV^2-\Kt^2}{\QV^2+\Kt^2}\right)-\frac{(\DV^2+\QV^2)^2}{\Kt^4}\ln\left(1+\frac{\Kt^2}{\QV^2}\right)\right]\right\}\,.
\end{align}
One then easily sees that 
\begin{equation}
    \tau_{T,2,\otimes}(\chi)=\left(\frac{1+\chi^2}{2}\right)\times\tau_{T,1,\otimes}(\chi) \,.
\end{equation}
We will return to the computation of the remaining $z_g$ integral once we have obtained all terms in $\Hcal^{\lambda= \rm T}_{\rm NLO,2}$. 

\paragraph{Computation of the $\Jcal_{\odot}$ term.} We turn now to the calculation of the term proportional to $\Jcal_{\odot}$ which is very similar to the term already computed in the longitudinally polarized case. The building block is the integral
\begin{align}
    \Ical_{\odot}^{ik}=\int\frac{\der^2\ut}{(2\pi)}e^{-i\Pt\cdot\ut}\frac{\ut^i\ut^k}{u_\perp}K_1(\bar{Q}_{\mathrm{V3}} u_\perp)e^{i\frac{z_g}{z_1}\Pt\cdot\ut}\Jcal_{\odot}\left(\ut,\left(1-\frac{z_g}{z_1}\right)\Pt,\Delta_{\rm V3}\right) \,.
\end{align}
We proceed in a similar fashion,
\begin{align}
    \Ical_{\odot}^{ik}&=\int\frac{\der^2 \ut}{(2\pi)}\frac{\ut^i\ut^k}{u_\perp}\textrm{K}_1(\bar Q_{\rm V3}\ut)\int\frac{\der^2\lt}{(2\pi)}\frac{2(\lt+\Kt) \cdot \Kt \ e^{i\lt \cdot \ut}}{(\lt+\Kt)^2\left[\lt^2-\Delta^2_{\rm V3} - i \epsilon\right]} \nonumber \\
    &=\int\frac{\der^2\lt}{(2\pi)}\frac{2(\lt+\Kt) \cdot \Kt}{(\lt+\Kt)^2\left[\lt^2-\Delta^2_{\rm V3} - i \epsilon\right]}\int\frac{\der^2 \ut}{(2\pi)}\frac{\ut^i\ut^k}{u_\perp}\textrm{K}_1(\bar Q_{\rm V3}\ut)e^{i\lt \cdot \ut} \nonumber \\
    &=\frac{1}{\QV}\int\frac{\der^2\lt}{(2\pi)}\frac{2(\lt+\Kt) \cdot \Kt}{(\lt+\Kt)^2\left[\lt^2-\Delta^2_{\rm V3} - i \epsilon\right]}\left[ \frac{\delta^{ik}}{(\lt^2+\bar{Q}_{\mathrm{V3}}^2)}-\frac{2\lt^i\lt^k}{(\lt^2+\bar{Q}_{\mathrm{V3}}^2)^2}\right] \,.
\end{align}
We need to be careful with the tensor structure of this integral. Only diagonal components of the tensor contribute, but the two coefficients of the diagonal are not the same for the $\lt^i\lt^k$ term.
\begin{align}
\Ical_{\odot}^{ik}
    &=\frac{2\delta^{ik}}{\QV}\int_0^\infty\der\ell\frac{\ell K_\perp}{(\ell^2+\Kt^2)\left[\ell^2-\Delta^2_{\rm V3} - i \epsilon\right](\ell^2+\bar Q_{\rm V3}^2)}\int_0^{2\pi}\frac{\der\theta}{(2\pi)}\frac{\ell\cos(\theta)+K_\perp}{1+\frac{2\ell K_\perp}{\ell^2+K_\perp^2}\cos(\theta)}\nonumber\\
    &-\frac{4\delta^{ik}}{\QV}\int_0^\infty\der\ell\frac{\ell^3K_\perp}{(\ell^2+\Kt^2)\left[\ell^2-\Delta^2_{\rm V3} - i \epsilon\right](\ell^2+\bar Q_{\rm V3}^2)^2}\int_0^{2\pi}\frac{\der\theta}{(2\pi)}\frac{\sin^2(\theta)(\ell\cos(\theta)+K_\perp)}{1+\frac{2\ell K_\perp}{\ell^2+K_\perp^2}\cos(\theta)}\nonumber\\
    &-\frac{4\Pt^i\Pt^k}{\QV\Pt^2}\int_0^\infty\der\ell\frac{\ell^3K_\perp}{(\ell^2+\Kt^2)\left[\ell^2-\Delta^2_{\rm V3} - i \epsilon\right](\ell^2+\bar Q_{\rm V3}^2)^2}\int_0^{2\pi}\frac{\der\theta}{(2\pi)}\frac{\cos(2\theta)(\ell\cos(\theta)+K_\perp)}{1+\frac{2\ell K_\perp}{\ell^2+K_\perp^2}\cos(\theta)} \,.
\end{align}
The angular integrals read
\begin{align}
    \int_0^{2\pi}\frac{\der\theta}{(2\pi)}\frac{\ell\cos(\theta)+K_\perp}{1+\frac{2\ell K_\perp}{\ell^2+K_\perp^2}\cos(\theta)}&=\frac{\ell^2+\Kt^2}{K_\perp}\Theta(K_\perp-\ell) \,, \\
    \int_0^{2\pi}\frac{\der\theta}{(2\pi)}\frac{\sin^2(\theta)(\ell\cos(\theta)+K_\perp)}{1+\frac{2\ell K_\perp}{\ell^2+K_\perp^2}\cos(\theta)}&=\frac{(\ell^2+\Kt^2)(2\Kt^2-\ell^2)}{4\Kt^3}\Theta(K_\perp-\ell)+\frac{K_\perp(\Kt^2+\ell^2)}{4\ell^2}\Theta(\ell-K_\perp) \,, \\
    \int_0^{2\pi}\frac{\der\theta}{(2\pi)}\frac{\cos(2\theta)(\ell\cos(\theta)+K_\perp)}{1+\frac{2\ell K_\perp}{\ell^2+K_\perp^2}\cos(\theta)}&=\frac{\ell^2(\ell^2+\Kt^2)}{2K_\perp^3}\Theta(K_\perp-\ell)-\frac{K_\perp(\Kt^2+\ell^2)}{2\ell^2}\Theta(\ell-K_\perp) \,.
\end{align}
The integral $\Ical_{\odot}^{ik}$ involves the four following integral
\begin{align}
    \Ical_{\odot,1}&=\int_0^{K_\perp}\der\ell\frac{\ell}{\left[\ell^2-\Delta^2_{\rm V3}\right](\ell^2+\bar Q_{\rm V3}^2)}\,,\quad
    \Ical_{\odot,2}=\int_0^{K_\perp}\der\ell\frac{\ell^3}{\left[\ell^2-\Delta^2_{\rm V3}\right](\ell^2+\bar Q_{\rm V3}^2)^2}\,,\nonumber\\
    \Ical_{\odot,3}&=\int_0^{K_\perp}\der\ell\frac{\ell^5}{\left[\ell^2-\Delta^2_{\rm V3}\right](\ell^2+\bar Q_{\rm V3}^2)^2}\,,\quad
    \Ical_{\odot,4}=\int_{K_\perp}^\infty\der\ell\frac{\ell}{\left[\ell^2-\Delta^2_{\rm V3} - i \epsilon\right](\ell^2+\bar Q_{\rm V3}^2)^2} \,.
\end{align}
The calculation of these integral over $\ell$ yields
\begin{align}
    \Ical_{\odot,1}&=\frac{1}{2(\QV^2+\DV^2)}\ln\left(\frac{\QV^2(\DV^2-\Kt^2)}{\DV^2(\Kt^2+\QV^2)}\right) \,, \\
    \Ical_{\odot,2}&=\frac{\Kt^2}{2(\Delta_{\rm V3}^2+\bar Q_{\rm V3}^2)(\Kt^2+\bar Q_{\rm V3}^2)}\left[1+\frac{\Delta^2_{\rm V3}(\Kt^2+\bar Q_{\rm V3}^2)}{\Kt^2(\Delta_{\rm V3}^2+\bar Q_{\rm V3}^2)}\ln\left(\frac{\bar Q_{\rm V3}^2(\Delta_{\rm V3}^2-\Kt^2)}{\Delta_{\rm V3}^2(\Kt^2+\bar Q_{\rm V3}^2)}\right)\right] \,, \\
    \Ical_{\odot,3}&=-\frac{\Kt^2\QV^2}{2(\DV^2+\QV^2)(\Kt^2+\QV^2)}\left\{1-\frac{(\Kt^2+\QV^2)\DV^4}{(\DV^2+\QV^2)\Kt^2\QV^2}\right.\nonumber\\
    &\left.\times\left[\ln\left(\frac{(\DV^2-\Kt^2)\QV^2}{(\QV^2+\Kt^2)\DV^2}\right)+\frac{(\DV^2+\QV^2)^2}{\DV^4}\ln\left(1+\frac{\Kt^2}{\QV^2}\right)\right]\right\} \,, \\
    \Ical_{\odot,4}&=-\frac{1}{2(\Delta_{\rm V3}^2+\bar Q_{\rm V3}^2)(\Kt^2+\bar Q_{\rm V3}^2)}\left[1+\frac{(\Kt^2+\bar Q_{\rm V3}^2)}{(\Delta_{\rm V3}^2+\bar Q_{\rm V3}^2)}\ln\left(\frac{\Delta_{\rm V3}^2-\Kt^2}{\Kt^2+\bar Q_{\rm V3}^2}\right)-\frac{i\pi(\Kt^2+\bar Q_{\rm V3}^2)}{(\Delta_{\rm V3}^2+\bar Q_{\rm V3}^2)}\right] \,.
\end{align}
We now write the contribution of this integral to $\Hcal^{\textrm{T},ij}_{\rm NLO,2}$ that we call $\Hcal^{\textrm{T},ij}_{\rm NLO,2,\odot}$:
\begin{align}
    \Hcal^{\textrm{T},ij}_{\rm NLO,2,\odot}&=\alpha_{\rm em}\alpha_se_f^2\deltatwo\times z_1z_2(z_1^2+z_2^2)\int_0^{z_1}\frac{\der z_g}{z_g}\frac{-\left[z_1(z_1-z_g)+z_2(z_2+z_g)\right]}{[z_1^2+z_2^2]}\left(1-\frac{z_g}{2z_1}+\frac{z_g}{2z_2}-\frac{z_g^2}{2z_1z_2}\right)\nonumber\\
    &\times\QV\Ical_{\odot}^{ik}\delta^{kl}\left[\frac{\delta^{lj}}{\Pt^2+\bar Q^2}-\frac{2\Pt^l\Pt^j}{(\Pt^2+\bar Q^2)^2}\right]+(1\leftrightarrow 2)+c.c. \,.
\end{align}
The calculation of the tensor structure is straightforward. In analogy with $\tau_{T,1,\otimes}$, we define $\tau_{T,1,\odot}$ and we find
after replacing by the values of the parameters:
\begin{align}
    &\tau_{T,1,\odot}(\chi)=\int_0^{z_1}\frac{\der z_g}{z_g}\frac{-z_1^2\left[z_1(z_1-z_g)+z_2(z_2+z_g)\right]}{(z_1-z_g)^2(z_1^2+z_2^2)}\left(1-\frac{z_g}{2z_1}+\frac{z_g}{2z_2}-\frac{z_g^2}{2z_1z_2}\right)\nonumber\\
    &\times\left\{-\frac{z_2(z_1-z_g)}{z_1(z_2+z_g)}+(1+\chi^2)\ln\left(1+\frac{z_2(z_1-z_g)}{z_1(z_2+z_g)\chi^2}\right)+\frac{z_1(z_2+z_g)+2\chi^2z_2(z_1-z_g)}{z_1(z_2+z_g)(1+\chi^2)}\ln(\chi^2)\right.\nonumber\\
    &+\left.\frac{(z_2^2(z_1-z_g)^2+z_1^2(z_2+z_g)^2)+2z_1z_2(z_1-z_g)(z_2+z_g)\chi^2}{(1+\chi^2)z_1^2(z_2+z_g)^2}\ln\left(\frac{z_g}{z_2(z_1-z_g)+z_1(z_2+z_g)\chi^2}\right)\right\}\nonumber\\
    &+(1\leftrightarrow 2)
\end{align}
for the factor associated with $\delta^{ij}$, and
\begin{align}
    &\tau_{T,2,\odot}(\chi)=\int_0^{z_1}\frac{\der z_g}{z_g}\frac{-z_1^2\left[z_1(z_1-z_g)+z_2(z_2+z_g)\right]}{2(z_1-z_g)^2(z_1^2+z_2^2)}\left(1-\frac{z_g}{2z_1}+\frac{z_g}{2z_2}-\frac{z_g^2}{2z_1z_2}\right)\nonumber\\
    &\left\{-\frac{z_2(z_1-z_g)(1+\chi^2)(2z_2(z_1-z_g)+\chi^2(z_2(z_g-z_1)+z_1(z_2+z_g)\chi^2))}{z_1(z_2+z_g)(z_2(z_1-z_g)+z_1(z_2+z_g)\chi^2)\chi^2}\right.\nonumber\\
    &+(1+\chi^2)^2\ln\left(1+\frac{z_2(z_1-z_g)}{z_1(z_2+z_g)\chi^2}\right)+\frac{2z_2(z_1-z_g)+z_1(z_2+z_g)}{z_1(z_2+z_g)}\ln(\chi^2)\nonumber\\
    &\left.+\frac{(z_2(z_1-z_g)+z_1(z_2+z_g))^2}{z_1^2(z_2+z_g)^2}\ln\left(\frac{z_g}{z_2(z_1-z_g)+z_1(z_2+z_g)\chi^2}\right)\right\} +(1\leftrightarrow 2) 
\end{align}
for the factor associated with $\Pt^i\Pt^j$.

\paragraph{Combining the three terms $K$, $\odot$ and $\otimes$ and checking the slow gluon limit.}
We now compute and simplify the two functions $\tau_{T,1}$ and $\tau_{T,2}$ and check that the counterterm of the slow gluon divergence coincides with the last term proportional to $\ln((z_gP_\perp u_\perp)/(c_0z_1z_2))$ in the hard factor $\Hcal_{\rm NLO,2}^{\lambda=\textrm{T},ij}(\Pt)$ (see Eq.\,\eqref{eq:HNLO2-transverse}). The latter can be obtained from the calculation of $\Hcal_{\rm NLO,1}^{\lambda=\textrm{T},ij}(\Pt)$ and the known $\kappa_{T,1}$ and $\kappa_{T,2}$ functions. In terms of equations, we have
\begin{align}
    \tau_{T,1}(\chi)&=\tau_{T,1,K}(\chi)+\tau_{T,1,\otimes}(\chi)+\tau_{T,1,\odot}(\chi)+\int_0^{z_1}\frac{\der z_g}{z_g}\left[\frac{1}{2}\kappa_{T,1}(\chi)-2\ln(c_0)+2\ln\left(\frac{z_g}{z_1z_2}\right)\right]+(1\leftrightarrow 2) \,, \\
    \tau_{T,2}(\chi)&=\tau_{T,2,K}(\chi)+\tau_{T,2,\otimes}(\chi)+\tau_{T,2,\odot}(\chi)+\int_0^{z_1}\frac{\der z_g}{z_g}\left[\frac{1}{2}\kappa_{T,2}(\chi)-2\ln(c_0)+2\ln\left(\frac{z_g}{z_1z_2}\right)\right]+(1\leftrightarrow 2) \,,
\end{align}
which should be finite as $z_g\to 0$. We write the function $\tau_{T,1}$ defined above as
\begin{align}
    \tau_{T,1}(\chi)&=\int_0^{z_1}\der z_g\left\{\frac{1+z_1(1+z_1^2+z_2^2+2z_2z_g+z_g)+(1+z_1z_g)\chi^2}{(z_1^2+z_2^2)(z_1-z_g)}\frac{\chi^2}{1+\chi^2}\ln(\chi^2)\right.\nonumber\\
    &-\frac{(z_1-z_g)(1-z_2z_g)+(1+z_g)^2\chi^2+(z_2+z_g)(1+z_1z_g)\chi^4}{(1+\chi^2)(z_1^2+z_2^2)(z_1-z_g)(z_2+z_g)}\ln\left(\frac{z_2(z_1-z_g)+z_1(z_2+z_g)\chi^2}{z_1z_2}\right)\nonumber\\
    &+\frac{1-\chi^2}{z_g}\left[\ln\left(\frac{z_2(z_1-z_g)+z_1(z_2+z_g)\chi^2}{z_1z_2}\right)-\ln\left(1+\chi^2\right)\right]+\frac{2z_1z_2}{(z_1^2+z_2^2)}\nonumber\\
    &+\frac{-2z_g+(z_1-z_2-2z_1z_2z_g+(z_2-z_1)z_g^2)(1+\chi^2)}{(1+\chi^2)(z_1^2+z_2^2)(z_1-z_g)(z_2+z_g)}\ln\left(\frac{z_g}{z_1z_2}\right)\nonumber\\
    &\left.+\frac{z_1(1+\chi^2)(z_1(z_1^2+z_2^2)-(z_1-z_2)^2z_g+(z_1-2z_2)z_g^2-z_g^3)}{(z_1^2+z_2^2)(z_1-z_g)^2z_g}\ln\left(1+\frac{z_g}{z_2}\right)\right\}+(1\leftrightarrow 2) \,.
\end{align}
One notices that indeed there is no logarithmic divergence in $z_g=0$ anymore.

Similarly, the function $\tau_{T,2}$ reads
\begin{align}
    &\tau_{T,2}(\chi)=\int_0^{z_1}\der z_g\left\{\frac{-1+z_1(2-2z_1(z_2+z_g)+z_g)+(1+z_1z_g)\chi^2(2+\chi^2)}{2(z_1^2+z_2^2)(z_1-z_g)}\ln(\chi^2)\right.\nonumber\\
    &+\frac{1-3z_1+z_g+z_g(-2z_g+3z_1(z_2+z_g))-(z_2+z_g)(1+z_1z_g)\chi^2(2+\chi^2)}{2(z_1^2+z_2^2)(z_2+z_g)(z_1-z_g)}\ln\left(\frac{z_2(z_1-z_g)+z_1(z_2+z_g)\chi^2}{z_1z_2}\right)
    \nonumber\\
    &+\frac{(1-\chi^2)(3+\chi^2)}{2z_g}\left[\ln\left(\frac{z_2(z_1-z_g)+z_1(z_2+z_g)\chi^2}{z_1z_2}\right)-\ln\left(1+\chi^2\right)\right]\nonumber\\
    &-\frac{1+z_g-2z_1^2z_g-z_g^2+2z_1(-1+z_g+z_g^2)}{(z_1^2+z_2^2)(z_2+z_g)(z_1-z_g)}\ln\left(\frac{z_g}{z_1z_2}\right)\nonumber\\
    &+\frac{z_1(2z_1^2-2z_1(1+z_g)+(1+z_g)^2)}{2(z_1^2+z_2^2)(z_1-z_g)z_g}(1+\chi^2)^2\ln\left(1+\frac{z_g}{z_2}\right)\nonumber\\
    &-\frac{z_2(-1+z_1(2-2z_1(z_2+z_g)+z_g))}{2\chi^2(z_1^2+z_2^2)(z_2(z_1-z_g)+z_1(z_2+z_g)\chi^2)}-\frac{2+z_1(3(z_g-1)+z_1(-2-7z_g+4z_1(1+z_2+z_g)))}{2(z_1^2+z_2^2)(z_2(z_1-z_g)+z_1(z_2+z_g)\chi^2)}\nonumber\\
    &\left.+\frac{1-z_1(1+z_g+z_1(-4-3z_g+2z_1(1+z_2+z_g)))}{2(z_1^2+z_2^2)(z_2(z_1-z_g)+z_1(z_2+z_g)\chi^2)}\chi^2+\frac{z_1(1+z_g-z_1(2+z_g))\chi^4}{2(z_1^2+z_2^2)(z_2(z_1-z_g)+z_1(z_2+z_g)\chi^2)}\right\}\nonumber\\
    &+(1\leftrightarrow 2) \,.
\end{align}
Once again, there is no slow gluon divergence which is a good cross-check of this expression. Another important feature of this integral is the cancellation of the apparent pole in $z_g=z_1$ between different terms.

\paragraph{Final analytic result.} The last step of the calculation is to analytically perform the regular $z_g$ integral. As for $\Hcal_{\rm NLO,1}^{\lambda=\rmT}$, we write the final result for $\Hcal_{\rm NLO,2}^{\lambda=\rmT}$ as
\begin{equation}
    \Hcal_{\rm NLO,2}^{\textrm{T},ij}(\Pt)=\alpha_{\rm em}\alpha_se_f^2\deltatwo\times z_1z_2(z_1^2+z_2^2)\left\{\frac{\delta^{ij}}{(\Pt^2+\bar Q^2)^2}\tau_{T,1}\left(\chi\right)-\frac{4\bar Q^2\Pt^i\Pt^j}{(\Pt^2+\bar Q^2)^4}\tau_{T,2}\left(\chi\right)\right\} \,, \label{eq:tau-def}
\end{equation}
with
\begin{align}
    \tau_{T,1}(\chi)&=-\frac{(1+z_1)^2+\chi^2(1+z_1^2)}{1-2z_1z_2}\frac{\chi^2}{1+\chi^2}\left[\frac{\pi^2}{6}-\textrm{Li}_2\left(\frac{z_1\chi^2-z_2}{\chi^2}\right)\right]\nonumber\\
    &+\frac{2+z_1(-2+z_1+\chi^2z_1)}{(1-2z_1z_2)(1+\chi^2)}\left[\textrm{Li}_2\left(\frac{z_2-z_1\chi^2}{z_2}\right)-\textrm{Li}_2(z_2-z_1 \chi^2)-\ln\left(\frac{z_1}{z_2}\right)\ln(z_2)\right]\nonumber\\
    &+\frac{1-z_1^2-\chi^2(2+z_1(3z_1-4))-\chi^4(1-2z_1z_2)}{(1+\chi^2)(1-2z_1z_2)}\textrm{Li}_2\left(-\frac{z_1}{z_2}\right)\nonumber\\
    &+\frac{(1+\chi^2)(1+z_1^2)}{1-2z_1z_2}\left[\textrm{Li}_2(z_2)+\ln(z_1)\ln(z_2)\right]+(\chi^2-1)\textrm{Li}_2\left(\frac{z_2-z_1\chi^2}{z_2(1+\chi^2)}\right)\nonumber\\
    &-\frac{z_1z_2(1+\chi^2)}{(1-2z_1z_2)}\ln(1+\chi^2)+\frac{z_1z_2(2+(3-2z_1)\chi^2+\chi^4)}{(1+\chi^2)(1-2z_1z_2)}\ln(\chi^2)\nonumber\\
    &+\frac{2(1+\chi^2)z_2^2z_1}{(z_2-z_1\chi^2)(1-2z_2z_1)}\ln\left(\frac{z_2(1+\chi^2)}{\chi^2}\right)+\frac{2z_1^2z_2}{1-2z_1z_2}+(1\leftrightarrow 2) \,, \label{eq:tau1-final}
\end{align}
and 
\begin{align}
    \tau_{T,2}(\chi)&=\frac{z_2^2-\chi^2(2+\chi^2)(1+z_1^2)}{2(1-2z_1z_2)}\left[\frac{\pi^2}{6}-\textrm{Li}_2\left(\frac{z_1\chi^2-z_2}{\chi^2}\right)\right]\nonumber\\
    &+\frac{1-z_1z_2}{1-2z_1z_2}\left[\textrm{Li}_2\left(\frac{z_2-z_1\chi^2}{z_2}\right)-\textrm{Li}_2(z_2-z_1\chi^2)-\ln\left(\frac{z_1}{z_2}\right)\ln(z_2)\right]\nonumber\\
    &+\frac{1-\chi^2(2+\chi^2)(1-2z_1z_2)}{2(1-2z_1z_2)}\textrm{Li}_2\left(-\frac{z_1}{z_2}\right)+\frac{(1+\chi^2)^2(1+z_1^2)}{2(1-2z_1z_2)}\left[\textrm{Li}_2(z_2)+\ln(z_1)\ln(z_2)\right]\nonumber\\
    &-\frac{(1-\chi^2)(3+\chi^2)}{2}\textrm{Li}_2\left(\frac{z_2-z_1\chi^2}{z_2(1+\chi^2}\right)-\frac{(1+\chi^2)^2z_1z_2}{2(1-2z_1z_2)}\ln(1+\chi^2)+\frac{z_1z_2(3z_2+z_1+2\chi^2+\chi^4)}{2(1-2z_1z_2)}\ln(\chi^2)\nonumber\\
    &+\left[\frac{z_2^4-z_2^2(2+z_1(z_1-3z_2))\chi^2+(1+z_1z_2(z_1-z_2)(z_1-4z_2))\chi^4}{2\chi^2(z_2-z_1\chi^2)^2(1-2z_1z_2)}\right.\nonumber\\
    &\left.+\frac{z_1(1-z_1(8+z_1(4z_1-9)))\chi^6+z_1^2(z_1^2-z_2)\chi^8}{2\chi^2(z_2-z_1\chi^2)^2(1-2z_1z_2)}\right]\ln\left(\frac{z_2(1+\chi^2)}{\chi^2}\right)-\frac{(1+\chi^2)z_1^2z_2(z_1-z_2-2\chi^2z_2+\chi^4)}{2\chi^2(z_2-z_1\chi^2)(1-2z_1z_2)}\nonumber\\
    &+(1\leftrightarrow 2) \,.
    \label{eq:tau2-final}
\end{align}

\subsection{Hard factor from dressed self-energy and vertex corrections}
\label{sub:hard-SE-V}

We finally turn to the calculation of the back-to-back limit of $\der \sigma^{(0),\lambda=\rm T}_{\rm other}$ given by Eq.\,\eqref{eq:V-no-sud-other-full}. As noticed in \cite{Caucal:2023nci}, the first step is to perform the change of variables in the integral in Eq.\,\eqref{eq:V-no-sud-other-full}:
\begin{align}
    \ut &=\left(1-\frac{z_g}{z_1}\right)\xt-\yt+\frac{z_g}{z_1}\zt \,,\nonumber \\
    \bt&=(z_1-z_g)\xt+z_2\yt+z_g\zt \,, \nonumber \\
    \rt&=\left(1-\frac{z_g}{z_1}\right)(\zt-\xt) \,.\label{eq:SE1-b2b-variable}
\end{align}
For the transverse coordinates in the complex conjugate amplitude, one makes the same change of variables as for the LO amplitude, namely $\ut'=\xt'-\yt'$ and $\bt'=z_1\xt'+z_2\yt'$.
This change of variables presents the following advantage: the expansion of the color correlator $\Xi_{\rm NLO,1}$ to leading power in $r_\perp,u_\perp\ll b_\perp$ enables one to get the leading power contribution of $\der \sigma^{(0),\lambda=\rm T}_{\rm other}$ in $q_\perp/P_\perp$. This expansion gives, at leading power,
\begin{align}
    \Xi_{\rm NLO,1}
    &=\left[C_F\ut^i\ut'^j+\left(\frac{N_c}{2}+\frac{1}{2N_c}\frac{z_g}{(z_1-z_g)}\right)\rt^i\ut'^j\right]\times\frac{\alpha_s}{2N_c}G^{ij}_{Y_f}(\bt,\bt')+\mathcal{O}\left(\frac{q_\perp^2}{P_\perp^2}\right)\,.\label{eq:NLO1-b2b}
\end{align}
The leading power contribution is therefore proportional to the WW gluon TMD. We emphasize that the change of variable Eq.\,\eqref{eq:SE1-b2b-variable} is crucial to ensure that the expansion Eq.\,\eqref{eq:NLO1-b2b} to leading power in $u_\perp$ and $r_\perp$ correctly account for the leading power in $q_\perp/P_\perp$ in the final expression, as shown in Appendix~(B) of \cite{Caucal:2023nci}.

Following \cite{Caucal:2023nci}, we also define the variable 
\begin{align}
    \omega=\frac{z_g}{z_2(z_1-z_g)}\,,
\end{align}
which appears in the NLO light-cone wave-function after the change of variable Eq.\,\eqref{eq:SE1-b2b-variable}. To simplify the calculation of Eq.\,\eqref{eq:V-no-sud-other-full}, we decompose it into three terms: the instantaneous, regularized self-energy and vertex corrections terms
\begin{align}
    \der \sigma^{\lambda=\rm T}_{\rm other} = \der\sigma^{\lambda=\rm T}_{\rm inst} + \der\sigma^{\lambda=\rm T}_{\rm SE_1} + \der\sigma^{\lambda=\rm T}_{\rm V_1} \,,
\end{align}
 corresponding respectively to instantaneous diagrams in light-cone perturbation theory, the UV finite piece of the dressed quark self-energy, and the dressed vertex correction. The corresponding Feynman diagrams are given in \cite{Caucal:2021ent,Caucal:2023nci}.

\subsubsection{Instantaneous terms}

For transversely polarized photons, instantaneous diagrams in light-cone perturbation theory contribute to the amplitude. In Eq.\,\eqref{eq:V-no-sud-other-full}, they correspond to the term
\begin{align}
    &\der\sigma^{\lambda=\rm T}_{\rm inst}=\frac{\alpha_{\rm em}e_f^2N_c\deltatwo}{(2\pi)^6}\int\der^2\xt\der^2\xt'\der^2\yt\der^2\yt' e^{-i\ktone\cdot\rxxtp-i\kttwo\cdot\ryytp} \ 2z_1^2z_2^2\frac{QK_1(\bar Qr_{x'y'})}{r_{x'y'}}\nonumber\\
    &\times\frac{(-\alpha_s)}{\pi} \int_0^{z_1}\frac{\der z_g}{z_g}\int\frac{\der^2\zt}{\pi}e^{-i\frac{z_g}{z_1}\ktone\cdot\rzxt}\frac{\bar Q K_1(QX_V)}{X_V}\Xi_{\rm NLO,1}\frac{\rzxt\cdot\rxytp}{\rzxt^2}\left[\frac{z_g(z_g-z_1)^2z_2}{2 z_1^3}+\frac{z_g(z_1-z_g)}{2(z_g+z_2)}\right]\,.\label{eq:V-no-sud-other-inst}
\end{align}
In the back-to-back limit, after performing the change of variables in Eq.\,\eqref{eq:SE1-b2b-variable} and performing the expansion of $\Xi_{\rm NLO,1}$ in Eq.\,\eqref{eq:NLO1-b2b}, the instantaneous diagrams factorize as
\begin{equation}
    \der\sigma_{\rm inst}^{\lambda=\rmT}=\Hcal_{\rm inst.}^{ij,\lambda=\rmT}(\Pt)\int\frac{\der^2\bt\der^2\bt'}{(2\pi)^4}e^{-i\qt\cdot\rbbpt}G^{ij}_{Y_f}(\bt,\bt')\,, \label{eq:inst-def}
\end{equation}
with
\begin{align}
    \Hcal_{\rm inst.}^{ij,\lambda=\rmT}(\Pt)&=\alpha_{\rm em}\alpha_se_f^2\deltatwo\times 2z_1z_2h^{ik}_{\rm inst.}h^{kj}_{\rm LO} \,, \\
    h^{kj}_{\rm LO}&=\frac{\delta^{kj}}{\Pt^2+\bar Q^2}-\frac{2\Pt^k\Pt^j}{(\Pt^2+\bar Q^2)^2}\,, \\
    h^{ik}_{\rm inst.}&=\frac{(-\alpha_s)}{\pi}\int_0^{z_1}\der z_g\left[\frac{(z_1-z_g)z_2}{2z_1^2}+\frac{z_1}{2(z_2+z_g)}\right]\left(\frac{N_c}{2}+\frac{1}{2N_c}\frac{z_g}{z_1-z_g}\right)I_{\rm inst}^{ik}\,,
\end{align}
and where the master integral $I_{\rm inst}^{ik}$ is defined by
\begin{align}
    I_{\rm inst}^{ik}&=\int\frac{\der^2\ut}{(2\pi)}\int\frac{\der^2\rt}{(2\pi)}e^{-i\Pt\cdot\ut}\frac{\rt^i\rt^k}{\rt^2}\frac{\bar QK_1\left(\bar Q\sqrt{\ut^2+\omega\rt^2}\right)}{\sqrt{\ut^2+\omega\rt^2}}\,.
\end{align}
This integral is computed using the same method as in \cite{Caucal:2023nci}, namely by writing the NLO wave-function expressed in terms of modified Bessel function back to momentum space, thanks to the identity
\begin{align}
    \frac{\bar QK_1\left(\bar Q\sqrt{\ut^2+\omega\rt^2}\right)}{\sqrt{\ut^2+\omega\rt^2}}=\int\frac{\der^2\ltone}{(2\pi)}\int\frac{\der^2\lttwo}{(2\pi)}\frac{e^{i\ltone\cdot\ut}e^{i\lttwo\cdot\rt}}{\lttwo^2+\omega(\ltone+\bar Q^2)}\,,
\end{align}
which "undoes" the integration over internal loop momenta performed in \cite{Caucal:2021ent}.
We end up with
\begin{align}
     I_{\rm inst}^{ik}=\int\frac{\der^2\rt}{(2\pi)}\frac{\rt^i\rt^k}{\rt^2}K_0\left[\sqrt{\omega(\Pt^2+\bar Q^2)}r_\perp\right]=\frac{\delta^{ik}}{2\omega(\Pt^2+\bar Q^2)}\,.
\end{align}
Plugging this result inside $h^{ik}_{\rm inst.}$, one realizes that a logarithmic divergence arises because of the $1/\omega\propto 1/z_g$ dependence of the integral $ I_{\rm inst}^{ik}$. This divergence cannot be canceled by the slow gluon subtraction since instantaneous diagrams do not contribute to the rapidity evolution of the LO cross-section. We shall see that this new slow gluon divergence of the instantaneous diagram in the back-to-back limit is a crucial element in the overall cancellation of slow gluon divergences once the counterterm from kinematically constrained rapidity evolution is subtracted.

In the end, we get
\begin{align}
    \Hcal_{\rm inst.}^{ij,\lambda=\rmT}(\Pt)&=\alpha_{\rm em}\alpha_se_f^2\deltatwo\times2z_1z_2\frac{(-\alpha_s)}{\pi}\int_0^{z_1}\der z_g\left[\frac{(z_1-z_g)z_2}{2z_1^2}+\frac{z_1}{2(z_2+z_g)}\right]\left(\frac{N_c}{2}+\frac{1}{2N_c}\frac{z_g}{z_1-z_g}\right)\frac{1}{2\omega}\nonumber\\
    &\times\left[\frac{\delta^{ij}}{(\Pt^2+\bar Q^2)^2}-\frac{2\Pt^i\Pt^j}{(\Pt^2+\bar Q^2)^3}\right]\,.\label{eq:Hinst-final}
\end{align}

\subsubsection{Regularized self-energy crossing the SW}

In Eq.\,\eqref{eq:V-no-sud-other-full}, the UV regular part of the dressed quark self energy is given by
\begin{align}
    &\der\sigma^{\lambda=\rm T}_{\rm SE_1}=\frac{\alpha_{\rm em}e_f^2N_c\deltatwo}{(2\pi)^6}\int\der^2\xt\der^2\xt'\der^2\yt\der^2\yt' e^{-i\ktone\cdot\rxxtp-i\kttwo\cdot\ryytp} \ 2z_1^2z_2^2(z_1^2+z_2^2)\frac{QK_1(\bar Qr_{x'y'})}{r_{x'y'}}\nonumber\\
    &\times\frac{\alpha_s}{\pi} \int_0^{z_1}\frac{\der z_g}{z_g}\int\frac{\der^2\zt}{\pi}\left\{e^{-i\frac{z_g}{z_1}\ktone\cdot\rzxt}\frac{\bar Q K_1(QX_V)}{X_V}\Xi_{\rm NLO,1}\left(1-\frac{z_g}{z_1}+\frac{z_g^2}{2z_1^2}\right)\frac{\RtS\cdot\rxytp}{\rzxt^2}\right.\nonumber\\
    &\left.-e^{-\frac{\rzxt^2}{\rxyt^2e^{\gamma_E}}}\left(1-\frac{z_g}{z_1}+\frac{z_g^2}{2z_1^2}\right)\frac{\rxyt\cdot\rxytp}{\rzxt^2} QK_1(\bar Qr_{xy})C_F\Xi_{\rm LO}\right\}\,.\label{eq:SE1-full}
\end{align}
Using Eqs\,\eqref{eq:SE1-b2b-variable} and \eqref{eq:NLO1-b2b}, this diagram is also factorized as
\begin{equation}
    \der\sigma_{\rm SE_1}^{\lambda=\rmT}=\Hcal_{\rm NLO,3}^{ij,\lambda=\rmT}(\Pt)\int\frac{\der^2\bt\der^2\bt'}{(2\pi)^4}e^{-i\qt\cdot\rbbpt}G^{ij}_{Y_f}(\bt,\bt')\,,\label{eq:HSE1-def}
\end{equation}
with the hard factor $\Hcal_{\rm NLO,3}^{ij,\lambda=\rmT}(\Pt)$ written as
\begin{align}
    \Hcal_{\rm NLO,3}^{ij,\lambda=\rmT}(\Pt)&=\alpha_{\rm em}\alpha_se_f^2\deltatwo\times2z_1z_2(z_1^2+z_2^2)h_{\rm SE_1}^{ik}h_{\rm LO}^{kj}\,, \\
    h^{kj}_{\rm LO}&=\frac{\delta^{kj}}{\Pt^2+\bar Q^2}-\frac{2\Pt^k\Pt^j}{(\Pt^2+\bar Q^2)^2}\,, \\
    h_{\rm SE_1}^{ik}&=\frac{\alpha_sC_F}{\pi}\int_0^{z_1}\frac{\der z_g}{z_g}\left(1-\frac{z_g}{z_1}+\frac{z_g^2}{2z_1^2}\right)(i\partial^i_{\Pt}I_0^k)\,,
\end{align}
and the integral $I_0^k$ defined by
\begin{align}
    I_0^k&\equiv\int\frac{\der^2\ut}{(2\pi)}\frac{\der^2\rt}{(2\pi)}e^{-i\Pt\cdot\ut}\frac{\ut^k}{\rt^2}\left\{\frac{\bar QK_1\left(\bar Q\sqrt{\ut^2+\omega\rt^2}\right)}{\sqrt{\ut^2+\omega\rt^2}}-e^{-\frac{\rt^2}{\ut^2e^{\gamma_E}}}\frac{\bar QK_1\left(\bar Qu_\perp\right)}{u_\perp}\right\}\,.
\end{align}
To compute this integral we follow the same strategy as in \cite{Caucal:2023nci} (see Section A.3 in the Appendix), we find
\begin{align}
    I_0^k&=\frac{(-i\Pt^k)}{2(\Pt^2+\bar Q^2)}\left[\ln\left(\frac{\bar Q^2}{\omega(\Pt^2+\bar Q^2)}\right)+\frac{\Pt^2-\bar Q^2}{\Pt^2}\ln\left(1+\frac{\Pt^2}{\bar Q^2}\right)\right]\,.
\end{align}
After taking the derivative of this result with respect to $\Pt$, we find
\begin{align}
    \Hcal_{\rm NLO,3}^{ij,\lambda=\rmT}(\Pt)&=\alpha_{\rm em}\alpha_se_f^2\deltatwo\times2z_1z_2(z_1^2+z_2^2)\frac{\alpha_sC_F}{\pi}\int_0^{z_1}\frac{\der z_g}{z_g}\left(1-\frac{z_g}{z_1}+\frac{z_g^2}{2z_1^2}\right)\nonumber\\
    &\times\left\{\frac{\delta^{ij}}{2(\Pt^2+\bar Q^2)^2}\left[-\ln(\omega)-\frac{\bar Q^2}{\Pt^2}\ln\left(1+\frac{\Pt^2}{\bar Q^2}\right)\right]\right.\nonumber\\
    &\left.+\frac{\bar Q^2\Pt^i\Pt^j}{(\Pt^2+\bar Q^2)^4}\left[1+2\ln(\omega)-\frac{\bar Q^2}{\Pt^2}+\left(-1+\frac{\bar Q^2(2\Pt^2+\bar Q^2)}{\Pt^4}\right)\ln\left(1+\frac{\Pt^2}{\bar Q^2}\right)\right]\right\}\,,\label{eq:HNLO3-final}
\end{align}
where the integral over $z_g$ remains undone.

\subsubsection{Vertex correction crossing the shock-wave}

Finally, we turn to the calculation of the hard factor arising from the dressed vertex correction, whose expression extracted from Eq.\,\eqref{eq:V-no-sud-other-full}, reads:
\begin{align}
    \der\sigma^{\lambda=\rm T}_{\rm V_1}&=\frac{\alpha_{\rm em}e_f^2N_c\deltatwo}{(2\pi)^6}\int\der^2\xt\der^2\xt'\der^2\yt\der^2\yt' e^{-i\ktone\cdot\rxxtp-i\kttwo\cdot\ryytp} \ 2z_1^2z_2^2\frac{QK_1(\bar Qr_{x'y'})}{r_{x'y'}}\nonumber\\
    &\times\frac{(-\alpha_s)}{\pi} \int_0^{z_1}\frac{\der z_g}{z_g}\int\frac{\der^2\zt}{\pi}e^{-i\frac{z_g}{z_1}\ktone\cdot\rzxt}\frac{\bar QK_1( Q X_V)}{X_V}\Xi_{\rm NLO,1}\nonumber\\
    &\times\left[[z_1(z_1-z_g)+z_2(z_2+z_g)]\left(1-\frac{z_g}{z_1}\right)\left(1+\frac{z_g}{z_2}\right)\left(1-\frac{z_g}{2z_1}-\frac{z_g}{2(z_2+z_g)}\right)\frac{(\RtV\cdot\rxytp)(\rzxt\cdot\rzyt)}{\rzxt^2\rzyt^2}\right.\nonumber\\
    &\left.+\frac{z_g(z_1-z_g)(z_g+z_2-z_1)^2}{2z_1^2z_2}\frac{(\RtV\times\rxytp)(\rzxt\times\rzyt)}{\rzxt^2\rzyt^2}\right]\label{eq:V1-full}\,.
\end{align}
The back-to-back limit of this contribution is by far the most difficult because of the intricate transverse coordinate integrals. As in the previous subsections, the vertex correction $\rm V_1$ can be cast into a factorized form using the same change of variables in Eq.\,\eqref{eq:SE1-b2b-variable} and the expansion in Eq.\,\eqref{eq:NLO1-b2b}:
\begin{equation}
    \der\sigma_{\rm V_1}^{\lambda=\rmT}=\Hcal_{\rm NLO,4}^{ij,\lambda=\rmT}(\Pt)\int\frac{\der^2\bt\der^2\bt'}{(2\pi)^4}e^{-i\qt\cdot\rbbpt}G^{ij}_{Y_f}(\bt,\bt')\,, \label{eq:HNLO4_a-def}
\end{equation}
with the hard factor decomposed into two pieces according to the two terms in Eq.\,\eqref{eq:V1-full} respectively proportional to $\rzxt\times\rzyt$ and $\rzxt\cdot\rxyt$:
\begin{align}
\Hcal_{\rm NLO,4}^{ij,\lambda=\rmT}(\Pt)=\Hcal_{\rm NLO,4,\otimes}^{ij,\lambda=\rmT}(\Pt)+\Hcal_{\rm NLO,4,\odot}^{ij,\lambda=\rmT}(\Pt)\,.
\end{align}
The hard factor $\Hcal_{\rm NLO,4,\otimes}^{ij,\lambda=\rmT}(\Pt)$ reads
\begin{align}
    \Hcal_{\rm NLO,4,\otimes}^{ij,\lambda=\rmT}(\Pt)&=\alpha_{\rm em}\alpha_se_f^2\deltatwo\times2z_1z_2 [h^{ik}_{\rm V_1,1}+h^{ik}_{\rm V_1,2}]\epsilon^{kl}h_{\rm LO}^{lj}\,, \\
    h^{lj}_{\rm LO}&=\frac{\delta^{lj}}{\Pt^2+\bar Q^2}-\frac{2\Pt^l\Pt^j}{(\Pt^2+\bar Q^2)^2}\,, \\
    h^{ik}_{\rm V_1,1}&=\frac{(-\alpha_s)}{\pi}\int_0^{z_1}\der z_g\frac{(z_g+z_2-z_1)^2}{2z_1(z_2+z_g)}\left(\frac{N_c}{2}+\frac{1}{2N_c}\frac{z_g}{z_1-z_g}\right)I_1^{ik}\,, \\
    h^{ik}_{\rm V_1,2}&=\frac{(-\alpha_s)C_F}{\pi}\int_0^{z_1}\der z_g\frac{(z_g+z_2-z_1)^2}{2z_1(z_2+z_g)}(i\partial^i_{\Pt})I_2^{k}\,,
\end{align}
where the master integrals are defined by
\begin{align}
    I_1^{ik}&=\int\frac{\der^2\ut}{(2\pi)}\int\frac{\der^2\rt}{(2\pi)}e^{-i\Pt\cdot\ut}\rt^i\frac{\rt\times(\rt+\ut)}{\rt^2(\rt+\ut)^2}[\ut^k-\omega\rt^k]\frac{\bar QK_1\left(\bar Q\sqrt{\ut^2+\omega\rt^2}\right)}{\sqrt{\ut^2+\omega\rt^2}}\,,\\
    I_2^{k}&=\int\frac{\der^2\ut}{(2\pi)}\int\frac{\der^2\rt}{(2\pi)}e^{-i\Pt\cdot\ut}\frac{\rt\times(\rt+\ut)}{\rt^2(\rt+\ut)^2}[\ut^k-\omega\rt^k]\frac{\bar QK_1\left(\bar Q\sqrt{\ut^2+\omega\rt^2}\right)}{\sqrt{\ut^2+\omega\rt^2}}\,.
\end{align}
Using the expression for $I_1^{ik}$
and $I_2^k$ found in section 5 of the supplemental material (see Eqs.\,\eqref{eq:I1-decomp}, \eqref{eq:I1A} and \eqref{eq:I1B}), and organizing the result according to the color factors, we obtain
\begin{align}
    &\Hcal_{\rm NLO,4,\otimes}^{ij,\lambda=\rmT}=\alpha_{\rm em}\alpha_se_f^2\deltatwo\times2z_1z_2\frac{(-\alpha_s)N_c}{2\pi}\int_0^{z_1}\der z_g\frac{(z_g+z_2-z_1)^2}{2z_1(z_2+z_g)}\nonumber\\
    &\times\left\{\frac{\delta^{ij}}{4(\Pt^2+\bar Q^2)^2}\left[1-\frac{\Pt^2+\bar Q^2}{\Pt^2}\ln\left(1+\frac{\Pt^2}{\bar Q^2}\right)+\frac{\Pt^2-(1+\omega)\bar Q^2}{\Pt^2+\bar Q^2}\ln\left(\frac{\Pt^2+(1+\omega)\bar Q^2}{\omega\bar Q^2}\right)\right]\right.\nonumber\\
    &\left.+\frac{\bar Q^2\Pt^i\Pt^j}{2(\Pt^2+\bar Q^2)^4}\left[-\frac{\Pt^2+\bar Q^2}{\Pt^2}+\frac{(\Pt^2+\bar Q^2)^2}{\Pt^4}\ln\left(1+\frac{\Pt^2}{\bar Q^2}\right)+\omega\ln\left(\frac{\Pt^2+(1+\omega)\bar Q^2}{\omega\bar Q^2}\right)]\right]\right\}\nonumber\\
    &+\alpha_{\rm em}\alpha_se_f^2\deltatwo\times2z_1z_2\times \frac{(-\alpha_s)}{2N_c\pi}\int_0^{z_1}\der z_g\frac{z_g(z_g+z_2-z_1)^2}{2z_1(z_2+z_g)(z_1-z_g)}\nonumber\\
    &\times\left\{\frac{\delta^{ij}}{4(\Pt^2+\bar Q^2)^2}\right.\left[1-\frac{(1+\omega)(\Pt^2+\bar Q^2)}{\Pt^2}\ln\left(\frac{\Pt^2+(1+\omega)\bar Q^2}{(1+\omega)\bar Q^2}\right)-\frac{\bar Q^2-(1+\omega)\Pt^2}{\Pt^2+\bar Q^2}\ln\left(\frac{\Pt^2+(1+\omega)\bar Q^2}{\omega\bar Q^2}\right)\right]\nonumber\\
    &\left.+\frac{\bar Q^2\Pt^i\Pt^j}{2(\Pt^2+\bar Q^2)^4}\left[-\frac{\Pt^2+\bar Q^2}{\Pt^2}+\left(1+\frac{(1+\omega)\bar Q^2(2\Pt^2+\bar Q^2)}{\Pt^4}\right)\ln\left(\frac{\Pt^2+(1+\omega)\bar Q^2}{(1+\omega)\bar Q^2}\right)-\omega\ln\left(\frac{1+\omega}{\omega}\right)\right]\right\}\nonumber\\
    &+\alpha_{\rm em}\alpha_se_f^2\deltatwo\times2z_1z_2\times \frac{(-\alpha_s)}{2N_c\pi}\int_0^{z_1}\der z_g\frac{z_g(z_g+z_2-z_1)^2}{2z_1z_2(z_2+z_g)(z_1-z_g)}\nonumber\\
    &\times\left\{\frac{\delta^{ij}}{4(\Pt^2+\bar Q^2)^2}\left[\ln\left(\frac{1+\omega}{\omega}\right)\right.-\frac{\bar Q^2}{\Pt^2}\ln\left(\frac{\Pt^2+(1+\omega)\bar Q^2}{(1+\omega)\bar Q^2}\right)-\frac{\Pt^2+\bar Q^2}{\omega\Pt^2}\ln\left(\frac{\Pt^2+(1+\omega)\bar Q^2}{(1+\omega)(\Pt^2+\bar Q^2)}\right)\right]\nonumber\\
    &+\frac{\bar Q^2\Pt^i\Pt^j}{2(\Pt^2+\bar Q^2)^4}\left[\frac{\bar Q^2(2\Pt^2+\bar Q^2)}{\Pt^4}\ln\left(\frac{\Pt^2+(1+\omega)\bar Q^2}{(1+\omega)\bar Q^2}\right)-\ln\left(\frac{(1+\omega)(\Pt^2+(1+\omega)\bar Q^2)}{\omega^2\bar Q^2}\right)\right.\nonumber\\
    &\left.\left.+\frac{(\Pt^2+\bar Q^2)^2}{\omega\Pt^4}\ln\left(\frac{\Pt^2+(1+\omega)\bar Q^2}{(1+\omega)(\Pt^2+\bar Q^2)}\right)\right]\right\}\,.\label{eq:HNLO4_a_final}
\end{align}

In a similar fashion, the $\odot$ hard factor reads
\begin{align}
    \Hcal_{\rm NLO,4,\odot}^{ij,\lambda=\rmT}(\Pt)&=\alpha_{\rm em}\alpha_se_f^2\deltatwo\times2z_1z_2 [h^{ik}_{\rm V_1,1}+h^{ik}_{\rm V_1,2}]h_{\rm LO}^{kj}\,, \\
    h^{kj}_{\rm LO}&=\frac{\delta^{kj}}{\Pt^2+\bar Q^2}-\frac{2\Pt^k\Pt^j}{(\Pt^2+\bar Q^2)^2}\,, \\
    h^{ik}_{\rm V_1,1}&=\frac{(-\alpha_s)}{\pi}\int_0^{z_1}\frac{\der z_g}{z_g}[z_1(z_1-z_g)+z_2(z_2+z_g)]\left(1-\frac{z_g}{2z_1}-\frac{z_g}{2(z_g+z_2)}\right)\nonumber\\
    &\times\left(\frac{N_c}{2}+\frac{1}{2N_c}\frac{z_g}{z_1-z_g}\right)I_3^{ik}\label{eq:hv1}\,, \\
    h^{ik}_{\rm V_1,2}&=\frac{(-\alpha_s)C_F}{\pi}\int_0^{z_1}\frac{\der z_g}{z_g}[z_1(z_1-z_g)+z_2(z_2+z_g)]\left(1-\frac{z_g}{2z_1}-\frac{z_g}{2(z_g+z_2)}\right)\nonumber\\
    &\times(i\partial^i_{\Pt})I_4^{k} \,,\label{eq:HNLO4_b_def}
\end{align}
with the master integrals
\begin{align}
     I_3^{ik}&=\int\frac{\der^2\ut}{(2\pi)}\int\frac{\der^2\rt}{(2\pi)}e^{-i\Pt\cdot\ut}\rt^i\frac{\rt\cdot(\rt+\ut)}{\rt^2(\rt+\ut)^2}[\ut^k-\omega\rt^k]\frac{\bar QK_1\left(\bar Q\sqrt{\ut^2+\omega\rt^2}\right)}{\sqrt{\ut^2+\omega\rt^2}}\,,\label{eq:I3ik-def} \\
    I_4^{k}&=\int\frac{\der^2\ut}{(2\pi)}\int\frac{\der^2\rt}{(2\pi)}e^{-i\Pt\cdot\ut}\frac{\rt\cdot(\rt+\ut)}{\rt^2(\rt+\ut)^2}[\ut^k-\omega\rt^k]\frac{\bar QK_1\left(\bar Q\sqrt{\ut^2+\omega\rt^2}\right)}{\sqrt{\ut^2+\omega\rt^2}}  \,.
\end{align}
Using the results the expressions for $I_{3}^{ik}$ in Eqs.\,\eqref{eq:I3-decomp}, \eqref{eq:I3ikA} and \eqref{eq:I3ikB}, we find 
\begin{align}
    &\Hcal_{\rm NLO,4,\odot}^{ij,\lambda=\rmT}=\alpha_{\rm em}\alpha_se_f^2\deltatwo\times2z_1z_2 \frac{(-\alpha_s)N_c}{2\pi}\int_0^{z_1}\frac{\der z_g}{z_g}[z_1(z_1-z_g)+z_2(z_2+z_g)]\left(1-\frac{z_g}{2z_1}-\frac{z_g}{2(z_g+z_2)}\right)\nonumber\\
    &\times\left\{\frac{\delta^{ij}}{4(\Pt^2+\bar Q^2)^2}\left[-1+\frac{\Pt^2+(1+\omega)\bar Q^2}{\Pt^2+\bar Q^2}\ln\left(\frac{\Pt^2+(1+\omega)\bar Q^2}{\omega\bar Q^2}\right)-\frac{\Pt^2+\bar Q^2}{\Pt^2}\ln\left(1+\frac{\Pt^2}{\bar Q^2}\right)\right]\right.\nonumber\\
    &\left.+\frac{\bar Q^2\Pt^i\Pt^j}{2(\Pt^2+\bar Q^2)^4}\left[\frac{(2\Pt^2-\bar Q^2)(\Pt^2+\bar Q^2)}{\Pt^2\bar Q^2}-(2+\omega)\ln\left(\frac{\Pt^2+(1+\omega)\bar Q^2}{\omega\bar Q^2}\right)+\frac{(\Pt^2+\bar Q^2)^2}{\Pt^4}\ln\left(1+\frac{\Pt^2}{\bar Q^2}\right)\right]\right\}\nonumber\\
    &+\alpha_{\rm em}\alpha_se_f^2\deltatwo\times2z_1z_2\times \frac{(-\alpha_s)}{2N_c\pi}\int_0^{z_1}\frac{\der z_g}{z_g}[z_1(z_1-z_g)+z_2(z_2+z_g)]\left(1-\frac{z_g}{2z_1}-\frac{z_g}{2(z_g+z_2)}\right)\frac{z_g}{z_1-z_g}\nonumber\\
    &\times\left\{\frac{\delta^{ij}}{4(\Pt^2+\bar Q^2)^2}\left[-1-\frac{(1+\omega)(\Pt^2+\bar Q^2)}{\Pt^2}\ln\left(\frac{1+\omega}{\omega}\right)\right.+\frac{\bar Q^2(\Pt^2(1+2\omega)+\bar Q^2(1+\omega))}{\Pt^2(\Pt^2+\bar Q^2)}\ln\left(\frac{\Pt^2+(1+\omega)\bar Q^2}{\omega\bar Q^2}\right)\right]\nonumber\\
    &+\frac{\bar Q^2\Pt^i\Pt^j}{2(\Pt^2+\bar Q^2)^4}\left[\frac{(2\Pt^4-\Pt^2\bar Q^2+(1+\omega)\bar Q^4)(\Pt^2+\bar Q^2)}{\bar Q^2\Pt^2(\Pt^2+(1+\omega)\bar Q^2)}+\frac{(1+\omega)(\Pt^2+\bar Q^2)^2}{\Pt^4}\ln\left(\frac{1+\omega}{\omega}\right)\right.\nonumber\\
    &\left.\left.+\frac{\Pt^4-\bar Q^2(1+\omega)(2\Pt^2+\bar Q^2)}{\Pt^4}\ln\left(\frac{\Pt^2+(1+\omega)\bar Q^2}{\omega\bar Q^2}\right)\right]\right\}\nonumber\\
    &+\alpha_{\rm em}\alpha_se_f^2\deltatwo\times2z_1z_2\times \frac{\alpha_s}{2N_c\pi}\int_0^{z_1}\frac{\der z_g}{z_g}[z_1(z_1-z_g)+z_2(z_2+z_g)]\left(1-\frac{z_g}{2z_1}-\frac{z_g}{2(z_g+z_2)}\right)\nonumber\\
    &\times\left\{\frac{\delta^{ij}}{4(\Pt^2+\bar Q^2)^2}\left[-\frac{\Pt^2+\bar Q^2}{\Pt^2}\ln\left(1+\frac{\Pt^2}{\bar Q^2}\right)+\frac{(1+\omega)(\Pt^2+\bar Q^2)}{\Pt^2}\ln\left(\frac{1+\omega}{\omega}\right)\right.\right.\nonumber\\
    &\left.+\frac{\Pt^2-(1+\omega)\bar Q^2}{\Pt^2}\ln\left(\frac{\Pt^2+(1+\omega)\bar Q^2}{\omega\bar Q^2}\right)\right]\nonumber\\
    &+\frac{\bar Q^2\Pt^i\Pt^j}{2(\Pt^2+\bar Q^2)^4}\left[\frac{2(1+\omega)(\Pt^4-\bar Q^4)}{\Pt^2(\Pt^2+(1+\omega)\bar Q^2)}+\frac{(\Pt^2+\bar Q^2)^2}{\Pt^4}\ln\left(1+\frac{\Pt^2}{\bar Q^2}\right)-\frac{(1+\omega)(\Pt^2+\bar Q^2)^2}{\Pt^4}\ln\left(\frac{1+\omega}{\omega}\right)\right.\nonumber\\
    &\left.\left.+\frac{(1+\omega)\bar Q^2(2\Pt^2+\bar Q^2)-(3+\omega)\Pt^4}{\Pt^4}\ln\left(\frac{\Pt^2+(1+\omega)\bar Q^2}{\omega \bar Q^2}\right)\right]\right\}\,.\label{eq:HNLO4_b_final}
\end{align}
This hard factor contains a logarithmic divergence in $z_g$ arising from the contribution of $I_{3}^{ik}=I_{3A}^{ik}-\omega I_{3B}^{ik}$ (see Eq.\,\eqref{eq:I3-decomp} and \eqref{eq:I3ikB}). The divergence coming from $I_{3B}^{ik}$ cannot be cured by slow gluon subtraction since the slow subtraction term does not contain the tensor structure $\rt^i\rt^k$. It is obtained by setting $\omega=0$ in Eq.\,\eqref{eq:I3ik-def}--see the next section. This divergence will cancel when it is combined with the instantaneous diagrams, as we will explicitly check.

\subsubsection{Subtracting the kinematically constrained slow gluon divergence}

The slow gluon subtraction term is defined by setting $z_g=0$ in the integral of the term $\der\sigma_{\rm other}^{(0),\lambda=\rmT}$ everywhere but in the $1/z_g$ singular factor and imposing the kinematic constraint. After this manipulation, the instantaneous terms do not contribute. We have then
\begin{align}
    \der\sigma_{\rm SE_1+V_1, slow}^{\lambda=\rmT}&\equiv   \frac{\alpha_{\rm em}e_f^2N_c\deltatwo}{(2\pi)^6}\int\der^2\xt\der^2\xt'\der^2\yt\der^2\yt' e^{-i\ktone\cdot\rxxtp-i\kttwo\cdot\ryytp} \ 2z_1z_2(z_1^2+z_2^2)\frac{\bar Q K_1(\bar Qr_{x'y'})}{r_{x'y'}}\nonumber\\
    &\times\frac{\alpha_s}{\pi}\left\{ \int_0^{z_f}\frac{\der z_g}{z_g}\int\frac{\der^2\zt}{\pi}\left[\frac{1}{\rzxt^2}-\frac{\rzxt\cdot\rzyt}{\rzxt^2\rzyt^2}\right] \frac{\rxyt\cdot\rxytp}{r_{xy}}\bar Q K_1(\bar Qr_{xy}) \Xi_{\rm NLO,1} \Theta\left(\frac{z_f}{z_gQ_f^2}-\mathrm{max}(\rzxt^2,\rzyt^2)\right) \right. \nonumber\\
    & - \left. \int_0^{z_f}\frac{\der z_g}{z_g}\int\frac{\der^2\zt}{\pi}e^{-\frac{\rzxt^2}{\rxyt^2e^{\gamma_E}}}\frac{1}{\rzxt^2}C_F\Xi_{\rm LO} \frac{\rxyt\cdot\rxytp}{r_{xy}}\bar Q K_1(\bar Qr_{xy}) +(1\leftrightarrow 2)\right\} +c.c.\,,\label{eq:V-no-sud-other-full-slow}
\end{align}
where the $\Theta$-function in the second line enforces the kinematic constraint on slow gluons. This particular choice follows from the analysis of the NLO light-cone wave-function $K_1(QX_V)$ in Eqs.\,\eqref{eq:SE1-full}-\eqref{eq:V1-full} since one can approximate $Q X_V = \bar{Q} r_{xy}$ if and only if $z_g \mathrm{max}(\rzxt^2,\rzyt^2) \ll z_1z_2 \rxyt^2$. Since the typical dipole size is inversely proportional to the invariant mass of the final state we have $z_1z_2\rxyt^2\sim 1/M_{q\bar{q}}^2$, then
\begin{equation}
    z_g \mathrm{max}(\rzxt^2,\rzyt^2)\ll 1/M_{q\bar{q}}^2\,.
\end{equation}
When imposing the kinematic constraint, we introduce the more general scale transverse $\frac{z_f}{Q_s^2}$, and we will verify that indeed it must be of order $\frac{1}{M_{q\bar{q}}^2}$.

After performing the change of variable Eq.\,\eqref{eq:SE1-b2b-variable} (with $z_g=0$) and the subsequent correlation expansion, the slow gluon subtraction term of $\rm SE_1$ and $\rm V_1$ factorizes as 
\begin{align}
    \der\sigma_{\rm SE_1+V_1, slow}^{\lambda=\rmT}=\Hcal_{\rm NLO,3+4,slow}^{ij,\lambda=\rmT}(\Pt)\int\frac{\der^2\bt\der^2\bt'}{(2\pi)^4}e^{-i\qt\cdot\rbbpt}G^{ij}_{Y_f}(\bt,\bt')\,,\label{eq:slow-def}
\end{align}
where the slow gluon counterterm hard factor is defined as
\begin{align}
    \Hcal_{\rm NLO,3+4,slow}^{ij,\lambda=\rmT}(\Pt)&=\alpha_{\rm em}\alpha_se_f^2\deltatwo\times2z_1z_2(z_1^2+z_2^2)[h^{ik}_{\rm SE_1 slow}+h^{ik}_{\rm V_1,slow}]h^{kj}_{\rm LO}\,, 
    \label{eq:Hard34}\\
    h^{ik}_{\rm SE_1 slow}&=\frac{\alpha_sC_F}{\pi}\int_0^{z_f}\frac{\der z_g}{z_g}(i\partial^i_{\Pt})I_{0,\rm slow}^{k}\,, 
    \label{eq:hSE1slow}\\
    h^{ik}_{\rm V_1 slow} &=\frac{(-\alpha_s)}{\pi}\int_0^{z_f}\frac{\der z_g}{z_g}\left[\frac{N_c}{2}I_{3,\rm slow}^{ik}+C_F(i\partial^i_{\Pt})I_{4,\rm slow}^k\right]\,,
    \label{eq:hV1slow}
\end{align}
with
\begin{align}
    I_{0,\rm slow}^{k}&=\int\frac{\der^2\ut}{(2\pi)}\int\frac{\der^2\rt}{(2\pi)}e^{-i\Pt\cdot\ut}\frac{\ut^k}{\rt^2}\frac{\bar QK_1\left(\bar Qu_\perp\right)}{u_\perp}\left\{\Theta\left(\frac{z_f}{z_gQ_f^2}-\rt^2\right)-e^{-\frac{\rt^2}{\ut^2e^{\gamma_E}}}\right\}\,, \\
    I_{3,\rm slow}^{ik}&=\int\frac{\der^2\ut}{(2\pi)}\int\frac{\der^2\rt}{(2\pi)}e^{-i\Pt\cdot\ut}\rt^i\frac{\rt\cdot(\rt+\ut)}{\rt^2(\rt+\ut)^2}\ut^k\frac{\bar QK_1\left(\bar Qu_\perp\right)}{u_\perp}\Theta\left(\frac{z_f}{z_gQ_f^2}-\rt^2\right)\,, \\
    I_{4,\rm slow}^{k}&=\int\frac{\der^2\ut}{(2\pi)}\int\frac{\der^2\rt}{(2\pi)}e^{-i\Pt\cdot\ut}\frac{\rt\cdot(\rt+\ut)}{\rt^2(\rt+\ut)^2}\ut^k\frac{\bar QK_1\left(\bar Qu_\perp\right)}{u_\perp}\Theta\left(\frac{z_f}{z_gQ_f^2}-\rt^2\right)\,.
\end{align}
These integrals have already been computed in the longitudinally polarized photon case in \cite{Caucal:2023nci} (see Eqs.\,(3.63)-(3.64)-(3.65)-(3.66) therein). They are known analytically up to powers of $z_g$ corrections. Once the integral over $z_g$ up to $z_f$ is performed in the subtraction term, these power corrections are suppressed by powers of $z_f$. Since we take $z_f\sim q_\perp/P_\perp$ as explained in section 3 of the supplemental material, these terms are suppressed for back-to-back kinematics and can safely be neglected.

In \cite{Caucal:2023nci} we found that
\begin{align}
    I_{0,\rm slow}^k=I_{4,\rm slow}^k+\mathcal{O}(z_g)\,,
\end{align}
thus only the integral $I_{3,\rm slow}^{ik}$ integral contributes to the slow gluon limit in Eq.\,\eqref{eq:Hard34} (note the overall minus sign between $\rm SE_1$ and $\rm V_1$ between Eqs.\,\eqref{eq:hSE1slow} and \eqref{eq:hV1slow}). For $I_{3,\rm slow}$, we find
\begin{align}
    I_{3,\rm slow}^{ik}&=\partial^k_{\Pt}\frac{\Pt^i}{4(\Pt^2+\bar Q^2)}\left[1+\ln\left(\frac{c_0z_gQ_f^2}{z_f\bar Q^2}\right)-\frac{\Pt^2-\bar Q^2}{\Pt^2}\ln\left(1+\frac{\Pt^2}{\bar Q^2}\right)\right]+\mathcal{O}(z_g) \,, \nonumber \\
    &=\frac{\delta^{ik}}{4(\Pt^2+\bar Q^2)}\left[1+\ln\left(\frac{c_0^2z_gQ_f^2}{z_f\bar Q^2}\right)-\frac{\Pt^2-\bar Q^2}{\Pt^2}\ln\left(1+\frac{\Pt^2}{\bar Q^2}\right)\right]\nonumber\\
    &+\frac{\Pt^i\Pt^k}{2(\Pt^2+\bar Q^2)^2}\left[-2-\ln\left(\frac{c_0^2z_gQ_f^2}{z_f\bar Q^2}\right)+\frac{\bar Q^2}{\Pt^2}+\left(1-\frac{\bar Q^2(2\Pt^2+\bar Q^2)}{\Pt^4}\right)\ln\left(1+\frac{\Pt^2}{\bar Q^2}\right)\right]+\mathcal{O}(z_g)\label{eq:kin-slow}\,.
\end{align}
On the other hand, the slow gluon divergence of the sum of the instantaneous terms in Eq.\,\eqref{eq:Hinst-final}, the regular part of $\rm SE_1$ in Eq.\,\eqref{eq:HNLO3-final} and $\rm V_1$ in Eqs.\,\eqref{eq:HNLO4_a_final}-\eqref{eq:HNLO4_b_final} reads
\begin{align}
    &\Hcal^{ij,\lambda=\rmT}_{\rm inst.}(\Pt)+\Hcal^{ij,\lambda=\rmT}_{\rm NLO,3}(\Pt)+\Hcal^{ij,\lambda=\rmT}_{\rm NLO,4}(\Pt)=\alpha_{\rm em}\alpha_se_f^2\deltatwo\times2z_1z_2(z_1^2+z_2^2)h_{\rm LO}^{kj}\nonumber\\
    &\times\frac{(-\alpha_s)N_c}{2\pi}\int_0^{z_1}\frac{\der z_g}{z_g}\left\{\frac{\delta^{ik}}{4(\Pt^2+\bar Q^2)}\left[\ln\left(\frac{z_g}{z_1z_2}\right)+1+\frac{\bar Q^2}{\Pt^2}\ln\left(1+\frac{\Pt^2}{\bar Q^2}\right)-1\right]\right.\nonumber\\
    &\left.+\frac{\Pt^i\Pt^k}{2(\Pt^2+\bar Q^2)^2}\left[-1+\frac{\bar Q^2}{\Pt^2}-\ln\left(\frac{z_g}{z_1z_2}\right)-\frac{\bar Q^2(2\Pt^2+\bar Q^2)}{\Pt^4}\ln\left(1+\frac{\Pt^2}{\bar Q^2}\right)\right]+\mathcal{O}(z_g)\right\}\,.\label{eq:slow-from-analytic}
\end{align}
As in the slow gluon subtraction term, the slow gluon divergence in the regular part of the self-energy term cancels with the divergence coming from the $I_{4}^k$ contribution in the vertex correction. The singular piece in Eq.\,\eqref{eq:slow-from-analytic} comes mostly from the $z_g\to0$ limit of the integral $I_{3A}^{ik}$ in the vertex correction $\mathrm{V}_1$, except for the $-1$ term highlighted in red which is the sum of the rapidity divergence of the instantaneous terms and the rapidity divergence associated with the integral $I_{3B}^{ik}$ in the vertex correction. For Eqs.\,\eqref{eq:kin-slow} and \eqref{eq:slow-from-analytic} to cancel each other out leaving no divergence,  one must choose
\begin{equation}
    \frac{z_f}{Q_f^2}=\frac{ec_0^2z_1z_2}{\Pt^2+\bar Q^2}=\frac{ec_0^2}{M_{q\bar q}^2+Q^2}\,.\label{eq:zf-Qf-from-slow}
\end{equation}
Since $z_f/Q_f^2=1/(2k_c^+q^-)$, the scale $k_c^+$ is fixed from the condition of the cancellation of the $z_g\to0$ divergence in the hard factors of virtual diagrams with gluon crossing the shock-wave.

Remarkably, the divergence of the instantaneous terms and the $I_{3B}$ term of $\rm V_1$ combine in such a way that the same value of $k_c^+$ cancels the slow gluon divergence in both tensor structure $\delta^{ik}$ and $\Pt^i\Pt^k$.
This is an important consistency check of the TMD factorization in the back-to-back limit for transversely polarized virtual photons. Also, it is crucial that the same value of $Q_f$ is obtained for both longitudinally and transversely polarized photons \cite{Caucal:2023nci}.

\section{Supplemental Material 5: Useful integrals}

This supplemental material sketches the calculation of the four master integrals necessary for the analytic calculation of the NLO hard factor from the vertex correction with virtual gluon crossing the shock-wave. These integrals are far from being trivial, and they can be computed following these steps:
\begin{enumerate}
    \item Write the NLO wave-function in momentum space using either of the two representations (see e.g. Appendix E in \cite{Caucal:2021ent}):
\begin{align}
        \frac{\bar QK_1\left(\bar Q\sqrt{\ut^2+\omega\rt^2}\right)}{\sqrt{\ut^2+\omega\rt^2}}&=\int\frac{\der^2\ltone}{(2\pi)}\int\frac{\der^2\lttwo}{(2\pi)}\frac{e^{i\ltone\cdot\ut}e^{i\lttwo\cdot\rt}}{\lttwo^2+\omega(\ltone+\bar Q^2)}\label{eq:inst-K1}\,,\\
        \ut^k\frac{\bar QK_1\left(\bar Q\sqrt{\ut^2+\omega\rt^2}\right)}{\sqrt{\ut^2+\omega\rt^2}}&=(-\rt^l)\int\frac{\der^2\ltone}{(2\pi)}\int\frac{\der^2\lttwo}{(2\pi)}\frac{\ltone^k\lttwo^l e^{i\ltone\cdot\ut}e^{i\lttwo\cdot\rt}}{(\ltone^2+\bar Q^2)[\lttwo^2+\omega(\ltone+\bar Q^2)]}\,.\label{eq:K1-rep}
\end{align}

\item Write the WW kernel in momentum space using
\begin{equation}
    \frac{(\rt^i+\ut^i)}{(\rt+\ut)^2}=(-i)\int\frac{\der^2\ltthre}{(2\pi)}\frac{\ltthre^i}{\ltthre^2}e^{i\ltthre\cdot(\rt+\ut)}\,.
\end{equation}

\item Perform the integral over the coordinate $\ut$ conjugate to $\Pt$. Since $\ut$ now only appears inside phases, this integral yields a simple $\delta$-function, which can be used to simplify the internal momentum integrals over $\ltone$ and $\lttwo$. 

\item Finally, perform the $\rt$ integral and then the $\ltthre$ integral in polar coordinates.

\end{enumerate}

\subsection{The master integral $I_1^{ik}$}

The first integral $I_1^{ik}$ is defined as
\begin{equation}
        I_1^{ik}=\int\frac{\der^2\ut}{(2\pi)}\int\frac{\der^2\rt}{(2\pi)}e^{-i\Pt\cdot\ut}\rt^i\frac{\rt\times(\rt+\ut)}{\rt^2(\rt+\ut)^2}[\ut^k-\omega\rt^k]\frac{\bar QK_1\left(\bar Q\sqrt{\ut^2+\omega\rt^2}\right)}{\sqrt{\ut^2+\omega\rt^2}}\,.
\end{equation}
We decompose $I_1^{ki}$ into two pieces:
\begin{align}
    I_1^{ik}=I_{1A}^{ik}-\omega I_{1B}^{ik}\,.
    \label{eq:I1-decomp}
\end{align}
The integral $I_{1A}^{ik}$ reads
\begin{align}
    I_{1A}^{ik}&=\int\frac{\der^2\ut}{(2\pi)}\int\frac{\der^2\rt}{(2\pi)}e^{-i\Pt\cdot\ut}\rt^i\ut^k\frac{\rt\times(\rt+\ut)}{\rt^2(\rt+\ut)^2}\frac{\bar QK_1\left(\bar Q\sqrt{\ut^2+\omega\rt^2}\right)}{\sqrt{\ut^2+\omega\rt^2}} \nonumber \\
    &=\epsilon^{ab}\partial^k_{\Pt}\int\frac{\der^2\ltthre}{(2\pi)}\frac{\ltthre^b}{\ltthre^2}\int\frac{\der^2\rt}{(2\pi)}\frac{\rt^i\rt^a}{\rt^2}e^{i\ltthre\rt}K_0(\Delta r_\perp) \nonumber \\
    &=\epsilon^{ib}\partial^k_{\Pt}\int\frac{\der^2\ltthre}{(2\pi)}\frac{\ltthre^b}{2\ltthre^4}\ln\left(1+\frac{\ltthre^2}{\Delta^2}\right)\,,
\end{align}
where $\Delta^2=\omega((\ltthre-\Pt)^2+\bar Q^2)$ depends on $\ltthre$. Using the polar integral (for $x<1$)
\begin{align}
    \int_0^{2\pi}\frac{\der\theta}{(2\pi)}\cos(\theta)\ln\left(1-x\cos(\theta)\right)=\frac{-1+\sqrt{1-x^2}}{x}\,,
\end{align}
we end up with
\begin{align}
    I_{1A}^{ik}&=\epsilon^{ib}\partial^k_{\Pt}\frac{\Pt^b}{\Pt^2}\left\{\frac{\ln(1+\omega)}{4}-\frac{\Pt^2\ln(\omega)}{4(\Pt^2+\bar Q^2)}+\frac{1}{4\omega}\ln\left(\frac{(1+\omega)(\Pt^2+\bar Q^2)}{\Pt^2+(1+\omega)\bar Q^2}\right)-\frac{\bar Q^2}{4(\Pt^2+\bar Q^2)}\ln\left(\frac{\Pt^2+(1+\omega)\bar Q^2}{\bar Q^2}\right)\right\}\,.
\end{align}
The final result for $I_{1A}^{ki}$ is
\begin{align}
    I_{1A}^{ik}&=\frac{\epsilon^{ik}}{4\Pt^2}\left\{\ln(1+\omega)-\frac{\Pt^2\ln(\omega)}{(\Pt^2+\bar Q^2)}+\frac{1}{\omega}\ln\left(\frac{(1+\omega)(\Pt^2+\bar Q^2)}{\Pt^2+(1+\omega)\bar Q^2}\right)-\frac{\bar Q^2}{\Pt^2+\bar Q^2}\ln\left(\frac{\Pt^2+(1+\omega)\bar Q^2}{\bar Q^2}\right)\right\}\nonumber\\
    &+\frac{\epsilon^{ib}\Pt^k\Pt^b}{2\Pt^4}\left\{\frac{\Pt^4\ln(\omega)}{(\Pt^2+\bar Q^2)^2}-\ln(1+\omega)+\frac{\bar Q^2(2\Pt^2+\bar Q^2)}{(\Pt^2+\bar Q^2)^2}\ln\left(\frac{\Pt^2+(1+\omega)\bar Q^2}{\bar Q^2}\right)-\frac{1}{\omega}\ln\left(\frac{(1+\omega)(\Pt^2+\bar Q^2)}{\Pt^2+(1+\omega)\bar Q^2}\right)\right\}\,.
    \label{eq:I1A}
\end{align}

The $B$ term is given by
\begin{align}
    I_{1B}^{ik}&=\int\frac{\der^2\ut}{(2\pi)}\int\frac{\der^2\rt}{(2\pi)}e^{-i\Pt\cdot\ut}\rt^i\rt^k\frac{\rt\times(\rt+\ut)}{\rt^2(\rt+\ut)^2}\frac{\bar QK_1\left(\bar Q\sqrt{\ut^2+\omega\rt^2}\right)}{\sqrt{\ut^2+\omega\rt^2}} \nonumber \\
    &=-i\epsilon^{ab}\int\frac{\der^2\ltthre}{(2\pi)}\frac{\ltthre^b}{\ltthre^2}\int\frac{\der^2\rt}{(2\pi)}\frac{\rt^i\rt^k\rt^a}{\rt^2}e^{i\ltthre\rt}K_0(\Delta r_\perp) \nonumber \\
    &=-\epsilon^{ib}\int\frac{\der^2\ltthre}{(2\pi)}\frac{\ltthre^b}{\ltthre^3} \ \partial^i_{\ltthre}\partial^k_{\ltthre}\left.\left[\frac{\ltthre^a}{2\ltthre^2}\ln\left(1+\frac{\ltthre^2}{\Delta^2}\right)\right]\right|_{\Delta=\omega((\ltthre-\Pt)^2+\bar Q^2)} \nonumber  \\
    &=-\epsilon^{kb}\int\frac{\der^2\ltthre}{(2\pi)}\frac{\ltthre^b\ltthre^i}{\ltthre^6(\ltthre^2+\Delta^2)}\left[\ltthre^2-(\ltthre^2+\Delta^2)\ln\left(1+\frac{\ltthre^2}{\Delta^2}\right)\right]+(k\leftrightarrow i)\,.
\end{align}
Note the nice cancellation of the logarithmic divergence in $\ltthre\to0$ between the two terms inside the square bracket.
This time, one needs the following polar integrals: 
\begin{align}
    \int_0^{2\pi}\frac{\der\theta}{(2\pi)}\cos^2(\theta)\ln\left(1-x\cos(\theta)\right)&=\frac{1}{4x^2}\left[-2+x^2+2\sqrt{1-x^2}+2x^2\ln\left(\frac{1+\sqrt{1-x^2}}{2}\right)\right] \,, \\
    \int_0^{2\pi}\frac{\der\theta}{(2\pi)}\ln\left(1-x\cos(\theta)\right)&=\ln\left(\frac{1+\sqrt{1-x^2}}{2}\right)\,.
\end{align}
The second one enables to get the polar integral with a $\sin^2(\theta)$ prefactor instead of $\cos^2(\theta)$ since $\sin^2(\theta)+\cos^2(\theta)=1$.
We find, after performing the tedious $\ltthre$ integral:
\begin{align}
    I^{ik}_{1B}&=\frac{1}{4}\left[\epsilon^{kb}\frac{\Pt^i\Pt^b}{\Pt^2}+\epsilon^{ib}\frac{\Pt^k\Pt^b}{\Pt^2}\right]\times\left\{\frac{1}{\omega(\Pt^2+\bar Q^2)}-\frac{\ln(\omega)\Pt^2}{(\Pt^2+\bar Q^2)^2}+\frac{\ln(1+\omega)}{\Pt^2}\right.\nonumber\\
    &\left.+\frac{1}{\omega^2\Pt^2}\ln\left(\frac{\Pt^2+(1+\omega)\bar Q^2}{(1+\omega)(\Pt^2+\bar Q^2)}\right)-\frac{\bar Q^2(2\Pt^2+\bar Q^2)}{\Pt^2(\Pt^2+\bar Q^2)^2}\ln\left(\frac{\Pt^2+(1+\omega)\bar Q^2}{\bar Q^2}\right)\right\}\,.
    \label{eq:I1B}
\end{align}

\subsection{The master integral $I^{k}_2$}

The second master integral is defined by
\begin{equation}
    I_2^{k}=\int\frac{\der^2\ut}{(2\pi)}\int\frac{\der^2\rt}{(2\pi)}e^{-i\Pt\cdot\ut}\frac{\rt\times(\rt+\ut)}{\rt^2(\rt+\ut)^2}[\ut^k-\omega\rt^k]\frac{\bar QK_1\left(\bar Q\sqrt{\ut^2+\omega\rt^2}\right)}{\sqrt{\ut^2+\omega\rt^2}}\,.
\end{equation}
We follow the same strategy and we decompose $I_2^k$ into two integrals
\begin{align}
    I^{k}_2&=I^k_{2A}-\omega I^k_{2B}\,.
\end{align}
Let us first compute the $A$ term.
\begin{align}
    I_{2A}^{k}&=\int\frac{\der^2\ut}{(2\pi)}\int\frac{\der^2\rt}{(2\pi)}e^{-i\Pt\cdot\ut}\ut^k\frac{\rt\times(\rt+\ut)}{\rt^2(\rt+\ut)^2}\frac{\bar QK_1\left(\bar Q\sqrt{\ut^2+\omega\rt^2}\right)}{\sqrt{\ut^2+\omega\rt^2}} \nonumber \\
    &=\partial^k_{\Pt}\int\frac{\der^2\ltthre}{(2\pi)}\frac{\ltthre^b\epsilon^{ab}}{\ltthre^2}\int\frac{\der^2\rt}{(2\pi)}\frac{\rt^a}{\rt^2}e^{i\ltthre\cdot\rt}K_0(\Delta \rt) \nonumber \\
    &=i\partial^k_{\Pt}\int\frac{\der^2\ltthre}{(2\pi)}\frac{\ltthre^a\ltthre^b\epsilon^{ab}}{2\ltthre^4}\ln\left(1+\frac{\ltthre^2}{\Delta^2}\right) \nonumber  \\
    &=0\,.
\end{align}
We could have anticipated this result given the dependence on $\rt\times \ut$ of the $\rt$ integrand which is odd.

Regarding the $B$ term, we have already computed it when we calculated $I^{ik}_{1A}$ since
\begin{align}
    (i\partial^i_{\Pt})I_{2B}^k=I_{1A}^{ki}\,.
\end{align}
Therefore
\begin{align}
    I_{2B}^{k}&=\int\frac{\der^2\ut}{(2\pi)}\int\frac{\der^2\rt}{(2\pi)}e^{-i\Pt\cdot\ut}\rt^k\frac{\rt\times(\rt+\ut)}{\rt^2(\rt+\ut)^2}\frac{\bar QK_1\left(\bar Q\sqrt{\ut^2+\omega\rt^2}\right)}{\sqrt{\ut^2+\omega\rt^2}} \nonumber  \\
    &=(-i)\frac{\epsilon^{kb}\Pt^b}{4\Pt^2}\left\{\ln(1+\omega)-\frac{\Pt^2\ln(\omega)}{\Pt^2+\bar Q^2}+\frac{1}{\omega}\ln\left(\frac{(1+\omega)(\Pt^2+\bar Q^2)}{\Pt^2+(1+\omega)\bar Q^2}\right)-\frac{\bar Q^2}{\Pt^2+\bar Q^2}\ln\left(\frac{\Pt^2+(1+\omega)\bar Q^2}{\bar Q^2}\right)\right\}\,.
\end{align}

\subsection{The master integral $I_3^{ik}$}

The third master integral is given by 
\begin{equation}
    I_3^{ik}=\int\frac{\der^2\ut}{(2\pi)}\int\frac{\der^2\rt}{(2\pi)}e^{-i\Pt\cdot\ut}\rt^i\frac{\rt\cdot(\rt+\ut)}{\rt^2(\rt+\ut)^2}[\ut^k-\omega\rt^k]\frac{\bar QK_1\left(\bar Q\sqrt{\ut^2+\omega\rt^2}\right)}{\sqrt{\ut^2+\omega\rt^2}}\,.
\end{equation}
We expect this master integral to be the most difficult one, since many simplifications that we have done for $I_1^{ik}$ coming from the antisymmetry of the $\epsilon$ tensor cannot be used here. As usual, we write
\begin{align}
    I_3^{ik}=I_{3A}^{ik}-\omega I_{3B}^{ik}\,.
    \label{eq:I3-decomp}
\end{align}

We start with the $A$ term. It gives
\begin{align}
    I_{3A}^{ik}&=\int\frac{\der^2\ut}{(2\pi)}\int\frac{\der^2\rt}{(2\pi)}e^{-i\Pt\cdot\ut}\rt^i\ut^k\frac{\rt\cdot(\rt+\ut)}{\rt^2(\rt+\ut)^2}\frac{\bar QK_1\left(\bar Q\sqrt{\ut^2+\omega\rt^2}\right)}{\sqrt{\ut^2+\omega\rt^2}} \nonumber  \\
    &=\partial^k_{\Pt}\int\frac{\der^2\ltthre}{(2\pi)}\frac{\ltthre^a}{\ltthre^2}\int\frac{\der^2\rt}{(2\pi)}\frac{\rt^i\rt^a}{\rt^2}e^{i\ltthre\rt}K_0(\Delta r_\perp) \nonumber \\
    &=\partial^k_{\Pt}\left\{-\int\frac{\der^2\ltthre}{(2\pi)}\frac{\ltthre^i}{2\ltthre^4}\ln\left(1+\frac{\ltthre^2}{\Delta^2}\right)+\int\frac{\der^2\ltthre}{(2\pi)}\frac{\ltthre^i}{\ltthre^2(\ltthre^2+\Delta^2)}\right\} \nonumber \\
    &=\partial^k_{\Pt}\left\{-\frac{\Pt^i}{4\Pt^2}\left[\ln(1+\omega)-\frac{\Pt^2\ln(\omega)}{\Pt^2+\bar Q^2}+\frac{1}{\omega}\ln\left(\frac{(1+\omega)(\Pt^2+\bar Q^2)}{\Pt^2+(1+\omega)\bar Q^2}\right)-\frac{\bar Q^2}{\Pt^2+\bar Q^2}\ln\left(\frac{\Pt^2+(1+\omega)\bar Q^2}{\bar Q^2}\right)\right]\right.\nonumber\\
    &\left.+\frac{\Pt^i}{2\Pt^2\omega}\ln\left(\frac{(1+\omega)(\Pt^2+\bar Q^2)}{\Pt^2+(1+\omega)\bar Q^2}\right)\right\} \nonumber \\
    &=\partial^k_{\Pt}\frac{(-\Pt^i)}{4\Pt^2}\left\{\ln(1+\omega)-\frac{\Pt^2\ln(\omega)}{\Pt^2+\bar Q^2}-\frac{1}{\omega}\ln\left(\frac{(1+\omega)(\Pt^2+\bar Q^2)}{\Pt^2+(1+\omega)\bar Q^2}\right)-\frac{\bar Q^2}{\Pt^2+\bar Q^2}\ln\left(\frac{\Pt^2+(1+\omega)\bar Q^2}{\bar Q^2}\right)\right\}\,.
\end{align}
In the end, the $\partial^k_{\Pt}$ derivative gives the following result
\begin{align}
    I_{3A}^{ik}&=\frac{\delta^{ik}}{4\Pt^2}\left\{-\ln(1+\omega)+\frac{\Pt^2\ln(\omega)}{\Pt^2+\bar Q^2}+\frac{1}{\omega}\ln\left(\frac{(1+\omega)(\Pt^2+\bar Q^2)}{\Pt^2+(1+\omega)\bar Q^2}\right)+\frac{\bar Q^2}{\Pt^2+\bar Q^2}\ln\left(\frac{\Pt^2+(1+\omega)\bar Q^2}{\bar Q^2}\right)\right\}\nonumber\\
    &+\frac{\Pt^i\Pt^k}{2\Pt^4}\left\{\frac{2\Pt^2\bar Q^2}{(\Pt^2+\bar Q^2)(\Pt^2+(1+\omega)\bar Q^2)}-\frac{\Pt^4\ln(\omega)}{(\Pt^2+\bar Q^2)^2}+\ln(1+\omega)\right.\nonumber\\
    &\left.-\frac{\bar Q^2(2\Pt^2+\bar Q^2)}{(\Pt^2+\bar Q^2)^2}\ln\left(\frac{\Pt^2+(1+\omega)\bar Q^2}{\bar Q^2}\right)-\frac{1}{\omega}\ln\left(\frac{(1+\omega)(\Pt^2+\bar Q^2)}{\Pt^2+(1+\omega)\bar Q^2}\right)\right\}\,.\label{eq:I3ikA}
\end{align}

The B term is computed as follows
\begin{align}
    I_{3B}^{ik}&=\int\frac{\der^2\ut}{(2\pi)}\int\frac{\der^2\rt}{(2\pi)}e^{-i\Pt\cdot\ut}\rt^i\rt^k\frac{\rt\cdot(\rt+\ut)}{\rt^2(\rt+\ut)^2}\frac{\bar QK_1\left(\bar Q\sqrt{\ut^2+\omega\rt^2}\right)}{\sqrt{\ut^2+\omega\rt^2}} \nonumber \\
    &=(-i)\int\frac{\der^2\ltthre}{(2\pi)}\frac{\ltthre^a}{\ltthre^2}\int\frac{\der^2\rt}{(2\pi)}\frac{\rt^i\rt^k\rt^a}{\rt^2}e^{i\ltthre\rt}K_0(\Delta r_\perp) \nonumber \\
    &=-\int\frac{\der^2\ltthre}{(2\pi)}\frac{\ltthre^a}{\ltthre^2} \ \partial^i_{\ltthre}\partial^k_{\ltthre}\left[\frac{\ltthre^a}{2\ltthre^2}\ln\left(1+\frac{\ltthre^2}{\Delta^2}\right)\right] \nonumber \\
    &=2\int\frac{\der^2\ltthre}{(2\pi)}\frac{\ltthre^i\ltthre^k}{\ltthre^6(\ltthre^2+\Delta^2)}\left[\ltthre^2-(\ltthre^2+\Delta^2)\ln\left(1+\frac{\ltthre^2}{\Delta^2}\right)+\frac{\ltthre^4}{\ltthre^2+\Delta^2}\right] \nonumber\\
    &-\delta^{ik}\int\frac{\der^2\ltthre}{(2\pi)}\frac{1}{\ltthre^4(\ltthre^2+\Delta^2)}\left[\ltthre^2-(\ltthre^2+\Delta^2)\ln\left(1+\frac{\ltthre^2}{\Delta^2}\right)\right]\,.
\end{align}
Using previous results for the $\ltthre$ integrals, we find
\begin{align}
    I_{3B}^{ik}&=\frac{\delta^{ik}}{4}\left\{\frac{1}{\omega(\Pt^2+\bar Q^2)}-\frac{\ln(\omega)\Pt^2}{(\Pt^2+\bar Q^2)^2}+\frac{\ln(1+\omega)}{\Pt^2}\right.\nonumber\\
    &\left.-\frac{1}{\omega^2\Pt^2}\ln\left(\frac{\Pt^2+(1+\omega)\bar Q^2}{(1+\omega)(\Pt^2+\bar Q^2)}\right)-\frac{\bar Q^2(2\Pt^2+\bar Q^2)}{\Pt^2(\Pt^2+\bar Q^2)^2}\ln\left(\frac{\Pt^2+(1+\omega)\bar Q^2}{\bar Q^2}\right)\right\}\nonumber\\
    &+\frac{\Pt^i\Pt^k}{2\Pt^2}\left\{\frac{2}{\omega(\Pt^2+(1+\omega)\bar Q^2)}-\frac{1}{\omega(\Pt^2+\bar Q^2)}-\frac{\ln(1+\omega)}{\Pt^2}+\frac{\Pt^2\ln(\omega)}{(\Pt^2+\bar Q^2)^2}\right.\nonumber\\
    &\left.+\frac{\bar Q^2(2\Pt^2+\bar Q^2)}{\Pt^2(\Pt^2+\bar Q^2)^2}\ln\left(\frac{\Pt^2+(1+\omega)\bar Q^2}{\bar Q^2}\right)-\frac{1}{\omega^2\Pt^2}\ln\left(\frac{(1+\omega)(\Pt^2+\bar Q^2)}{\Pt^2+(1+\omega)\bar Q^2}\right)\right\}\,.\label{eq:I3ikB}
\end{align}

\subsection{The master integral $I^{k}_4$}

Finally, the master integral $I^{k}_4$ is defined as
\begin{equation}
    I_4^{k}=\int\frac{\der^2\ut}{(2\pi)}\int\frac{\der^2\rt}{(2\pi)}e^{-i\Pt\cdot\ut}\frac{\rt\cdot(\rt+\ut)}{\rt^2(\rt+\ut)^2}[\ut^k-\omega\rt^k]\frac{\bar QK_1\left(\bar Q\sqrt{\ut^2+\omega\rt^2}\right)}{\sqrt{\ut^2+\omega\rt^2}}  \,.
\end{equation}
After decomposing $I_4^k$ into two integrals
\begin{align}
    I^{k}_4&=I^k_{4A}-\omega I^k_{4B}\,,
\end{align}
we compute the $A$ term using Eq.\,\eqref{eq:K1-rep}:
\begin{align}
    I_{4A}^{k}&=\int\frac{\der^2\ut}{(2\pi)}\int\frac{\der^2\rt}{(2\pi)}e^{-i\Pt\cdot\ut}\ut^k\frac{\rt\cdot(\rt+\ut)}{\rt^2(\rt+\ut)^2}\frac{\bar QK_1\left(\bar Q\sqrt{\ut^2+\omega\rt^2}\right)}{\sqrt{\ut^2+\omega\rt^2}} \nonumber  \\
    &=-\int\frac{\der^2\ltthre}{(2\pi)}\frac{\ltthre^a(\Pt^k-\ltthre^k)}{\ltthre^2[(\ltthre-\Pt)^2+\bar Q^2]}\int\frac{\der^2\rt}{(2\pi)}\frac{\Delta\rt^a}{r_\perp}e^{i\ltthre\cdot\rt}K_1(\Delta \rt) \nonumber \\
    &=i\int\frac{\der^2\ltthre}{(2\pi)}\frac{\ltthre^k}{(\ltthre^2+\bar Q^2)[(\ltthre+\Pt)^2+\omega(\ltthre^2+\bar Q^2)]}\,.
\end{align}
After performing the remaining $\ltthre$ integral, we find
\begin{equation}
    I_{4A}^k=\frac{i\Pt^k\bar Q^2}{2\Pt^2(\Pt^2+\bar Q^2)}\left[\ln\left(\frac{\Pt^2+(1+\omega)\bar Q^2}{(1+\omega)\bar Q^2}\right)+\frac{\Pt^2}{\bar Q^2}\ln\left(\frac{\omega}{1+\omega}\right)\right]\,.
\end{equation}
The B term is obtained from $I_{3A}^{ik}$ using 
\begin{align}
    (i\partial^i_{\Pt})I^k_{4B}=I^{ki}_{3A}=I^{ik}_{3A}\,,
\end{align}
we find
\begin{align}
    I_{4B}^{k}&=\int\frac{\der^2\ut}{(2\pi)}\int\frac{\der^2\rt}{(2\pi)}e^{-i\Pt\cdot\ut}\rt^k\frac{\rt\cdot(\rt+\ut)}{\rt^2(\rt+\ut)^2}\frac{\bar QK_1\left(\bar Q\sqrt{\ut^2+\omega\rt^2}\right)}{\sqrt{\ut^2+\omega\rt^2}} \nonumber \\
    &=\frac{i\Pt^i}{4\Pt^2}\left[\ln(1+\omega)-\frac{\Pt^2\ln(\omega)}{\Pt^2+\bar Q^2}-\frac{1}{\omega}\ln\left(\frac{(1+\omega)(\Pt^2+\bar Q^2)}{\Pt^2+(1+\omega)\bar Q^2}\right)-\frac{\bar Q^2}{\Pt^2+\bar Q^2}\ln\left(\frac{\Pt^2+(1+\omega)\bar Q^2}{\bar Q^2}\right)\right]\,.
\end{align}

 \end{widetext}

\end{document}